\documentclass[aps,prx,superscriptaddress,twocolumn,nofootinbib,nobibnotes,floatfix,showpacs,reprint,longbibliography]{revtex4-2}
\usepackage{siunitx}
\usepackage[utf8]{inputenc}
\usepackage[T1]{fontenc}
\usepackage{lmodern}
\usepackage{orcidlink}
\usepackage{amsmath,amsfonts,amssymb,amsthm}
\usepackage{bm} 
\usepackage{xcolor} 
\usepackage{xfrac} 
\usepackage{enumitem} 
\usepackage[normalem]{ulem} 
\usepackage{soul} 
\usepackage{subfigure}
\usepackage{multirow}
\usepackage{dcolumn} 
\usepackage{graphicx} 
\usepackage{hyperref} 
\hypersetup{
    unicode={true},
    colorlinks={true},
    linkcolor={blue},
    citecolor={blue},
    urlcolor={blue}
}
\usepackage[sectionbib]{bibunits}

\newtheoremstyle{def_break}
    {}{}
    {}{}
    {\bfseries}{.}
    {\newline}{}
\theoremstyle{def_break}
\newtheorem*{lemma}{Lemma}
\theoremstyle{remark}

\DeclareFontEncoding{LS1}{}{}
\DeclareFontSubstitution{LS1}{stix2}{m}{n}
\DeclareMathAlphabet{\mathscr}{LS1}{stix2scr}{m}{n}
\SetMathAlphabet{\mathscr}{bold}{LS1}{stix2scr}{b}{n}

\makeatletter
\def\maketitle{
\@author@finish
\title@column\titleblock@produce
\suppressfloats[t]}
\makeatother

\newcommand{\nocontentsline}[3]{}
\let\origcontentsline\addcontentsline
\newcommand\stoptoc{\let\addcontentsline\nocontentsline}
\newcommand\resumetoc{\let\addcontentsline\origcontentsline}

\makeatletter
\newcounter{savesection}
\newcounter{apdxsection}
\renewcommand\appendix{\par
  \setcounter{savesection}{\value{section}}%
  \setcounter{section}{\value{apdxsection}}%
  \setcounter{subsection}{0}%
  \gdef\thesubsection{\@Alph\c@section}}
\newcommand\unappendix{\par
  \setcounter{apdxsection}{\value{subsection}}%
  \setcounter{section}{\value{savesection}}%
  \setcounter{subsection}{0}%
  \gdef\thesubsection{\@arabic\c@section}}
\makeatother

\newcommand{\ssg}{\pmb{\mathscr{G}}}
\newcommand{\gspin}{\pmb{\mathbb{S}}}
\newcommand{\gspinnorm}{\bm{\mathcal{S}}}
\newcommand{\gspace}{\pmb{\mathbb{X}}}
\newcommand{\gspacenorm}{\mathbf{X}}
\newcommand{\sog}{\bm{\mathfrak{G}}_{\bm{\mathrm{S}}}}
\newcommand{\ntsg}{\pmb{\mathbb{G}}_{\bm{\mathrm{S}}}}
\newcommand{\ntsgspin}{\pmb{\widetilde{\mathcal{S}}}}
\newcommand{\csg}{\mathbf{G}}
\newcommand{\hsg}{\mathbf{H}}
\newcommand{\hsgone}{\mathbf{H_1}}
\newcommand{\hsgtwo}{\mathbf{H_2}}
\newcommand{\hsgthree}{\mathbf{H_3}}
\newcommand{\hsgfour}{\mathbf{H_4}}
\newcommand{\hsgfive}{\mathbf{H_5}}
\newcommand{\inv}{\mathcal{P}}
\newcommand{\trs}{\mathcal{T}}

\begin{document}

\title{The Fundamental Lemma of Altermagnetism: Emergence of Alterferrimagnetism}

\author{Chanchal K. Barman$^\bigstar$}
\email{arckb2@gmail.com}
\affiliation{Dipartimento di Fisica, Universit\`a di Cagliari, Cittadella Universitaria, Monserrato (CA) 09042, Italy}

\author{Bishal Das$^\bigstar$}
\email{dasbishal98@gmail.com}
\affiliation{Department of Physics, Indian Institute of Technology Bombay, Mumbai 400076, India}

\author{Alessio Filippetti}
\affiliation{Dipartimento di Fisica, Universit\`a di Cagliari, Cittadella Universitaria, Monserrato (CA) 09042, Italy}

\author{Aftab Alam}
\affiliation{Department of Physics, Indian Institute of Technology Bombay, Mumbai 400076, India}

\author{Fabio Bernardini}
\affiliation{Dipartimento di Fisica, Universit\`a di Cagliari, Cittadella Universitaria, Monserrato (CA) 09042, Italy}

\def\thefootnote{$\bigstar$}\footnotetext{These authors contributed equally to this work.}
\def\thefootnote{\arabic{footnote}}

\begin{abstract} 
    Recent years have seen a proliferation in investigations on \textit{Altermagnetism} due to its exciting prospects both from an applications perspective and theoretical standpoint. Traditionally, altermagnets are distinguished from collinear antiferromagnets using the central concept of \textit{halving} subgroups within the spin space group formalism. In this work, we propose the Fundamental Lemma of Altermagnetism (FLAM) deriving the exact conditions required for the existence of altermagnetic phase in a magnetic material on the basis of \textit{site-symmetry} groups and \textit{halving} subgroups for a given crystallographic space group. The spin group formalism further clubs ferrimagnetism with ferromagnetism since the same-spin and opposite-spin sublattices lose their meaning in the presence of multiple magnetic species. As a consequence of FLAM, we further propose a class of fully compensated ferrimagnets, termed as \textit{Alterferrimagnets} (AFiMs), which can show alternating momentum-dependent spin-polarized non-relativistic electronic bands within the first Brillouin zone. We show that alterferrimagnetism is a generalization of traditional collinear altermagnetism where  multiple magnetic species are allowed to coexist forming fully compensated magnetic-sublattices, each with individual up-spin and down-spin sublattices.    
\end{abstract}
\date{\today}
\maketitle
\stoptoc

\section{Introduction} \label{intro}

Since N\'eel's \cite{Neel1932a, Neel1932b, Neel1934, Neel1936a, Neel1936b, Neel1948, Neel1949, Neel1952, Smart1955, Neel1971} seminal formulation of antiferromagnetism and ferrimagnetism within the framework of local Weiss molecular field, the study of magnetism in condensed matter has uncovered a remarkable diversity of phenomena and substantially broadened our understanding of the possible magnetic orders in crystalline solids. Magnetic order has long been classified into ferromagnetic (FM), antiferromagnetic (AFM), and ferrimagnetic (FiM) phases, defined by spontaneous magnetization patterns that may be collinear or non-collinear. More recently, understanding of these different magnetic phases of matter has seen a paradigm shift with the advent of altermagnetism (AM). Initially regarded as an unconventional collinear AFM \cite{Noda2016, Suzuki2017, Okugawa2018, Hayami2019, Naka2019, Smejkal2020, Yuan2020, Hayami2020, Naka2021, Mazin2021, Yuan2021, Naka2022, Liu2022}, AM was later established by \v{S}mejkal and collaborators \cite{Smejkal2022a, Smejkal2022b} as a distinct collinear magnetic phase, on the basis of spin space groups (SSGs) \cite{Litvin1974, Litvin1977}, which are generalizations of the conventional magnetic space groups (MSGs). The SSG classification leads to three distinct types of collinear magnetic phases in magnetically ordered materials \cite{Smejkal2022a, Smejkal2022b}. Ferromagnets with a spontaneous net magnetization break the time-reversal ($\trs$) symmetry producing Zeeman-type spin splitting of electronic bands. These fall under the Type-I SSG category. In antiferromagnets, the staggered arrangement of local magnetic moments gives rise to two distinct spin sublattices yielding zero net magnetization and spin-degenerate electronic bands. The spin-degeneracy is enforced by the mapping of the opposite-spin sublattices into one another either by spatial inversion ($\inv$) or non-trivial lattice translation ($t$) which preserves $\trs$ symmetry and constitute the Type-II SSG category. By contrast, altermagnets exhibit momentum-dependent non-relativistic spin splitting despite zero net magnetization, with opposite-spin sublattices connected by symmorphic/non-symmorphic rotations or reflections rather than by $\inv$ or $t$ leading to the third distinct Type-III SSG category. Importantly, the spin splitting in altermagnets exists even in the absence of spin-orbit coupling (SOC) and is unlike Rashba~\cite{Rashba1955a, Rashba1955b, *[{English translation available in \href{https://iopscience.iop.org/article/10.1088/1367-2630/17/5/050202/data}{supplementary data} of: }] Bihlmayer2015, Bychkov1984}~or~Dresselhaus-type~\cite{Dresselhaus1955} spin splitting which typically is a SOC-induced phenomenon and arises from broken $\inv$ symmetry in a material. The present excitement surrounding AM is due to the amalgamation of properties exclusive to AFM and FM respectively in a single material which opens up interesting avenues in spintronics. 

Ferrimagnetism, conventionally described as N\'eel's `imperfect antiferromagnetism' \cite{Neel1948, Neel1949, Neel1952, Smart1955}, arises when inequivalent sublattices with opposite spin orientations fail to compensate each other, either due to different oxidation states of the same species or due to distinct magnetic ions. Such inequivalence precludes any opposite-spin sublattice transformations, restricting the classification again to the Type-I SSG category, identical to FM order \cite{Smejkal2022a, Smejkal2022b}.  While most FiMs are uncompensated and thus display FM-like isotropic spin splitting, there exists a special subclass, termed Luttinger-compensated ferrimagnets \cite{Mazin2022, Wei2024, Kawamura2024, Guo2025, Liu2025}, in which sublattice moments are fully compensated. According to Luttinger's theorem \cite{KohnLuttinger1960, LuttingerWard1960, Luttinger1960}, the volume enclosed by the Fermi surface of a many-particle system is conserved independent of interactions and is proportional to the particle density. This constrains the net magnetization of a FiM semiconductor/insulator to an integer and consequently Luttinger-compensated FiMs can realize strictly zero net magnetization. Despite this similarity of zero net magnetization to AFM or AM phases, the inequivalence of opposite-spin sublattices in general forces Luttinger-compensated FiMs to fall into Type-I SSG \cite{Smejkal2022a, Smejkal2022b} with isotropic spin splitting. This natural assimilation of FiM and FM obscures whether such `imperfect antiferromagnets' could ever host anisotropic spin splitting akin to altermagnets, i.e., whether there exists a distinct class of ferrimagnets which may be considered a more general version of altermagnets. In particular, while uncompensated FiMs are excluded by their finite magnetization, the possibility of fully compensated FiMs realizing AM-like spin-polarization remains unexplored.

Several known collinear antiferromagnets in well-studied classes of materials such as rutiles, perovskites, chalcogenides, pnictides etc., have now been revealed to be altermagnets \cite{Noda2016, Suzuki2017, Okugawa2018, Hayami2019, Naka2019, Smejkal2020, Yuan2020, Hayami2020, Naka2021, Mazin2021, Yuan2021, Naka2022, Liu2022, Smejkal2022a, Smejkal2022b, Yuan2023, Bai2024, Fender2025}. Among these, perovskites form a particularly important class, long known for their versatile functional properties ranging from metal-insulator transitions and orbital ordering to photovoltaics, multiferroics etc \cite{Imada1998, Tokura2000, Maekawa2004, Cheong2007, Green2014, Tilley2016, Spaldin2019}. Discovering altermagnetism in such class of materials shows their potential for being a platform hosting interplay of various functional properties \cite{Smejkal2024, Naka2025, Duan2025, Bernardini2025}. Further it was found that several orthorhombic perovskites (with the cooperative Jahn-Teller effect and GdFeO$_3$-type distortion) hosts a variety of collinear AFM order (A, C and G-type) which are now understood to be AM semiconductors/insulators \cite{Okugawa2018, Fender2025, Naka2025}. Interestingly, all the three fully compensated staggered collinear spin arrangement, with propagation vector $\vec{\bm{q}}=0$, possible in a \textit{primitive} orthorhombic lattice leads to the AM phase and this observation warrants more attention \cite{Okugawa2018}. Nevertheless, these subtle aspects have been largely overlooked in favour of relativistic manifestations of AM order, such as the anomalous Hall effect (AHE) which dominates current experimental focus \cite{Naka2022, Naka2025}.

In this work, we begin by relating the abstract concept of \textit{halving} subgroups of the parent crystallographic space group of a material to the more tangible and practical concept of \textit{site-symmetry} groups corresponding to the Wyckoff positions of a magnetically ordered crystal in the form of a mathematical lemma using discrete group theory in the context of crystals. Our proposed ``Fundamental Lemma of Altermagnetism'', which introduces the notion of \textit{compatibility} of a Wyckoff position with the existence of altermagnetic phase in a material, generalizes the concept of \textit{halving} subgroups which have been extensively used for the SSG classification of magnetic materials with collinear order \cite{Smejkal2022a, Smejkal2022b} and can additionally predict in a very simple way whether a material with $\vec{\bm{q}}=0$ collinear spin arrangement is capable of showing altermagnetism or not based only on its crystallographic space group. We additionally pose the question what would happen if there are more than one magnetic species that are fully compensated individually and occupy these \textit{compatible} Wyckoff positions, i.e., a ferrimagnet composed of interpenetrating altermagnets. This leads us to the realization of a distinct class of fully compensated ferrimagnets that exhibit symmetry-protected alternating spin splitting analogous to altermagnets. We term this class of ferrimagnetism as ``Alterferrimagnetism'' (AFiM), which may be viewed as a natural generalization of the AM phase.
This paper is organized as follows. Section~\ref{AM} summarizes the existing theoretical frameworks of \textit{altermagnetism}. Section~\ref{Wyckoff} highlights the role of Wyckoff positions and their corresponding \textit{site-symmetry} groups in dictating the SSG symmetries finally leading to the fundamental lemma of altermagnetism explained in Section~\ref{FLAM}. Building on these principles, Section~\ref{AFiM} demonstrates that the family of metal-oxide iron phosphates MFePO$_5$ (M=Fe,Cu,Ni,Co), which are fully compensated ferrimagnets\cite{Warner1992, Touaiher1994, Khayati2000, Khayati2001}, exhibit alternating anisotropic spin splitting with interpenetrating AM-type M and Fe sublattices, thereby establishing them as AFiMs. Beyond their fundamental significance, AFiMs provide a materials platform where multiple magnetic species can be exploited to achieve novel forms of tunability for practical spintronic applications. Since different magnetic species may respond differently to external fields---an advantage that the current FM, AFM or AM materials do not possess---AFiMs can become the ideal choice where directionality and versatility is required alongside effective spin-polarization and ultra-fast switching dynamics.

\section{Spin group classification of Ferrimagnets revisited}

Ferrimagnets are generally categorized under Type-I SSG as mentioned in Sec.~\ref{intro}. In this section, we propose a distinct class of FiMs that can exhibit alternating spin splitting, analogous to altermagnets. This arises because the magnetic species in such FiMs occupy Wyckoff positions \textit{compatible} with the AM phase, as will be detailed in the following sections.

\subsection{Altermagnetism: A brief recap} \label{AM}

In a magnetic material, the interacting magnetic moments on atoms, primarily arising due to spins of unpaired electrons, give rise to a magnetic order with a definite spin arrangement. The spin arrangement dictates how spins in a magnetic material are spatially arranged, essentially providing a mapping between the vector spaces--- \textit{spin-space} and \textit{real-space} (encompassing both position and momentum spaces). Magnetic space groups are used to describe the symmetries of magnetic materials which leave the spin arrangement in a crystal invariant. Within the MSG framework, the symmetry transformations act simultaneously on the basis atoms in the \textit{real-space} as well as on the spins in the \textit{spin-space} and is evidently a consequence of SOC, meaning both of these vector spaces no longer remain independent of each other. But in materials with weak or negligible SOC, \textit{real-space} and \textit{spin-space} may be regarded as decoupled, giving rise to certain symmetry transformations which act separately on these two vector spaces and also keeps the spin arrangement in a crystal invariant. This leads to a generalization of the MSGs where the symmetry transformations acting on \textit{real-space} and \textit{spin-space} are not necessarily the same resulting in the spin space groups. Since SSGs also keep the spin arrangement in a crystal invariant, MSGs form subgroups of SSGs. Henceforth, these groups shall be denoted by boldfaced letters.

A spin space group $\ssg$ may be defined as the \textit{internal} direct product of a spin-only group $\sog$ and a non-trivial spin group $\ntsg$, i.e., $\ssg = \sog \times \ntsg$ with group elements denoted by $[\,\mathfrak{R}\,||\,R\,]$. The transformations $\mathfrak{R}$ and $R$ act exclusively on the \textit{spin-space} and \textit{real-space} respectively. The spin-only group $\sog$ contains transformations of the \textit{spin-space} alone. By contrast, the non-trivial spin group $\ntsg$ contains transformations of both \textit{spin-space} and \textit{real-space} except those present in $\sog$, with the exception of SSG identity $[\,\mathfrak{I}\,||\,\mathbb{I}\,]$ which is common to both (more details available in Appendix~\ref{ssg}). Here, $\mathfrak{I}$ is the \textit{spin-space} identity and $\mathbb{I}$ is the \textit{real-space} identity.
For collinear spin arrangements, the group of underlying \textit{spin-space} transformations of $\sog$ is given by $\gspinnorm = \bm{\mathcal{C_{\infty}}} \cup \overline{\mathfrak{C}}_2\,\bm{\mathcal{C_{\infty}}}$ \cite{Kitz1965, Litvin1974, Smejkal2022a, Smejkal2022b}. Here, $\bm{\mathcal{C_{\infty}}}$ is the special orthogonal group of rotations about the common axis of all collinear magnetic moments in the \textit{spin-space} which keeps any collinear spin arrangement invariant. The \textit{spin-space} transformation $\overline{\mathfrak{C}}_2 = \trs\mathfrak{C}_2$ is a 2-fold rotation in the \textit{spin-space} about an axis perpendicular to the common axis followed by time-reversal. Both $\mathfrak{C}_2$ and $\trs$ act as spin inversion flipping the collinear spins, and their combined action keeps the spin arrangement invariant. Time-reversal additionally acts on the \textit{real-space} since $\trs$ also reverses momenta although $[\,\overline{\mathfrak{C}}_2\,||\,\trs\,] \in \sog$ where $\overline{\mathfrak{C}}_2 \in \gspinnorm$. Then for collinear spin arrangement, there arises two possibilities of choosing $\ntsgspin$, the group of underlying \textit{spin-space} transformations of $\ntsg$. The first possibility is that all \textit{spin-space} transformations keep the collinear spin arrangement invariant irrespective of the presence of any \textit{real-space} transformations, implying $\gspinnorm$ constitutes the group of all \textit{spin-space} transformations and $\ntsgspin = \{\mathfrak{I}\}$ is the trivial group. Here, there is no operation which can explicitly flip spins leading to only same-spin sublattice transformations that keep the collinear spin arrangement invariant. Thus $\ntsg$ can be constructed only from \textit{isomorphic} one-coset decompositions of \textit{spin-space} transformation $\ntsgspin = \{\mathfrak{I}\}$ and crystallographic space group $\csg$ defining Type-I SSGs which explicitly break $\trs$. These transformations produce a momentum-independent isotropic spin splitting throughout the Brillouin zone, characteristic of the ferromagnetic phase. The other possibility is to consider a \textit{spin-space} operation that can flip spins, yet keep the spin arrangement invariant. The natural choice is either $\mathfrak{C}_2$ or $\trs$ and we choose $\mathfrak{C}_2$ since it exclusively acts on the \textit{spin-space} unlike $\trs$, although both choices ultimately lead to the same $\ssg$. But it is important to note that flipping spins in a collinear spin arrangement can never keep it invariant unless the \textit{spin-space} transformation acts in conjunction with some \textit{real-space} transformation such that the original spin arrangement is recovered. Thus, $\ntsg$ can be constructed from $\ntsgspin = \{\mathfrak{I},\mathfrak{C}_2\}$ and $\csg$ (see eq.~\eqref{eqA2}). Interestingly, there are two different \textit{isomorphic} coset decompositions possible because now $\ntsgspin$ is no longer the trivial group. For an \textit{isomorphic} one-coset decomposition, the resulting $\ntsg$ and consequently the full SSG $\ssg$ preserves $\trs$ defining Type-II SSGs. These transformations produce fully spin-degenerate dispersion throughout the Brillouin zone, characteristic of the antiferromagnetic phase. But for an \textit{isomorphic} two-coset decomposition, there arises two further possibilities. Let us assume that $\csg$ contains a subgroup $\hsg$ of \textit{index} 2, also called a \textit{halving} subgroup, with the coset decomposition $\csg = \hsg \cup R\,\hsg = \hsg \cup (\csg - \hsg)$, where $R$ is a coset-representative such that $R \notin \hsg$ but $R \in \csg$. In general, $\csg$ can either be centrosymmetric $(\inv \in \csg)$ or non-centrosymmetric $(\inv \notin \csg)$ and depending on $R$, the two-coset decomposition leads to two distinct $\ntsg$. For either $R = \inv$ (\textit{real-space} inversion) or $R = t$ (translation), the resulting $\ntsg$ reduces to that of the Type-II SSGs. But if $R \neq \inv$ and $R \neq t$ ($R$ could be non-symmorphic, i.e, may contain $t$ combined with rotations or reflections), $\trs$ is explicitly broken in the resulting $\ssg$, yielding Type-III SSGs. These transformations produce momentum-dependent anisotropic spin splitting with alternating spin-polarized dispersions in the Brillouin zone, characteristic of the altermagnetic phase \cite{Smejkal2022a, Smejkal2022b}. Thus, the \textit{halving} subgroup of a parent crystallographic space group plays a central role in determining the AM phase.

The first and foremost condition for obtaining an AM phase (Type-III SSG) is the existence of a \textit{halving} subgroup $\hsg$, which by construction is a \textit{normal} subgroup, such that an \textit{isomorphic} two-coset decomposition is possible and which generates all the same-spin sublattice transformations. The coset $\csg - \hsg$ will then generate all the opposite-spin sublattice transformations. In general, there could be more than one \textit{halving} subgroup and if $\csg$ does not have any \textit{halving} subgroup, then AM phase cannot be achieved in such crystallographic space groups. Any space group with the underlying point group as $\bm{1}$ ($\bm{C_{1}}$), $\bm{3}$ ($\bm{C_{3}}$) or $\bm{23}$ ($\bm{T}$) falls under this category. As noted earlier, AM phase will be achieved in a centrosymmetric space group \textit{iff} $\inv,t \notin (\csg-\hsg)$ or equivalently $\inv,t \in \hsg$. The centrosymmetric space groups with the underlying point group as $\bm{\overline{1}}$ ($\bm{C_{i}}$), $\bm{\overline{3}}$ ($\bm{C_{3i}}$ or $\bm{S_{6}}$) or $\bm{m\overline{3}}$ ($\bm{T_{h}}$) do not satisfy these conditions and hence cannot yield AM phase. For non-centrosymmetric space groups, the condition $\inv \notin (\csg-\hsg)$ is satisfied trivially and hence AM phase may be achieved \textit{iff} $t \notin (\csg-\hsg)$. This final condition is trivially satisfied for space groups with \textit{primitive} Bravais lattices, which are denoted starting with $P$ in the Hermann-Mauguin notation \cite{IUCr_volA2016}. But for \textit{centered} Bravais lattices, i.e., space groups beginning with $F$ (face-centered), $I$ (body-centered), $R$ (rhombohedral) or $A,B,C$ (base-centered) in the Hermann-Mauguin notation \cite{IUCr_volA2016}, $t \notin (\csg-\hsg)$ may not be satisfied due to the existence of non-trivial \textit{centering} lattice translations. This shall be discussed further in Sec.~\ref{FLAM}. In the following discussions, we shall be primarily using the Sch\"onflies notation \cite{IUCr_volA2016} for denoting the crystallographic space groups and point groups since it is compact and occasionally use the Hermann-Mauguin notation \cite{IUCr_volA2016} when the space group needs to be distinguished explicitly.

\subsection{Wyckoff positions and \textit{site-symmetry} groups} \label{Wyckoff}

Before examining how the \textit{site-symmetry} groups corresponding to Wyckoff positions of a crystallographic space group play an important role in determining SSG symmetries and \textit{compatibility} of Wyckoff positions with the AM phase, it shall be instructive to recapitulate a few important definitions from crystallography \cite{IUCr_volA2016}. Each and every point in a crystalline lattice is subject to the symmetry operations of its crystallographic space group $\csg$. These operations either map the point onto itself or onto another symmetry-equivalent point within the lattice. The \textit{set} of distinct lattice points to which a particular lattice point is mapped to by all the symmetry operations in $\csg$ is known as the \textit{crystallographic orbit} of that lattice point. The \textit{site-symmetry} group $\mathbf{W}$ of a lattice point is a group that leaves that particular point invariant in the lattice, i.e., maps the point onto itself (up to a trivial lattice translation) and is necessarily \textit{isomorphic} to a subgroup of the underlying point group corresponding to $\csg$. This means no fractional lattice translations can exist in $\mathbf{W}$. It is important to note here that two distinct lattice points may have the same $\mathbf{W}$ but different \textit{crystallographic orbits}. This leads to the notion of Wyckoff positions which are classes of \textit{crystallographic orbits}. Within a given Wyckoff position, the \textit{site-symmetry} group $\mathbf{W}$ of every lattice point is conjugate to each other, and the number of such distinct lattice points in the \textit{crystallographic orbit} defines the multiplicity of that Wyckoff position. 
Different Wyckoff positions may share the same $\mathbf{W}$ and hence have the same multiplicity but they will always correspond to different \textit{crystallographic orbits}. While lattice points within a single Wyckoff position are symmetry-equivalent, points belonging to different Wyckoff positions cannot be related by any space group operations. In some cases, translations which belong to the \textit{affine normalizer} of the crystallographic space group may connect points across different Wyckoff positions and a detailed discussion on this lies beyond the scope of the present work.
For space groups with \textit{primitive} Bravais lattices, the Wyckoff positions with multiplicity 2 have \textit{site-symmetry} groups \textit{isomorphic} to a \textit{halving} subgroup of the parent crystallographic space group, denoted by $\mathbf{W}\cong\hsg$. But for space groups with \textit{centered} Bravais lattices, the multiplicity of such Wyckoff positions will be twice of the number of \textit{centering} translations because the \textit{centering} translations can never be in $\mathbf{W}$.

\subsection{The Fundamental Lemma of Altermagnetism} \label{FLAM}

Now, we shall state the Fundamental Lemma of Altermagnetism, which defines the notion of \textit{compatibility} with the altermagnetic phase, along with its proof.
\begin{lemma}
    Let $\csg$ be a crystallographic space group and $\hsg$ be one of its subgroups with \textit{index} 2 such that $\inv,t\in\hsg$ if $\inv,t\in\csg$. Further, let $\mathbf{W} \subseteq \csg$ be the \textit{site-symmetry} group corresponding to a Wyckoff position in $\csg$. Then that Wyckoff position is said to be \textit{compatible} with the altermagnetic phase \textit{iff} its multiplicity is \textit{even} and $\mathbf{W} \subseteq \hsg \triangleleft \csg$.
\end{lemma}

\begin{proof}
    $ $\newline
    The altermagnetic phase in a material with crystallographic space group $\csg$ requires the existence of a proper subgroup $\hsg \subset \csg$ such that $\csg$ can be partitioned into two disjoint subsets (cosets), i.e., $\csg = \hsg \,\cup\, (\mathbf{G-H})$ from which a non-trivial spin group $\ntsg$ may be constructed (also see Appendix~\ref{ssg}).
    Then the \textit{index} of $\hsg$ in $\csg$ is 2 and \textit{index} 2 subgroups are \textit{normal} by construction, denoted by $\hsg \triangleleft \csg$.
    This \textit{halving} subgroup $\hsg$ itself forms one of the subsets and consists of the same-spin sublattice transformations.
    The other subset $\mathbf{G-H}$ forms the only coset of $\hsg$ in $\csg$ and consists of opposite-spin sublattice transformations. Crucially, $\mathbf{G-H}$ does not form a group since $\hsg \,\cap\, (\mathbf{G-H})=\varnothing$ and $\mathbb{I}\in\hsg$, where $\mathbb{I}$ is the group identity. Further, if $\inv,t\in\csg$, then one must necessarily have $\inv,t\in\hsg$ for the altermagnetic phase to exist, else extra symmetry conditions due to the spin-only group $\sog$ reduce the above coset decomposition to that of the antiferromagnetic phase \cite{Smejkal2022a, Smejkal2022b}. To ensure existence of the altermagnetic phase, it shall be henceforth assumed that $\hsg$ satisfies all the above necessary conditions because not all $\hsg$ lead to the altermagnetic phase.
    
    \noindent In a crystal, the atoms forming the basis of a crystalline lattice occupy different Wyckoff positions within the lattice. The \textit{site-symmetry} group associated with each such Wyckoff position will map the individual atoms onto themselves as explained earlier. In the case of a magnetic crystal with collinear staggered order, the number of distinct lattice points in the \textit{crystallographic orbit} defined by the Wyckoff position of a magnetic species must be \textit{even} for perfect compensation so that two distinct spin sublattices can be formed. In other words, the multiplicity of the Wyckoff position must be \textit{even}. This is required for the existence of any compensated magnetic phase, such as the antiferromagnetic or altermagnetic phases. By construction, the \textit{site-symmetry} group associated with the Wyckoff position occupied by a magnetic species always generates the same-spin sublattice transformations. This shall form the basic criteria for determining the existence of altermagnetic phase.

    \noindent For simplicity, let us assume a magnetic material with single magnetic species that occupies a specific Wyckoff position with the \textit{site-symmetry} group $\mathbf{W}$ and multiplicity \textit{even}. A \textit{site-symmetry} group need not be a proper subgroup and thus $\mathbf{W}\subseteq\csg$. It is also important to realize that the conditions $\hsg \triangleleft \csg$ and $\mathbf{W}\subseteq\csg$ do not imply $\mathbf{W} \subseteq \hsg$ and requires an explicit proof. For a distinct lattice point in the \textit{crystallographic orbit} defined by the Wyckoff position, transformations in $\mathbf{W}$ maps the point onto itself. Any other transformations absent in $\mathbf{W}$ maps the point to another lattice point in the same \textit{crystallographic orbit}. If $\mathbf{W}$ contains \textit{all} the same-spin sublattice transformations, then the \textit{site-symmetry} group is \textit{isomorphic} to the \textit{halving} subgroup $(\mathbf{W}\cong\hsg)$, and every magnetic atom in the same \textit{crystallographic orbit} only gets mapped to itself. But in general, this is not to be expected since there may be certain same-spin sublattice transformations absent in $\mathbf{W}$. In such cases, if a distinct lattice point is mapped to another point in the same \textit{crystallographic orbit} and this second lattice point belongs to the same-spin sublattice as the first lattice point, then $\hsg$ can still be constructed from $\mathbf{W}$ and the set of same-spin sublattice transformations absent in $\mathbf{W}$. This yields a minimal constraint on $\mathbf{W}$, namely $\hsg$ must be \textit{constructible} from $\mathbf{W}$, which will also hold for magnetic materials with multiple magnetic species occupying different Wyckoff positions. In general, $\mathbf{W}$ and $\hsg$ may not have group elements in common but they will have \textit{at least} one common element which is the group identity $\mathbb{I}$ because both are subgroups of $\csg$. Based on this, it is easily seen that if $\exists\,w\in\mathbf{W}$ such that $w\notin\hsg$, then this implies $w\in(\csg-\hsg)$ because $\hsg$ is a \textit{halving} subgroup. And since $\hsg \,\cap\, (\mathbf{G-H})=\varnothing$, it will \textit{not} be possible to construct $\hsg$ from $\mathbf{W}$.

\noindent Thus, for a Wyckoff position to be \textit{compatible} with the altermagnetic phase, its corresponding $\mathbf{W}$ can only contain elements already in $\hsg$, i.e., $\mathbf{W}$ must be a proper or improper subgroup of $\hsg$, i.e., $\mathbf{W} \subseteq \hsg$. Together with the condition $\hsg \triangleleft \csg$, one finally obtains $\mathbf{W} \subseteq \hsg \triangleleft \csg$ completing the proof.
\end{proof}

The above FLAM is clearly reflected in the case of verified and proposed altermagnets in the literature. The established altermagnets MnTe \cite{Lee2024, Krempasky2024, Osumi2024} and CrSb \cite{Reimers2024, Zeng2024, Ding2024} belong to the hexagonal NiAs-type structure \cite{Aminoff1923, Furberg1953, Willis1953} with space group $\bm{P6_3/mmc}~(\bm{D_{6h}^{4}})$ and underlying point group $\bm{6/mmm}~(\bm{D_{6h}})$. In these two examples, the magnetic Mn and Cr atoms occupy the 2a Wyckoff position having the \textit{site-symmetry} group $\bm{\overline{3}m}$ $(\bm{D_{3d}})$ with multiplicity 2. This incidentally is a \textit{halving} subgroup of $\bm{D_{6h}}$ and thereby generates the same-spin sublattice transformations. A similar feature appears in the proposed altermagnet GdAlSi, which crystallizes in a body-centered tetragonal structure with space group $\bm{I4_1md}~(\bm{C_{4v}^{11}})$ and underlying point group $\bm{4mm}~(\bm{C_{4v}})$. The Gd atoms occupy 4a Wyckoff position having the \textit{site-symmetry} group $\bm{mm2}$ $(\bm{C_{2v}})$, again a \textit{halving} subgroup of $\bm{C_{4v}}$, with multiplicity 4 \cite{Bobev2005, Nag2024}.
The previous examples are trivial, in the sense that the \textit{site-symmetry} group of the Wyckoff position occupied by the magnetic species is \textit{isomorphic} to the \textit{halving} subgroup of the parent crystallographic space group $(\mathbf{W} \cong \hsg \triangleleft \csg)$. A non-trivial example would be LaMO$_3$ (M=Transition metal) which are orthorhombic perovskites belonging to the space group $\bm{Pnma}~(\bm{D_{2h}^{16}})$, with underlying point group $\bm{mmm}~(\bm{D_{2h}})$. In LaMnO$_3$ (M=Mn), the Mn atoms occupy 4a Wyckoff position \cite{Elemans1971, Norby1995} with the \textit{site-symmetry} group $\bm{\overline{1}}$ $(\bm{C_{i}})$. Although $\bm{C_{i}}$ is not a \textit{halving} subgroup of $\bm{D_{2h}}$, it is a subgroup of $\bm{C_{2h}}$ which in turn is one of the \textit{halving} subgroups of $\bm{D_{2h}}$ containing $\inv$. Crucially, $\bm{C_{i}}$ also contains $\inv$ and hence, opposite-spin sublattices cannot be related by $\inv$. Thus, a compensated spin arrangement in $\bm{Pnma}~(\bm{D_{2h}^{16}})$ where the magnetic species reside at a Wyckoff position with \textit{site-symmetry} group $\bm{C_{i}}$ will be \textit{compatible} with the AM phase. Here, the \textit{site-symmetry} group of the Wyckoff position occupied by the magnetic species is \textit{isomorphic} to a proper subgroup of the \textit{halving} subgroup of the parent crystallographic space group $(\mathbf{W} \subset \hsg \triangleleft \csg)$, in accordance with the Fundamental Lemma of Altermagnetism.

\begin{figure*}[!htb]
    \centering
    \includegraphics[width=\linewidth]{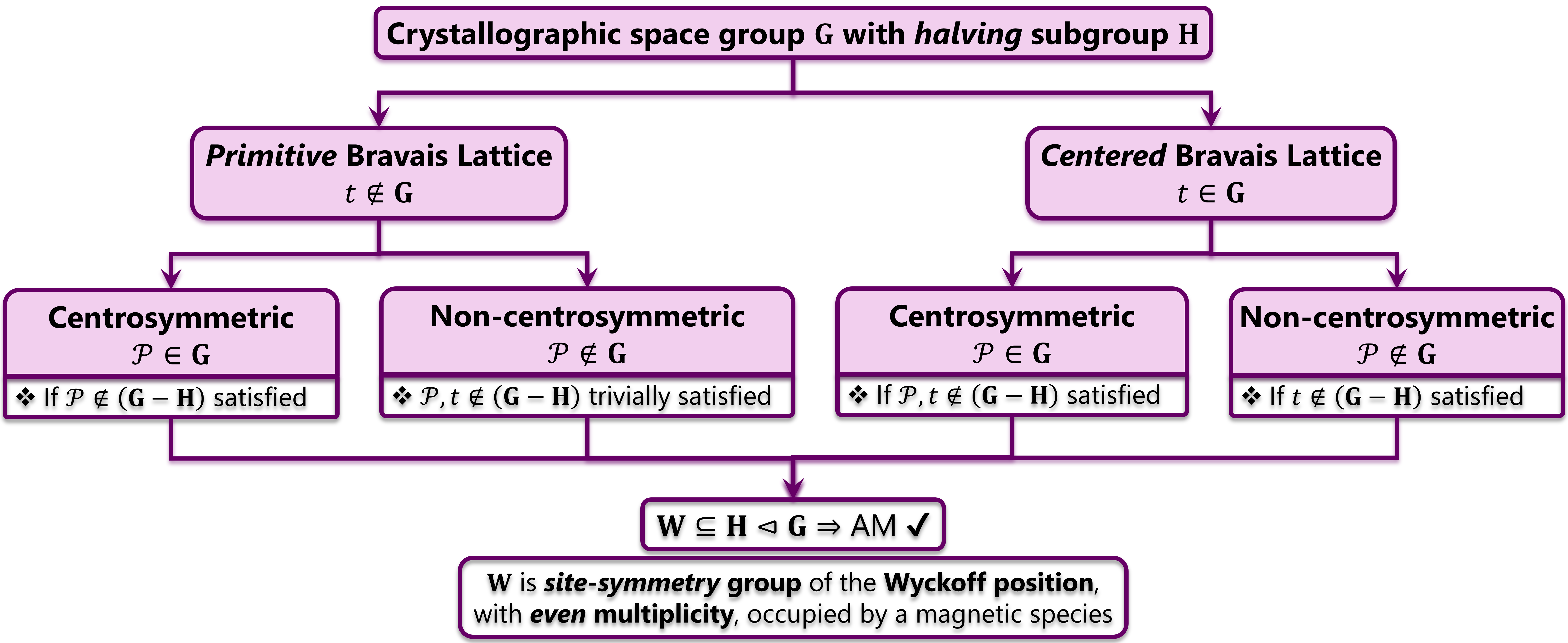}     
    \caption{Hierarchy of conditions required for the existence of altermagnetic (AM) phase in accordance with the fundamental lemma of altermagnetism. For a collinear fully compensated spin arrangement with $\vec{\bm{q}}=0$, if AM phase cannot exist, then antiferromagnetic (AFM) phase will exist. Here, $\mathcal{P}$ is spatial inversion and $t$ is non-trivial \textit{centering} lattice translation.}
    \label{fig1}
\end{figure*}

For a given space group $\csg$ with a \textit{halving} subgroup $\hsg$ and \textit{site-symmetry} group $\mathbf{W}$ of the Wyckoff position occupied by a magnetic species, known examples of altermagnets clearly reveal $\mathbf{W} \subseteq \hsg \triangleleft \csg$ with \textit{even} multiplicity, in accordance with FLAM. Such Wyckoff positions are then referred to as \textit{compatible} with the AM phase, else they are \textit{incompatible} and yield the AFM phase instead. Although \textit{even} multiplicity and $\mathbf{W} \subseteq \hsg \triangleleft \csg$ are necessary conditions, they are not sufficient to completely assert the existence of the AM phase even if magnetic species occupy the \textit{compatible} Wyckoff positions with a compensated spin arrangement. One must additionally take into account the underlying Bravais lattice of $\csg$ and whether $\csg$ is centrosymmetric or not which may instead yield the AFM phase resulting in the following distinct cases---

\begin{enumerate}[wide,left=0pt] 

    \item Space groups with \textit{primitive} Bravais lattices $(t \notin \csg)$: \\
    These space groups, beginning with $P$ in the Hermann-Mauguin notation \cite{IUCr_volA2016}, consist of only trivial ($primtive$) lattice translations.
    \begin{enumerate}[label=\roman*)] 
        \item Centrosymmetric space groups $(\inv \in \csg)$ will have the following two subcases---
        \begin{itemize}
            \item If $\inv \in \mathbf{W}$, then distinct lattice points in the same \textit{crystallographic orbit} can never be mapped by $\inv$. The condition $\inv,t \notin (\mathbf{G-H})$ is satisfied trivially meaning $\hsg$ can always be constructed from $\mathbf{W}$ for any compensated spin arrangement at the corresponding Wyckoff position such that existence of the AM phase is ensured. Wyckoff positions like these will always be \textit{compatible} with the AM phase.
            
            \item If $\inv \notin \mathbf{W}$, then distinct lattice points in the same \textit{crystallographic orbit} will be mapped by $\inv$. Although the condition $t \notin (\mathbf{G-H})$ is satisfied trivially, the AM phase will exist \textit{iff} a $\hsg$ can be constructed from $\mathbf{W}$ and the set of same-spin sublattice transformations absent in $\mathbf{W}$ such that $\inv \notin (\mathbf{G-H})$. Not all compensated spin arrangement at the corresponding Wyckoff position will yield the AM phase. Wyckoff positions like these are not always \textit{compatible} with the AM phase.
        \end{itemize}
        
        \item For non-centrosymmetric space groups $(\inv \notin \csg)$, the condition $\inv,t \notin (\mathbf{G-H})$ is satisfied trivially meaning $\hsg$ can always be constructed from $\mathbf{W}$ for any compensated spin arrangement at the corresponding Wyckoff position such that existence of the AM phase is ensured. Wyckoff positions like these will always be \textit{compatible} with the AM phase.
    \end{enumerate} 
    
    \item Space groups with \textit{centered} Bravais lattices $(t \in \csg)$: \\
    These space groups, beginning with $F$ (face-centered), $I$ (body-centered), $R$ (rhombohedral) or $A,B,C$ (base-centered) in the Hermann-Mauguin notation \cite{IUCr_volA2016}, contain \textit{centering} translations $t$ which are non-trivial lattice translations. These translations can never map a distinct lattice point to itself and hence $t \notin \mathbf{W}$ for any $\mathbf{W} \subseteq \csg$.
    \begin{enumerate}[label=\roman*)] 
        \item Centrosymmetric space groups $(\inv \in \csg)$ will have the following two subcases---
        \begin{itemize}
            \item If $\inv \in \mathbf{W}$, then distinct lattice points in the same \textit{crystallographic orbit} can never be mapped by $\inv$. Although the condition $\inv \notin (\mathbf{G-H})$ is satisfied trivially, the \textit{centering} translations $t$ will map distinct lattice points in the same \textit{crystallographic orbit}. The AM phase will exist \textit{iff} a $\hsg$ can be constructed from $\mathbf{W}$ and the set of same-spin sublattice transformations absent in $\mathbf{W}$ such that $t \notin (\mathbf{G-H})$. Not all compensated spin arrangement at the corresponding Wyckoff position will yield the AM phase. Wyckoff positions like these are not always \textit{compatible} with the AM phase.
            
            \item If $\inv \notin \mathbf{W}$, then distinct lattice points in the same \textit{crystallographic orbit} will be mapped by $\inv$, as well as the \textit{centering} translations $t$. The AM phase will exist \textit{iff} a $\hsg$ can be constructed from $\mathbf{W}$ and the set of same-spin sublattice transformations absent in $\mathbf{W}$ such that $\inv,t \notin (\mathbf{G-H})$. Not all compensated spin arrangement at the corresponding Wyckoff position will yield the AM phase. Wyckoff positions like these are not always \textit{compatible} with the AM phase.
        \end{itemize}
        \item For non-centrosymmetric space groups $(\inv \notin \csg)$, the condition $\inv \notin (\mathbf{G-H})$ is satisfied trivially. But the \textit{centering} translations $t$ will map distinct lattice points in the same \textit{crystallographic orbit}. The AM phase will exist \textit{iff} a $\hsg$ can be constructed from $\mathbf{W}$ and the set of same-spin sublattice transformations absent in $\mathbf{W}$ such that $t \notin (\mathbf{G-H})$. Not all compensated spin arrangement at the corresponding Wyckoff position will yield the AM phase. Wyckoff positions like these are not always \textit{compatible} with the AM phase.
    \end{enumerate} 
    
\end{enumerate} 

\noindent A summary of the above cases is shown in Fig.~\ref{fig1}. Based on these cases, Wyckoff positions that are \textit{compatible} with the AM phase for any compensated spin arrangement shall be referred to as being \textit{fully compatible} with the AM phase. Those Wyckoff positions that are \textit{compatible} with the AM phase for specific compensated spin arrangements shall be referred to as being \textit{partially compatible} with the AM phase. It should be noted that in the above description, the Wyckoff positions of the magnetic species and their corresponding \textit{site-symmetry} groups now play a pivotal role in determining the existence of the AM phase. As an exemplary case, Table~\ref{tab1} illustrates these classifications for the space group $\bm{Pnma}$.

Along with the aforementioned conditions, FLAM thus provides an intuitive and practical way of determining the existence of AM phase in a magnetic material. Simply from the knowledge of the crystallographic space group and the \textit{site-symmetry} groups associated with the Wyckoff positions, which is readily obtained from sample characterization via X-ray diffraction, the existence of the AM phase can be easily inferred.
Our FLAM rigorously generalizes the importance of \textit{site-symmetry} groups in defining altermagnets which has only been preliminarily discussed in the literature \cite{Smejkal2022b, Roig2024}  till now.

\begin{table*}[!htb]
    \begin{ruledtabular}
    \caption {List of space group operations, Wyckoff positions and corresponding \textit{site-symmetry} groups in $\bm{Pnma}$ \cite{Aroyo2006a, Aroyo2006b} and their \textit{compatibility} with the altermagnetic (AM) phase. Pure rotations are denoted by $C_{n\gamma}$ where $\gamma$ is the $n$-fold rotation axis, pure reflections are denoted by $m_{\alpha\beta}$ where $\alpha\beta$ is the reflection plane and the overhead `$\sim$' symbol denotes screw rotations or glide reflections. The corresponding fractional lattice translations are given after `$|$' symbol in direct coordinates. The orthogonal crystallographic axes $a$, $b$ and $c$ are taken to be oriented along the Cartesian $x$, $y$ and $z$ directions respectively. Groups are denoted in boldface by Hermann-Mauguin notations with Sch\"onflies notations given in parentheses \cite{IUCr_volA2016}. The \textit{crystallographic orbits} are given in direct coordinates.}
    \label{tab1}
        \begin{tabular}{c c c c}
            \multicolumn{4}{l}{Crystallographic Space Group $\csg$: $\bm{Pnma}\;(\bm{D_{2h}^{16}})$} \\
            \multicolumn{4}{l}{Crystallographic Point Group: $\bm{mmm}\;(\bm{D_{2h}})$} \\
            \multicolumn{4}{l}{Possible \textit{halving} subgroups $\hsg$: $\bm{P2_{1}/m}\;(\bm{C_{2h}^{2}})$, $\bm{P2_{1}/c}\;(\bm{C_{2h}^{5}})$, $\bm{Pmc2_{1}}\;(\bm{C_{2v}^{2}})$, $\bm{Pmn2_{1}}\;(\bm{C_{2v}^{7}})$, $\bm{Pna2_{1}}\;(\bm{C_{2v}^{9}})$, $\bm{P2_{1}2_{1}2_{1}}\;(\bm{D_{2}^{4}})$} \\
            \multicolumn{4}{l}{Point groups \textit{isomorphic} to \textit{halving} subgroups: $\bm{2/m}\;(\bm{C_{2h}})$, $\bm{mm2}\;(\bm{C_{2v}})$, $\bm{222}\;(\bm{D_{2}})$} \\
            \noalign{\smallskip}
            \hline\hline
            \noalign{\smallskip}
            \multicolumn{4}{c}{\textbf{Space group operations}} \\
            \hline\hline
            \noalign{\smallskip}
            \multicolumn{4}{c}{Identity: $\mathbb{I}$ , Inversion: $\inv$} \\
            \multicolumn{4}{c}{Screw rotations: $\tilde{C}_{2x} = (C_{2x}\,|\,\sfrac{1}{2},\sfrac{1}{2},\sfrac{1}{2})\;,\;\tilde{C}_{2y} = (C_{2y}\,|\,0,\sfrac{1}{2},0)\;,\;\tilde{C}_{2z} = (C_{2z}\,|\,\sfrac{1}{2},0,\sfrac{1}{2})$} \\
            \multicolumn{4}{c}{Glide reflections: $\tilde{m}_{yz} = (m_{yz}\,|\,\sfrac{1}{2},\sfrac{1}{2},\sfrac{1}{2})\;,\;\tilde{m}_{zx} = (m_{zx}\,|\,0,\sfrac{1}{2},0)\;,\;\tilde{m}_{xy} = (m_{xy}\,|\,\sfrac{1}{2},0,\sfrac{1}{2})$} \\
            \noalign{\smallskip}
            \hline\hline
            \noalign{\smallskip}
            \multirow{2}{*}{\textbf{\shortstack{Wyckoff \\ position}}} &  \multirow{2}{*}{\textbf{\shortstack{\textit{Site-symmetry} \\ group $\mathbf{W}$}}} & \multirow{2}{*}{\textit{\textbf{Crystallographic orbit}}} & \multirow{2}{*}{\textbf{AM-\textit{compatibility}}} \\
            & & & \\
            \noalign{\smallskip}
            \hline\hline
            & & & \\
            \multirow{2}{*}{8d} & \multirow{2}{*}{$\{\mathbb{I}\} \cong \bm{1}\,(\bm{C_1}$)} & \multirow{2}{*}{\shortstack{(\text{x},\text{y},\text{z}) , (-\text{x}+\sfrac{1}{2},-\text{y},\text{z}+\sfrac{1}{2}) , (-\text{x},\text{y}+\sfrac{1}{2},-\text{z}) , (\text{x}+\sfrac{1}{2},-\text{y}+\sfrac{1}{2},-\text{z}+\sfrac{1}{2}) , \\ (-\text{x},-\text{y},-\text{z}) , (\text{x}+\sfrac{1}{2},\text{y},-\text{z}+\sfrac{1}{2}) , (\text{x},-\text{y}+\sfrac{1}{2},\text{z}) , (-\text{x}+\sfrac{1}{2},\text{y}+\sfrac{1}{2},\text{z}+\sfrac{1}{2})}} & \multirow{2}{*}{\textit{Partial}} \\
            & & & \\
            & & & \\
            \multirow{2}{*}{4c} & \multirow{2}{*}{$\{\mathbb{I}, \tilde{m}_{zx}\} \cong \bm{m}\,(\bm{C_s})$} & \multirow{2}{*}{\shortstack{(\text{x},\sfrac{1}{4},\text{z}) , (-\text{x}+\sfrac{1}{2},\sfrac{3}{4},\text{z}+\sfrac{1}{2}) , \\ (-\text{x},\sfrac{3}{4},-\text{z}) , (\text{x}+\sfrac{1}{2},\sfrac{1}{4},-\text{z}+\sfrac{1}{2})}} & \multirow{2}{*}{\textit{Partial}} \\
            & & & \\
            & & & \\
            4b & $\{\mathbb{I}, \inv\} \cong \bm{\overline{1}}\,(\bm{C_i})$ & (0,0,\sfrac{1}{2}) , (\sfrac{1}{2},0,0) , (0,\sfrac{1}{2},\sfrac{1}{2}) , (\sfrac{1}{2},\sfrac{1}{2},0) & \textit{Full} \\
            & & & \\
            4a & $\{\mathbb{I}, \inv\} \cong \bm{\overline{1}}\,(\bm{C_i})$ & (0,0,0) , (\sfrac{1}{2},0,\sfrac{1}{2}) , (0,\sfrac{1}{2},0) , (\sfrac{1}{2},\sfrac{1}{2},\sfrac{1}{2}) & \textit{Full} \\
            & & & \\
        \end{tabular}
    \end{ruledtabular}
\end{table*}

\subsection{Momentum-dependent spin-splitting in Altermagnets}

The defining signature of the AM phase is the presence of momentum-dependent anisotropic spin-splitting of electronic bands. Since our aim is to understand symmetries responsible for such spin-split dispersion in the momentum $(\vec{k})$ space, it suffices to consider only the point group symmetry operations without any loss of generality, i.e., screw operations reduce to simple rotations and glide reflections reduce to simple reflections. The \textit{real-space} transformations are provided by the \textit{isomorphic} point group of the crystallographic space group $\csg$ whereas the \textit{spin-space} transformations are provided by underlying $\ntsgspin$ of $\ntsg$ such that $\ntsgspin = \{\mathfrak{I},\mathfrak{C}_2\}$ for the AM phase (see Sec.~\ref{AM}) \cite{Smejkal2022a, Smejkal2022b}. Since AMs show momentum-dependent spin-splitting, it is imperative to find the subset of symmetries that enforces spin degeneracies. This can only arise from transformations that exchange atoms belonging to opposite-spin sublattice. Considering a generic crystal momentum $\vec{k} = (k_x,k_y,k_z)$, the action of $R \in \csg - \hsg$ leads to $\vec{k}' = R\,\vec{k}$. Labeling the energy eigenvalues by $\varepsilon_{j}(\sigma,\vec{k})$, with $j$ the band index and $\sigma$ the spin, the action of opposite-spin sublattice transformations takes the form:
\begin{equation}
\text{\small{$\left[\mathfrak{C}_2\;||\;R\right]\;\varepsilon_{j}(\sigma,\vec{k}) = \varepsilon_{j}(-\sigma,\vec{k}') \Rightarrow \varepsilon_{j}(\sigma,\vec{k}) = \varepsilon_{j}(-\sigma,\vec{k}')$}}
    \label{eq1}
\end{equation}
If $R$ leaves $\vec{k}$ invariant up to a reciprocal lattice vector $\vec{g}$, i.e., $\vec{k}' = \vec{k} + \vec{g}$ implying $R$ belongs to the little group $(\mathcal{L}_{\vec{k}})$ of $\vec{k}$, then Eq.\,\eqref{eq1} enforces spin-degeneracy of energy bands. Else the energy bands become spin-split along momentum directions related by $R$ leading to alternating spin polarization, hallmark of the AM phase.

\section{Alterferrimagnetism --- A generalization of Altermagnetism} \label{AFiM}

Based on the current understanding and existing literature on the AM phase, the above discussions should apply only to materials with a single magnetic species. It has been mentioned in Sec.~\ref{intro} that materials with multiple magnetic species are incapable of showing AFM and AM phases since both these phases require vanishing net magnetization, and in general the magnetic moment of different species do not necessarily fully compensate each other. Although Luttinger-compensated FiMs exhibit a vanishing net magnetization enforced by symmetry, they still lack opposite-spin sublattice transformations and hence display isotropic spin-splitting. This raises an important question: what happens if the multiple magnetic species in a ferrimagnet are each fully compensated individually, as in antiferromagnets or altermagnets ? In such FiMs, the magnetic moments of different species are generally unequal; however, each magnetic species forms its own collinear staggered spin arrangement, resulting in interpenetrating, fully compensated magnetic sublattices.
Consequently, same-spin and opposite-spin sublattice transformations still remain well defined within each magnetic-sublattices. If even one of the magnetic species in such FiMs occupies a Wyckoff position that is \textit{compatible} with the AM phase, an altermagnetic-like phase can, in principle, emerge in a ferrimagnetic material. This is because each magnetic species resides on a distinct Wyckoff position whose associated \textit{site-symmetry} group generates the same-spin sublattice transformations, while opposite-spin sublattice transformations may relate distinct lattice points within the same \textit{crystallographic orbit}.
The non-trivial spin group $\ntsg$ is then constructed as the \textit{external} direct product of the group of \textit{spin-space} transformations $\ntsgspin = \{\mathfrak{I},\mathfrak{C}_2\}$ and the parent crystallographic space group $\csg$ (see Sec.~\ref{AM}), analogous to the AM phase \cite{Smejkal2022a, Smejkal2022b}. 
When more than one magnetic species occupy Wyckoff positions that are individually \textit{compatible} with the AM phase, the resulting $\ntsg$ will be reduced to the \textit{external} direct product of $\ntsgspin$ and some subgroup of $\csg$. This reduction arises because there is a possibility that certain opposite-spin sublattice transformations for a particular type of magnetic species could become the same-spin sublattice transformations for another type of magnetic species (or vice-versa).
Such transformations therefore fail to distinguish same-spin and opposite-spin sublattices in a \textit{global} sense and cannot be retained as symmetries of the combined spin space group. The \textit{real-space} part of the same-spin or opposite-spin sublattice transformations, \textit{common} to all the different types of magnetic species, can at most be \textit{isomorphic} to $\csg$ and will form a proper/improper subgroup of $\csg$ with lower symmetry in general. The full SSG $\ssg$ would then simply be given by the \textit{internal} direct product of spin-only group $\sog$ and the reduced non-trivial spin group $\ntsg$. FiMs belonging to this category therefore constitute a natural generalization of the AM phase and can be classified under type-III SSG. This distinguishes them from conventional FiMs, which fall under type-I SSG.
We propose to call such special class of fully compensated FiMs as \textit{Alterferrimagnets} (AFiMs), which will exhibit momentum-dependent anisotropic spin-splitting akin to altermagnets. We emphasize that our proposal is formulated in terms of the Wyckoff positions and their associated \textit{site-symmetry} groups for a given crystallographic space group. This framework is more general and naturally incorporates the central concept of \textit{halving} subgroups within the broader perspective of altermagnetism.
In this section, we explicitly explain the phenomenon of AFiMs in more detail through an example of orthorhombic material family MFePO$_5$ (M=Fe,Cu,Ni,Co). To best of our knowledge, there are no current reports of AFiMs in literature primarily because FiMs are inherently considered to fall under type-I SSG, as also outlined in Sec.~\ref{intro}.

\begin{figure}[!htb]
    \centering
    \includegraphics[width=\linewidth]{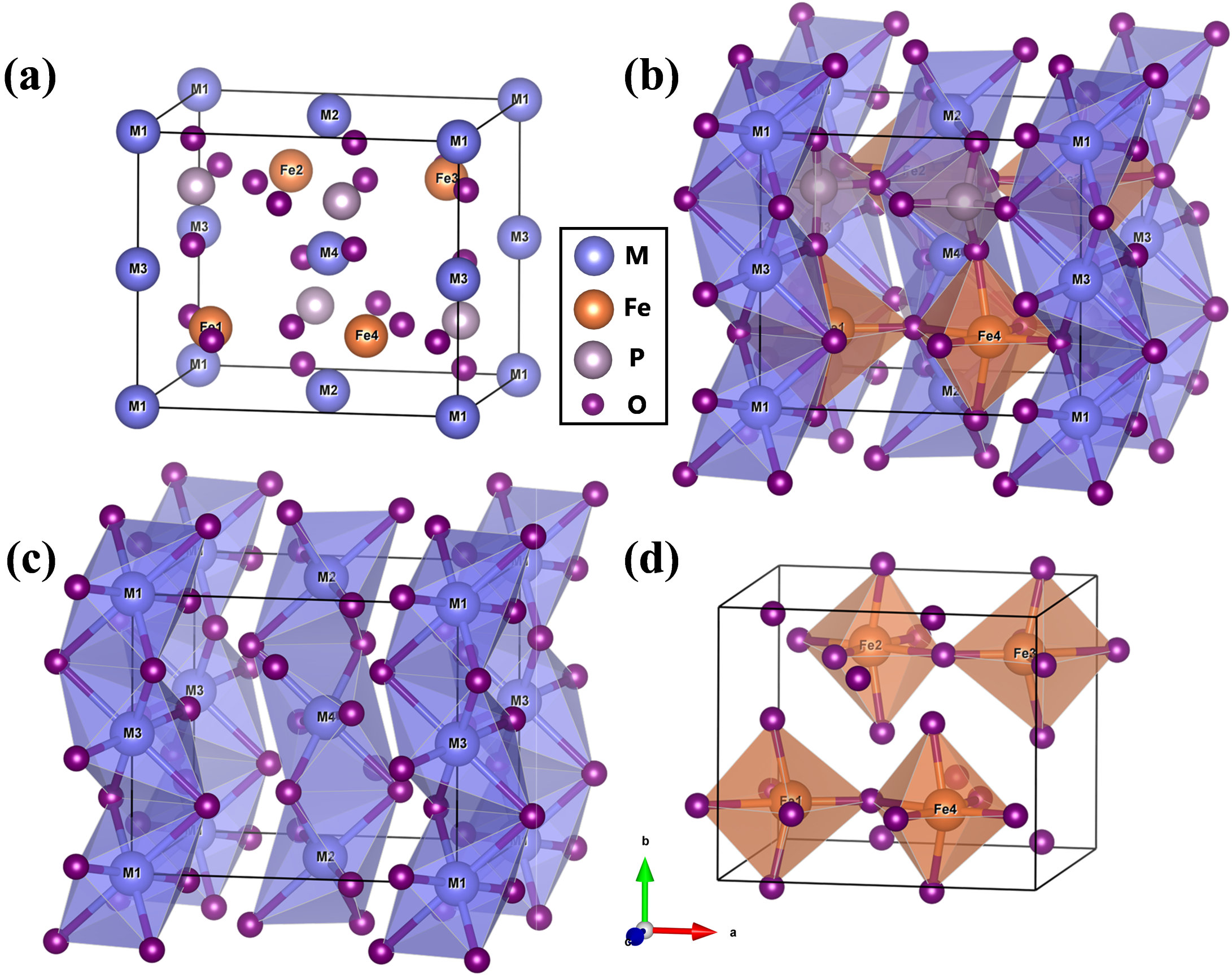}     
    \caption{(a) Crystal structure of MFePO$_5$ (M=Fe,Cu,Ni,Co). (b) The coordination polyhedra formed by the different atoms within the crystal structure. (c) The M sublattice shown along with the coordination octahedra with O atoms. The Fe atoms and their coordination octahedra are not shown. (d) The Fe sublattice shown along with the coordination octahedra with O atoms. The M atoms and their coordination octahedra are not shown. In both (c) and (d), the P atoms and their coordination tetrahedra with O atoms have been suppressed for better visualization.}
    \label{fig2}
\end{figure}

The orthorhombic metal-oxide iron-phosphates MFePO$_5$ or more suggestively MO$\cdot$FePO$_4$ (M=Fe,Cu,Ni,Co), crystallize in the $\bm{Pnma}~(\bm{D_{2h}^{16}})$ space group, see \textcolor{black}{Fig.~\ref{fig2}(a)}, with the underlying point group $\bm{D_{2h}}$ and are fully compensated ferrimagnets comprising of two magnetic species (M and Fe) occupying the 4a and 4c Wyckoff positions respectively (also refer Table~\ref{tab1}) \cite{Warner1992, Touaiher1994, Khayati2000, Khayati2001}. \textcolor{black}{Figure~\ref{fig2}(b)} displays the different coordination polyhedra formed within the crystal structure of MFePO$_5$. The M atoms form coordination octahedra with the O atoms and these octahedra show Jahn-Teller distortion and GdFeO$_3$--type tilting, similar to that observed in LaMnO$_3$ \cite{Elemans1971, Norby1995}. The Fe atoms also form similarly distorted coordination octahedra with the O atoms. \textcolor{black}{Figures~\ref{fig2}(c) and \ref{fig2}(d)} display these coordination polyhedra within the unit cell. The P atoms form the phosphate (PO$_{4}^{3-}$) coordination tetrahedra of MO$\cdot$FePO$_4$ and thus Fe exists in +3 oxidation state whereas M exists in +2 oxidation state to maintain charge neutrality in MFePO$_5$.

Henceforth, we shall primarily focus on CuFePO$_5$ as a representative of the isostructural MFePO$_5$ family. In the subsequent sections, the spin arrangement of Cu and Fe atoms within the two distinct magnetic-sublattices in CuFePO$_5$ shall be explained in detail. By considering the hypothetical compounds CuGaPO$_5$ and CaFePO$_5$ respectively, the effect of the Cu and Fe sublattices will be discussed separately in the context of altermagnetism followed by an elaboration of the ground state magnetic structure of CuFePO$_5$ from the perspective of alterferrimagnetism. Finally, we shall look at how the spin arrangements of Cu and Fe atoms help tune the possible magnetic structures of CuFePO$_5$ which provide a deeper insight into the phenomenon of alterferrimagnetism.

\subsection{The spin arrangement of Cu atoms in \texorpdfstring{CuFePO$_5$}{CuFePO5}} \label{Cu_CuFePO5}

\noindent The Cu atoms in CuFePO$_5$ form a fully compensated magnetic-sublattice and occupy the 4a Wyckoff position which is \textit{fully compatible} with AM phase (see Sec.~\ref{FLAM}, Fig.~\ref{fig2}(c) and Table~\ref{tab1}). We consider the hypothetical compound CuGaPO$_5$ by replacing Fe with Ga so as to maintain charge neutrality and obtain a single magnetic-sublattice. This will help us explore the various spin arrangements at only the 4a Wyckoff position occupied by Cu atoms. Considering magnetic order at the 4a Wyckoff position that can be accommodated within the crystallographic unit cell, there are only three possibilities for fully compensated staggered collinear $\vec{\bm{q}} = 0$ spin arrangements in a \textit{primitive} orthorhombic unit cell, namely A-type, C-type and G-type, which shall be explained in the following subsections.

\subsubsection{A-type spin arrangement} \label{Cu_A}
\begin{figure*}
    \centering
    \includegraphics[width=\linewidth]{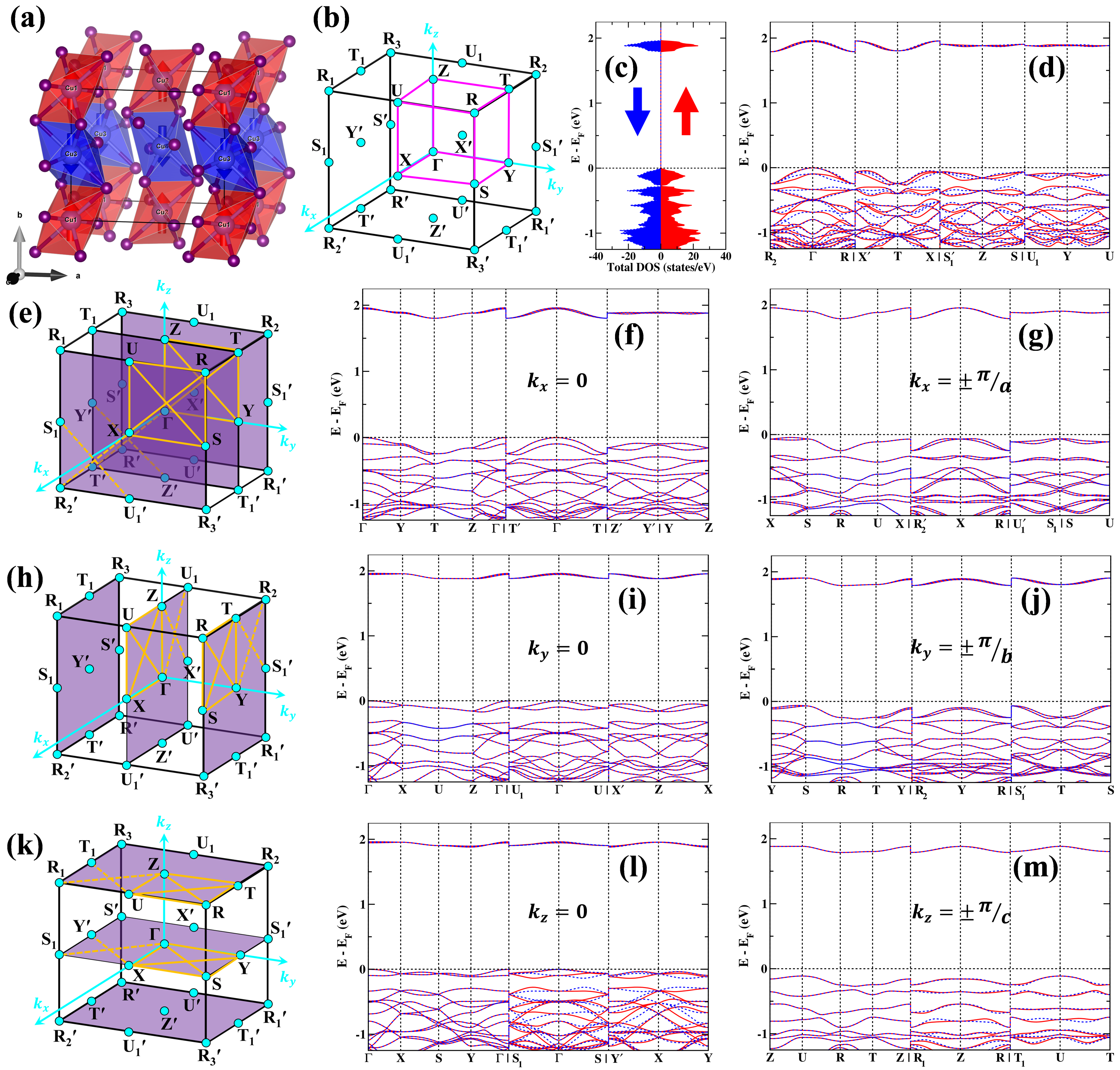}
    \caption{(a) The Cu sublattice with A-type spin arrangement. The red octahedra denote the up-spin sublattice and the blue octahedra denote the down-spin sublattice. (b) \textit{Primitive} orthorhombic Brillouin zone (BZ) with the high-symmetry points shown as cyan dots and the irreducible Brillouin zone (IBZ) marked by magenta lines. (c) Spin-polarized electronic total density of states (DOS) of CuGaPO$_5$. (d) Spin-polarized electronic band structure of CuGaPO$_5$ along body diagonals of the IBZ showing spin-splitting. (e) BZ with the $k_x=0$ and $k_x=\pm\frac{\pi}{a}$ planes highlighted in purple. The high-symmetry directions on the respective planes have been highlighted by orange lines and the spin-polarized electronic band structures of CuGaPO$_5$ along these lines have been shown in (f) and (g). (h-j) The same for $k_y=0$ and $k_y=\pm\frac{\pi}{b}$ planes. (k-m) The same for $k_z=0$ and $k_z=\pm\frac{\pi}{c}$ planes. In (e), (h) and (k), the solid orange lines denote paths within the IBZ while the broken orange lines denote paths outside the IBZ.}
    \label{fig3}
\end{figure*}
The A-type spin arrangement at the 4a Wyckoff position is formed when the Cu atoms in CuFePO$_5$ align ferromagnetically in-plane and antiferromagnetically out-of-plane as depicted in \textcolor{black}{Fig.~\ref{fig3}(a)}. Within the Cu sublattice, let us consider Cu1 \& Cu2 atoms to be up-spin and Cu3 \& Cu4 atoms to be down-spin. The Cu1 \& Cu2 atoms occupy $(0,0,0)$ \& $(\sfrac{1}{2},0,\sfrac{1}{2})$ lattice points respectively whereas the Cu3 \& Cu4 atoms occupy $(0,\sfrac{1}{2},0)$ \& $(\sfrac{1}{2},\sfrac{1}{2},\sfrac{1}{2})$ lattice points respectively, under the 4a Wyckoff position (also see Table~\ref{tab1}). The \textit{site-symmetry} group $\mathbf{W} \cong \bm{C_i}$ maps every lattice point in 4a Wyckoff position to itself. Additionally, the space group operations $\tilde{C}_{2z}$ and $\tilde{m}_{xy}$ (refer Table~\ref{tab1}) exchange the lattice points occupied by the up-spin atoms Cu1 \& Cu2 and the down-spin atoms Cu3 \& Cu4 respectively:
\begin{eqnarray*}
    \text{Cu1: }(0,0,0)\,&\overset{\tilde{C}_{2z}}{\underset{\tilde{m}_{xy}}\longleftrightarrow}&\,\text{Cu2: }(\sfrac{1}{2},0,\sfrac{1}{2}) \\
    \text{Cu3: }(0,\sfrac{1}{2},0)\,&\overset{\tilde{C}_{2z}}{\underset{\tilde{m}_{xy}}\longleftrightarrow}&\,\text{Cu4: }(\sfrac{1}{2},\sfrac{1}{2},\sfrac{1}{2})
\end{eqnarray*}
Therefore, the \textit{halving} subgroup of $\bm{Pnma}$ with the same-spin sublattice transformations can be constructed using $\mathbf{W}$ and a subset of space group operations $\{\tilde{C}_{2z},\tilde{m}_{xy}\}$ so that $\mathbf{W}\,\cup\,\{\tilde{C}_{2z}\,,\,\tilde{m}_{xy}\} = \{\,\mathbb{I}\,,\,\inv\,,\,\tilde{C}_{2z}\,,\,\tilde{m}_{xy}\,\} = \bm{\tilde{C}_{2h}^{z}} \cong \bm{C_{2h}^{5}}$ (also see Table~\ref{tab1}), where the superscript of $\bm{\tilde{C}_{2h}^{z}}$ denotes the orientation of principal axis of rotation along $z$-axis and `$\sim$' denotes presence of screw rotations and glide reflections. Thus, the \textit{halving} subgroup for A-type spin arrangement at the 4a Wyckoff position in $\bm{Pnma}$ is $\mathbf{H_1} = \bm{\tilde{C}_{2h}^{z}}$.

To understand the effect of SSG symmetries in the momentum $(\vec{k})$ space, only the point group symmetry operations shall be sufficient; screw rotations reduce to only rotations and glide reflections reduce to only reflections. The underlying point group of $\bm{Pnma}$ can then be decomposed as $\bm{D_{2h}} = \bm{C_{2h}^{z}} \cup (\bm{D_{2h}} - \bm{C_{2h}^{z}}) = \bm{C_{2h}^{z}} \cup \{C_{2x}, C_{2y}, m_{yz}, m_{zx}\}$, where $\bm{C_{2h}^{z}}$ is the \textit{isomorphic} point group of $\bm{\tilde{C}_{2h}^{z}}$. The corresponding non-trivial spin point group derived from \textit{isomorphic} two-coset decomposition is constructed as:
\begin{equation}
    \ntsg^{\text{point}} = [\mathfrak{I}\;||\;\bm{C_{2h}^{z}}] \cup [\mathfrak{C}_2\;||\;\bm{D_{2h}} - \bm{C_{2h}^{z}}]
    \label{eq2}
\end{equation}
For $R \in (\bm{D_{2h}} - \bm{C_{2h}^{z}})$, the transformed momenta $\vec{k}' = R\vec{k}$ takes the form:
\begin{equation}
    \vec{k}' =
    \begin{cases}
        (k_x, -k_y, -k_z) & R = C_{2x} \\
        (-k_x, k_y, -k_z) & R = C_{2y} \\
        (-k_x, k_y, k_z) & R = m_{yz} \\
        (k_x, -k_y, k_z) & R = m_{zx}
    \end{cases}
    \label{eq3}
\end{equation}
Now, if $R \in \mathcal{L}_{\vec{k}}$ , then Eq.~\eqref{eq3} yields the following constraints on the momentum components:
\begin{equation}
    \vec{k}' =
    \begin{cases}
        \left(k_x,\frac{n\pi}{b},\frac{n\pi}{c}\right) & R = C_{2x} \\
        \left(\frac{n\pi}{a},k_y,\frac{n\pi}{c}\right) & R = C_{2y} \\
        \left(\frac{n\pi}{a},k_y,k_z\right) & R = m_{yz} \\
        \left(k_x,\frac{n\pi}{b},k_z\right) & R = m_{zx}
    \end{cases}
    \quad (n\in\mathbb{Z})
    \label{eq4}
\end{equation}
Here, $a$, $b$, and $c$ are lattice constants of the \textit{primitive} orthorhombic unit cell (see \textcolor{black}{Fig.~\ref{fig3}(a)}). Eqs.~\eqref{eq1} and \eqref{eq4} together map out all the possible spin-degenerate nodal lines and nodal planes throughout the Brillouin zone (BZ) for A-type spin arrangement at the 4a Wyckoff position in $\bm{Pnma}$. \textcolor{black}{Figure~\ref{fig3}(b)} depicts the BZ along with the irreducible Brillouin zone (IBZ). First-principles calculations reveal a fully compensated Cu sublattice as evident from the spin-polarized total density of states (DOS) shown in \textcolor{black}{Fig.~\ref{fig3}(c)} with a spin magnetic moment of $\pm$0.67~$\mu_B$, slightly less than that of Cu$^{2+}$ indicating its covalent character. \textcolor{black}{Figure~\ref{fig3}(d)} shows the altermagnetic spin-splitting in the electronic band structure of CuGaPO$_5$ for A-type spin arrangement along body diagonals of the IBZ, in agreement with eqs.~\eqref{eq1} and \eqref{eq4}. The spin-degeneracy and spin splitting along the other momentum directions as a consequence of these SSG symmetries, namely the $k_x=0,\pm\sfrac{\pi}{a}$, $k_y=0,\pm\sfrac{\pi}{b}$ and $k_z=0,\pm\sfrac{\pi}{c}$ planes, are shown in \textcolor{black}{Figs.~\ref{fig3}(e–m)}. The SSG for A-type spin arrangement at 4a Wyckoff position is $\bm{P\,^{\overline{1}}n\,^{\overline{1}}m\,^{1}a}$ (S62.446) \cite{Litvin1977, Aroyo2006a, Chen2025}.

\subsubsection{C-type spin arrangement} \label{Cu_C}
\begin{figure*}
    \centering
    \includegraphics[width=\linewidth]{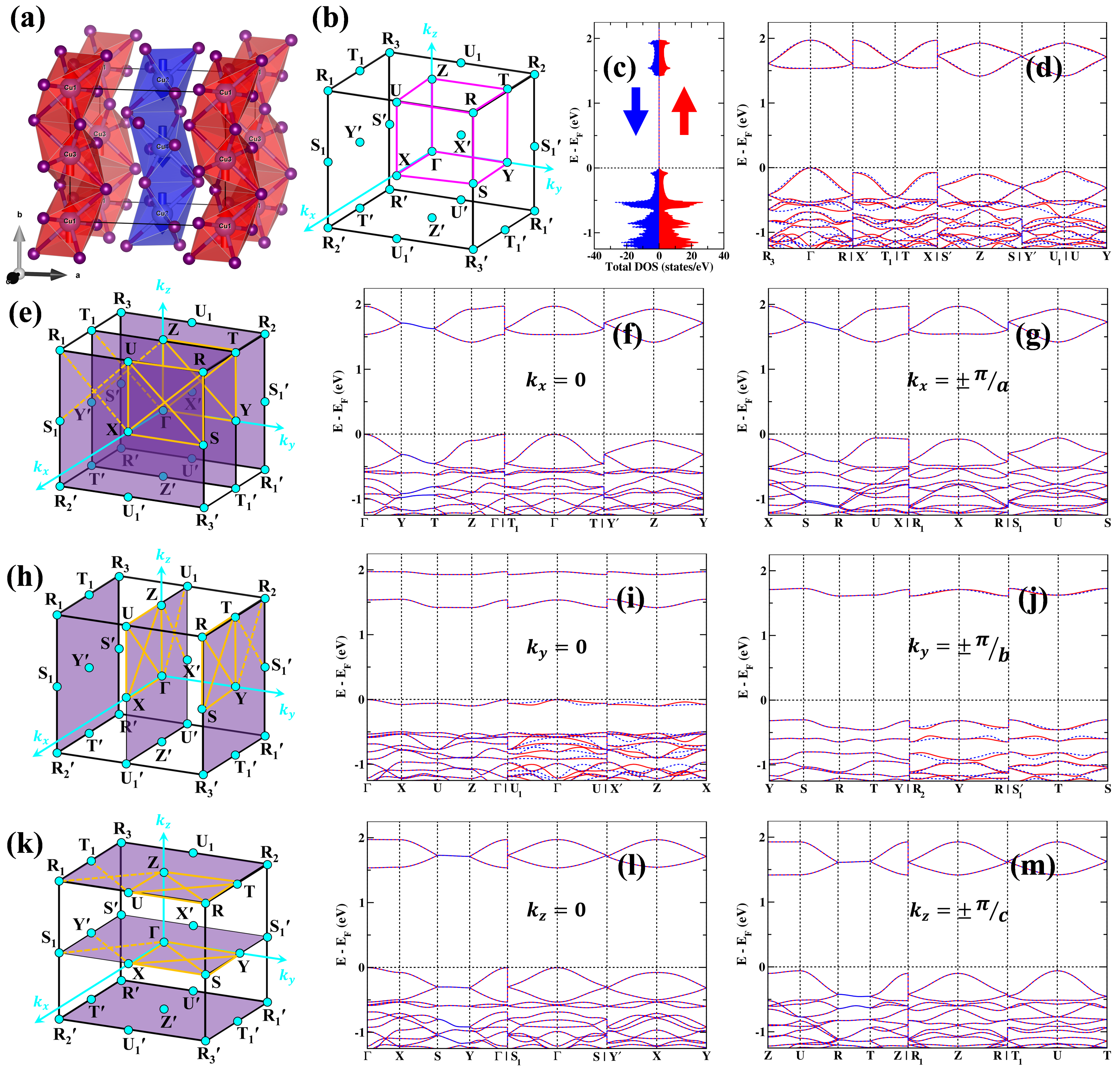}
    \caption{(a) The Cu sublattice with C-type spin arrangement. The red octahedra denote the up-spin sublattice and the blue octahedra denote the down-spin sublattice. (b) \textit{Primitive} orthorhombic Brillouin zone (BZ) with the high-symmetry points shown as cyan dots and the irreducible Brillouin zone (IBZ) marked by magenta lines. (c) Spin-polarized electronic total density of states (DOS) of CuGaPO$_5$. (d) Spin-polarized electronic band structure of CuGaPO$_5$ along body diagonals of the IBZ showing spin-splitting. (e) BZ with the $k_x=0$ and $k_x=\pm\frac{\pi}{a}$ planes highlighted in purple. The high-symmetry directions on the respective planes have been highlighted by orange lines and the spin-polarized electronic band structures of CuGaPO$_5$ along these lines have been shown in (f) and (g). (h-j) The same for $k_y=0$ and $k_y=\pm\frac{\pi}{b}$ planes. (k-m) The same for $k_z=0$ and $k_z=\pm\frac{\pi}{c}$ planes. In (e), (h) and (k), the solid orange lines denote paths within the IBZ while the broken orange lines denote paths outside the IBZ.}
    \label{fig4}
\end{figure*}
When the Cu atoms in CuFePO$_5$ are aligned antiferromagnetically in-plane but ferromagnetically out-of-plane as shown in \textcolor{black}{Fig.~\ref{fig4}(a)}, it leads to the C-type spin arrangement at the 4a Wyckoff position. Within the Cu sublattice, let us consider Cu1 \& Cu3 atoms to be up-spin and Cu2 \& Cu4 atoms to be down-spin. The Cu1 \& Cu3 atoms occupy $(0,0,0)$ \& $(0,\sfrac{1}{2},0)$ lattice points respectively whereas the Cu2 \& Cu4 atoms occupy $(\sfrac{1}{2},0,\sfrac{1}{2})$ \& $(\sfrac{1}{2},\sfrac{1}{2},\sfrac{1}{2})$ lattice points respectively, under the 4a Wyckoff position (also refer Table~\ref{tab1}). The \textit{site-symmetry} group $\mathbf{W} \cong \bm{C_i}$ maps every lattice point in 4a Wyckoff position to itself. Additionally, the space group operations $\tilde{C}_{2y}$ and $\tilde{m}_{zx}$ (refer Table~\ref{tab1}) exchange the lattice points occupied by the up-spin atoms Cu1 \& Cu3 and the down-spin atoms Cu2 \& Cu4 respectively:
\begin{eqnarray*}
    \text{Cu1: }(0,0,0)\,&\overset{\tilde{C}_{2y}}{\underset{\tilde{m}_{zx}}\longleftrightarrow}&\,\text{Cu3: }(0,\sfrac{1}{2},0) \\
    \text{Cu2: }(\sfrac{1}{2},0,\sfrac{1}{2})\,&\overset{\tilde{C}_{2y}}{\underset{\tilde{m}_{zx}}\longleftrightarrow}&\,\text{Cu4: }(\sfrac{1}{2},\sfrac{1}{2},\sfrac{1}{2})
\end{eqnarray*}
Therefore, the \textit{halving} subgroup of $\bm{Pnma}$ with the same-spin sublattice transformations can be constructed using $\mathbf{W}$ and the subset of space group operations $\{\tilde{C}_{2y},\tilde{m}_{zx}\}$ so that $\mathbf{W}\,\cup\,\{\tilde{C}_{2y}\,,\,\tilde{m}_{zx}\} = \{\,\mathbb{I}\,,\,\inv\,,\,\tilde{C}_{2y}\,,\,\tilde{m}_{zx}\,\} = \bm{\tilde{C}_{2h}^{y}} \cong \bm{C_{2h}^{2}}$ (also see Table~\ref{tab1}), where the principal axis of rotation of $\bm{\tilde{C}_{2h}^{y}}$ is oriented along the $y$-axis. Thus in $\bm{Pnma}$, the \textit{halving} subgroup for C-type spin arrangement at the 4a Wyckoff position is $\mathbf{H_2} = \bm{\tilde{C}_{2h}^{y}}$. 

The underlying point group of $\bm{Pnma}$ can then be decomposed as $\bm{D_{2h}} = \bm{C_{2h}^{y}} \cup (\bm{D_{2h}} - \bm{C_{2h}^{y}}) = \bm{C_{2h}^{y}} \cup \{C_{2z}, C_{2x}, m_{xy}, m_{yz}\}$, where $\bm{C_{2h}^{y}}$ is the \textit{isomorphic} point group of $\bm{\tilde{C}_{2h}^{y}}$. The corresponding non-trivial spin point group derived from \textit{isomorphic} two-coset decomposition is constructed as:
\begin{equation}
    \ntsg^{\text{point}} = [\mathfrak{I}\;||\;\bm{C_{2h}^{y}}] \cup [\mathfrak{C}_2\;||\;\bm{D_{2h}} - \bm{C_{2h}^{y}}]
    \label{eq5}
\end{equation}
For $R \in (\bm{D_{2h}} - \bm{C_{2h}^{y}})$, the transformed momenta $\vec{k}' = R\vec{k}$ takes the form: 
\begin{equation}
    \vec{k}' =
    \begin{cases}
        (-k_x, -k_y, k_z) & R = C_{2z} \\
        (k_x, -k_y, -k_z) & R = C_{2x} \\
        (k_x, k_y, -k_z) & R = m_{xy} \\
        (-k_x, k_y, k_z) & R = m_{yz}
    \end{cases}
    \label{eq6}
\end{equation}
Now, if $R \in \mathcal{L}_{\vec{k}}$ , then Eq.~\eqref{eq6} yields the following constraints on the momentum components:
\begin{equation}
    \vec{k}' =
    \begin{cases}
        \left(\frac{n\pi}{a},\frac{n\pi}{b},k_z\right) & R = C_{2z} \\
        \left(k_x,\frac{n\pi}{b},\frac{n\pi}{c}\right) & R = C_{2x} \\
        \left(k_x,k_y,\frac{n\pi}{c}\right) & R = m_{xy} \\
        \left(\frac{n\pi}{a},k_y,k_z\right) & R = m_{yz}
    \end{cases}
    \quad (n\in\mathbb{Z})
    \label{eq7}
\end{equation}
Eqs.~\eqref{eq1} and \eqref{eq7} together map out all the possible spin-degenerate nodal lines and nodal planes throughout the BZ for C-type spin arrangement at the 4a Wyckoff position in $\bm{Pnma}$. \textcolor{black}{Figure~\ref{fig4}(b)} depicts the BZ along with the IBZ. First-principles calculations reveal a fully compensated Cu sublattice as evident from the spin-polarized total density of states (DOS) shown in \textcolor{black}{Fig.~\ref{fig4}(c)} with a spin magnetic moment of $\pm$0.68~$\mu_B$, slightly less than that of Cu$^{2+}$ indicating its covalent character. \textcolor{black}{Figure~\ref{fig4}(d)} shows the altermagnetic spin-splitting in the electronic band structure of CuGaPO$_5$ for C-type spin arrangement along body diagonals of the IBZ, in agreement with eqs.~\eqref{eq1} and \eqref{eq7}. The spin-degeneracy and spin splitting along the other momentum directions as a consequence of these SSG symmetries, namely the $k_x=0,\pm\sfrac{\pi}{a}$, $k_y=0,\pm\sfrac{\pi}{b}$ and $k_z=0,\pm\sfrac{\pi}{c}$ planes, are shown in \textcolor{black}{Figs.~\ref{fig4}(e–m)}. The SSG for C-type spin arrangement at 4a Wyckoff position is $\bm{P\,^{\overline{1}}n\,^{1}m\,^{\overline{1}}a}$ (S62.448) \cite{Litvin1977, Aroyo2006a, Chen2025}.

\subsubsection{G-type spin arrangement} \label{Cu_G}
\begin{figure*}
    \centering
    \includegraphics[width=\linewidth]{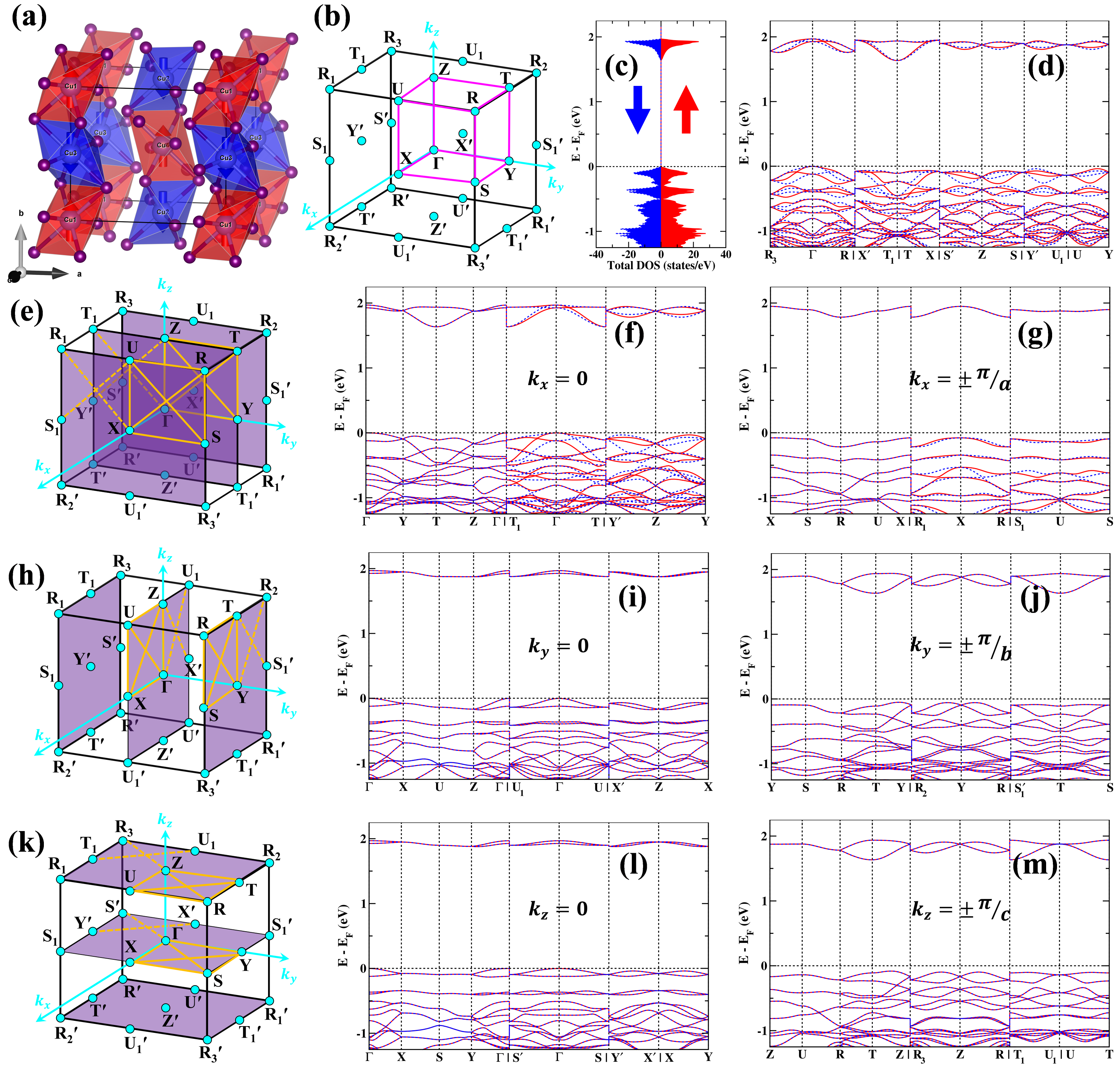}
    \caption{(a) The Cu sublattice with G-type spin arrangement. The red octahedra denote the up-spin sublattice and the blue octahedra denote the down-spin sublattice. (b) \textit{Primitive} orthorhombic Brillouin zone (BZ) with the high-symmetry points shown as cyan dots and the irreducible Brillouin zone (IBZ) marked by magenta lines. (c) Spin-polarized electronic total density of states (DOS) of CuGaPO$_5$. (d) Spin-polarized electronic band structure of CuGaPO$_5$ along body diagonals of the IBZ showing spin-splitting. (e) BZ with the $k_x=0$ and $k_x=\pm\frac{\pi}{a}$ planes highlighted in purple. The high-symmetry directions on the respective planes have been highlighted by orange lines and the spin-polarized electronic band structures of CuGaPO$_5$ along these lines have been shown in (f) and (g). (h-j) The same for $k_y=0$ and $k_y=\pm\frac{\pi}{b}$ planes. (k-m) The same for $k_z=0$ and $k_z=\pm\frac{\pi}{c}$ planes. In (e), (h) and (k), the solid orange lines denote paths within the IBZ while the broken orange lines denote paths outside the IBZ.}
    \label{fig5}
\end{figure*}
If the Cu atoms in CuFePO$_5$ are aligned antiferromagnetically both in-plane and out-of-plane as shown in \textcolor{black}{Fig.~\ref{fig5}(a)}, then the G-type spin arrangement is obtained at the 4a Wyckoff position. Within the Cu sublattice, let us consider Cu1 \& Cu4 atoms to be up-spin and Cu2 \& Cu3 atoms to be down-spin. The  Cu1 \& Cu4 atoms occupy $(0,0,0)$ \& $(\sfrac{1}{2},\sfrac{1}{2},\sfrac{1}{2})$ lattice points respectively whereas the  Cu2 \& Cu3 atoms occupy $(0,\sfrac{1}{2},0)$ \& $(\sfrac{1}{2},0,\sfrac{1}{2})$ lattice points repsectively, under the 4a Wyckoff position (also see Table~\ref{tab1}). The \textit{site-symmetry} group $\mathbf{W} \cong \bm{C_i}$ maps every lattice point in 4a Wyckoff position to itself. Additionally, the space group operations $\tilde{C}_{2x}$ and $\tilde{m}_{yz}$ (refer Table~\ref{tab1}) exchange the lattice points occupied by the up-spin atoms Cu1 \& Cu4 and the down-spin atoms Cu2 \& Cu3 respectively:
\begin{eqnarray*}
    \text{Cu1: }(0,0,0)\,&\overset{\tilde{C}_{2x}}{\underset{\tilde{m}_{yz}}\longleftrightarrow}&\,\text{Cu4: }(\sfrac{1}{2},\sfrac{1}{2},\sfrac{1}{2}) \\
    \text{Cu2: }(\sfrac{1}{2},0,\sfrac{1}{2})\,&\overset{\tilde{C}_{2x}}{\underset{\tilde{m}_{yz}}\longleftrightarrow}&\,\text{Cu3: }(0,\sfrac{1}{2},0)
\end{eqnarray*}
Therefore, the \textit{halving} subgroup of $\bm{Pnma}$ with the same-spin sublattice transformations can be constructed using $\mathbf{W}$ and the subset of space group operations $\{\tilde{C}_{2x},\tilde{m}_{yz}\}$ so that $\mathbf{W}\,\cup\,\{\tilde{C}_{2x}\,,\,\tilde{m}_{yz}\} = \{\,\mathbb{I}\,,\,\inv\,,\,\tilde{C}_{2x}\,,\,\tilde{m}_{yz}\,\} = \bm{\tilde{C}_{2h}^{x}} \cong \bm{C_{2h}^{5}}$, where the principal axis of rotation of $\bm{\tilde{C}_{2h}^{x}}$ is oriented along the $x$-axis. Thus in $\bm{Pnma}$, the \textit{halving} subgroup for G-type spin arrangement at the 4a Wyckoff position is $\mathbf{H_3} = \bm{\tilde{C}_{2h}^{x}}$. 

The underlying point group of $\bm{Pnma}$ can then be decomposed as $\bm{D_{2h}} = \bm{C_{2h}^{x}} \cup (\bm{D_{2h}} - \bm{C_{2h}^{x}}) = \bm{C_{2h}^{x}} \cup \{C_{2y}, C_{2z}, m_{zx}, m_{xy}\}$, where $\bm{C_{2h}^{x}}$ is the \textit{isomorphic} point group of $\bm{\tilde{C}_{2h}^{x}}$. The corresponding non-trivial spin point group derived from \textit{isomorphic} two-coset decomposition is constructed as:
\begin{equation}
    \ntsg^{\text{point}} = [\mathfrak{I}\;||\;\bm{C_{2h}^{x}}] \cup [\mathfrak{C}_2\;||\;\bm{D_{2h}} - \bm{C_{2h}^{x}}]
    \label{eq8}
\end{equation}
For $R \in (\bm{D_{2h}} - \bm{C_{2h}^{x}})$, the transformed momenta $\vec{k}' = R\vec{k}$ takes the form: 
\begin{equation}
    \vec{k}' =
    \begin{cases}
        (-k_x, k_y, -k_z) & R = C_{2y} \\
        (-k_x, -k_y, k_z) & R = C_{2z} \\
        (k_x, -k_y, k_z) & R = m_{zx} \\
        (k_x, k_y, -k_z) & R = m_{xy}
    \end{cases}
    \label{eq9}
\end{equation}
Now, if $R \in \mathcal{L}_{\vec{k}}$ , then Eq.~\eqref{eq9} yields the following constraints on the momentum components:
\begin{equation}
    \vec{k}' =
    \begin{cases}
        \left(\frac{n\pi}{a},k_y,\frac{n\pi}{c}\right) & R = C_{2y} \\
        \left(\frac{n\pi}{a},\frac{n\pi}{b},k_z\right) & R = C_{2z} \\
        \left(k_x,\frac{n\pi}{b},k_z\right) & R = m_{zx} \\
        \left(k_x,k_y,\frac{n\pi}{c}\right) & R = m_{xy}
    \end{cases}
    \quad (n\in\mathbb{Z})
    \label{eq10}
\end{equation}
Eqs.~\eqref{eq1} and \eqref{eq10} together map out all the possible spin-degenerate nodal lines and nodal planes throughout the Brillouin zone for G-type spin arrangement at the 4a Wyckoff position in $\bm{Pnma}$. \textcolor{black}{Figure~\ref{fig5}(b)} depicts the BZ along with the IBZ. First-principles calculations reveal a fully compensated Cu sublattice as evident from the spin-polarized total density of states (DOS) shown in \textcolor{black}{Fig.~\ref{fig5}(c)} with a spin magnetic moment of $\pm$0.67~$\mu_B$, slightly less than that of Cu$^{2+}$ indicating its covalent character. \textcolor{black}{Figure~\ref{fig5}(d)} shows the altermagnetic spin-splitting in the electronic band structure of CuGaPO$_5$ for G-type spin arrangement along body diagonals of the IBZ, in agreement with eqs.~\eqref{eq1} and \eqref{eq10}. The spin-degeneracy and spin splitting along the other momentum directions as a consequence of these SSG symmetries, namely the $k_x=0,\pm\sfrac{\pi}{a}$, $k_y=0,\pm\sfrac{\pi}{b}$ and $k_z=0,\pm\sfrac{\pi}{c}$ planes, are shown in \textcolor{black}{Figs.~\ref{fig5}(e–m)}. The SSG for G-type spin arrangement at 4a Wyckoff position is $\bm{P\,^{1}n\,^{\overline{1}}m\,^{\overline{1}}a}$ (S62.447) \cite{Litvin1977, Aroyo2006a, Chen2025}. \\

\noindent Thus, it is seen that any collinear staggered spin arrangement at the 4a Wyckoff position in $\bm{Pnma}$ will lead to the AM phase, albeit with different orientational symmetry. This further shows that 4a Wyckoff position is \textit{fully compatible} with the AM phase, in accordance with FLAM (Sec.~\ref{FLAM}). The different possible orientations of the \textit{halving} subgroup of $\bm{Pnma}$ is capable of tuning the symmetries responsible for spin-degeneracies and the momenta direction along which such degeneracies will occur leading to different SSG. The results of different spin arrangements of Cu atoms in the hypothetical CuGaPO$_5$ complement the earlier results by Okugawa and colleagues \cite{Okugawa2018} who derived the spin-degeneracy conditions for LaMO$_3$ (M=Cr,Mn,Fe), prior to the SSG formalism in the context of altermagnetism introduced by \cite{Smejkal2022a, Smejkal2022b}. In LaMnO$_3$, the Mn atoms occupy the 4a Wyckoff position \cite{Elemans1971, Norby1995} and hence LaMnO$_3$ shows altermagnetism for A-type, C-type and G-type spin arrangements, the same being true for LaCrO$_3$ and LaFeO$_3$ \cite{Okugawa2018}.

\subsection{The spin arrangement of Fe atoms in \texorpdfstring{CuFePO$_5$}{CuFePO5}} \label{Fe_CuFePO5}

\noindent The Fe atoms in CuFePO$_5$ also form a fully compensated magnetic-sublattice similar to the Cu atoms, but occupy the 4c Wyckoff position which is \textit{partially compatible} with AM phase (see Sec.~\ref{FLAM}, Fig.~\ref{fig2}(d) and Table~\ref{tab1}). We consider the hypothetical compound CaFePO$_5$ by replacing Cu with Ca so as to maintain charge neutrality and obtain a single magnetic-sublattice. This will help us explore the various spin arrangements at only the 4c Wyckoff position occupied by Fe atoms. Considering magnetic order at the 4c Wyckoff position that can be accommodated within the crystallographic unit cell, there are again three possibilities for fully compensated staggered collinear $\vec{\bm{q}} = 0$ spin arrangements in a \textit{primitive} orthorhombic unit cell which shall be explained in the following subsections.

\subsubsection{(\texorpdfstring{$+$$-$$+$$-$}{+-+-})-type spin arrangement} \label{Fe_pnpn}
\begin{figure*}
    \centering
    \includegraphics[width=\linewidth]{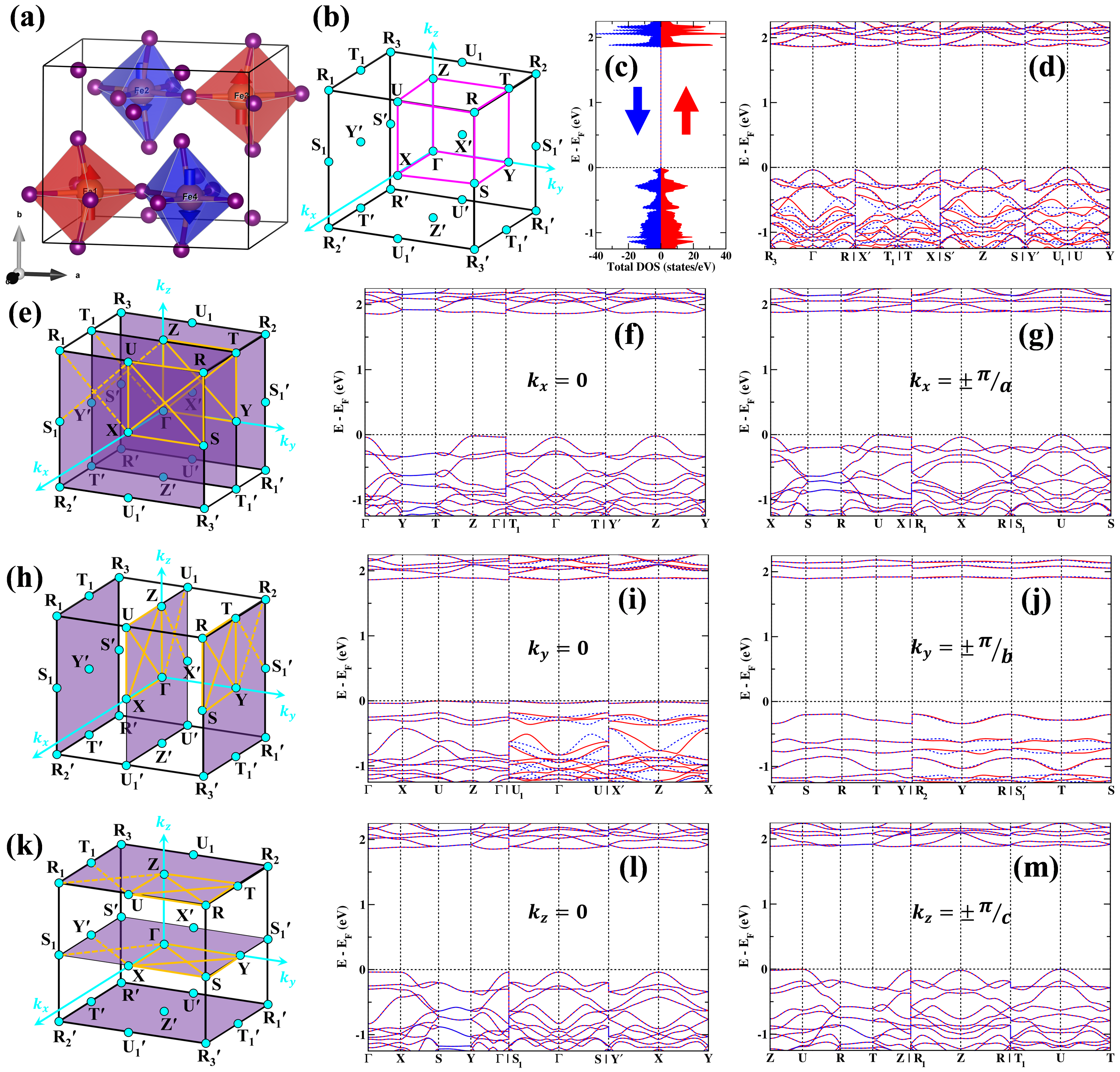}
    \caption{(a) The Fe sublattice with ($+$$-$$+$$-$)-type spin arrangement. The red octahedra denote the up-spin sublattice and the blue octahedra denote the down-spin sublattice. (b) \textit{Primitive} orthorhombic Brillouin zone (BZ) with the high-symmetry points shown as cyan dots and the irreducible Brillouin zone (IBZ) marked by magenta lines. (c) Spin-polarized electronic total density of states (DOS) of CaFePO$_5$. (d) Spin-polarized electronic band structure of CaFePO$_5$ along body diagonals of the IBZ showing spin-splitting, characteristic of the altermagnetic phase. (e) BZ with the $k_x=0$ and $k_x=\pm\frac{\pi}{a}$ planes highlighted in purple. The high-symmetry directions on the respective planes have been highlighted by orange lines and the spin-polarized electronic band structures of CaFePO$_5$ along these lines have been shown in (f) and (g). (h-j) The same for $k_y=0$ and $k_y=\pm\frac{\pi}{b}$ planes. (k-m) The same for $k_z=0$ and $k_z=\pm\frac{\pi}{c}$ planes. In (e), (h) and (k), the solid orange lines denote paths within the IBZ while the broken orange lines denote paths outside the IBZ.}
    \label{fig6}
\end{figure*}
Within the Fe sublattice, let us consider Fe1 \& Fe3 atoms to be up-spin and Fe2 \& Fe4 atoms to be down-spin. As shown in \textcolor{black}{Fig.~\ref{fig6}(a)}, the Fe1 \& Fe3 atoms occupy $(\text{x},\sfrac{1}{4},\text{z})$ \& $(-\text{x},\sfrac{3}{4},-\text{z})$ lattice points respectively whereas the Fe2 \& Fe4 atoms occupy $(-\text{x}+\sfrac{1}{2},\sfrac{3}{4},\text{z}+\sfrac{1}{2})$ \& $(\text{x}+\sfrac{1}{2},\sfrac{1}{4},-\text{z}+\sfrac{1}{2})$ lattice points respectively, under the 4c Wyckoff position (also refer Table~\ref{tab1}) with $\text{x} = 0.1732$ and $\text{z} = 0.7123$ \cite{Khayati2001}. We denote this spin arrangement by ($+$$-$$+$$-$)$\equiv$(Up, Down, Up, Down)-spin. The \textit{site-symmetry} group $\mathbf{W} \cong \bm{C_s}$ maps every lattice point in 4c Wyckoff position to itself . Additionally, the space group operations $\tilde{C}_{2y}$ and $\inv$ (refer Table~\ref{tab1}) exchange the lattice points occupied by the up-spin atoms Fe1 \& Fe3 and the down-spin atoms Fe2 \& Fe4 respectively:
\begin{eqnarray*}
    \text{\small{Fe1: (x, $\sfrac{1}{4}$, z)}}\!\! &\overset{\tilde{C}_{2y}}{\underset{\inv}\longleftrightarrow}& \!\!\text{\small{Fe3: ($-$x, $\sfrac{3}{4}$, $-$z)}} \\
    \text{\small{Fe2: ($-$x$+\sfrac{1}{2}$, $\sfrac{3}{4}$, z$+\sfrac{1}{2}$)}}\!\! &\overset{\tilde{C}_{2y}}{\underset{\inv}\longleftrightarrow}& \!\!\text{\small{Fe4: (x$+\sfrac{1}{2}$, $\sfrac{1}{4}$, $-$z$+\sfrac{1}{2}$)}}
\end{eqnarray*}
Then the \textit{halving} subgroup providing the same-spin sublattice transformations can be constructed using $\mathbf{W}$ and the space group operations $\inv$ and $\tilde{C}_{2y}$. This leads to $\mathbf{W}\,\cup\,\{\tilde{C}_{2y}\,,\,\inv\} = \{\,\mathbb{I}\,,\,\tilde{m}_{zx}\,,\,\tilde{C}_{2y}\,,\,\inv\,\} = \bm{\tilde{C}_{2h}^{y}} \cong \bm{C_{2h}^{2}}$ (also see Table~\ref{tab1}) which is a \textit{halving} subgroup of $\bm{Pnma}$. Thus, ($+$$-$$+$$-$)-type spin arrangement at the 4c Wyckoff position becomes analogous to the case of C-type spin arrangement at the 4a Wyckoff position as outlined in Sec.~\ref{Cu_C}, with the SSG $\bm{P\,^{\overline{1}}n\,^{1}m\,^{\overline{1}}a}$ (S62.448) \cite{Litvin1977, Aroyo2006a, Chen2025}. But it should be noted that $\inv \notin \mathbf{W}$ and appears as a separate same-spin sublattice transformation. Eqs.~\eqref{eq1} and \eqref{eq7} together map out all the possible spin-degenerate nodal lines and nodal planes throughout the BZ for ($+$$-$$+$$-$)-type spin arrangement at the 4c Wyckoff position in $\bm{Pnma}$. \textcolor{black}{Figure~\ref{fig6}(b)} depicts the BZ along with the IBZ. First-principles calculations reveal a fully compensated Fe sublattice as evident from the spin-polarized total density of states (DOS) shown in \textcolor{black}{Fig.~\ref{fig6}(c)} with a spin magnetic moment of $\pm$4.15~$\mu_B$, slightly less than that of Fe$^{3+}$ indicating its covalent character. \textcolor{black}{Figure~\ref{fig6}(d)} shows the altermagnetic spin-splitting in the electronic band structure of CaFePO$_5$ for ($+$$-$$+$$-$)-type spin arrangement along body diagonals of the IBZ, in agreement with eqs.~\eqref{eq1} and \eqref{eq7}. The spin-degeneracy and spin splitting along the other momentum directions as a consequence of these SSG symmetries, namely the $k_x=0,\pm\sfrac{\pi}{a}$, $k_y=0,\pm\sfrac{\pi}{b}$ and $k_z=0,\pm\sfrac{\pi}{c}$ planes, are shown in \textcolor{black}{Figs.~\ref{fig6}(e–m)}.

\subsubsection{(\texorpdfstring{$+$$+$$-$$-$}{++--})-type spin arrangement} \label{Fe_ppnn}
\begin{figure*}
    \centering
    \includegraphics[width=\linewidth]{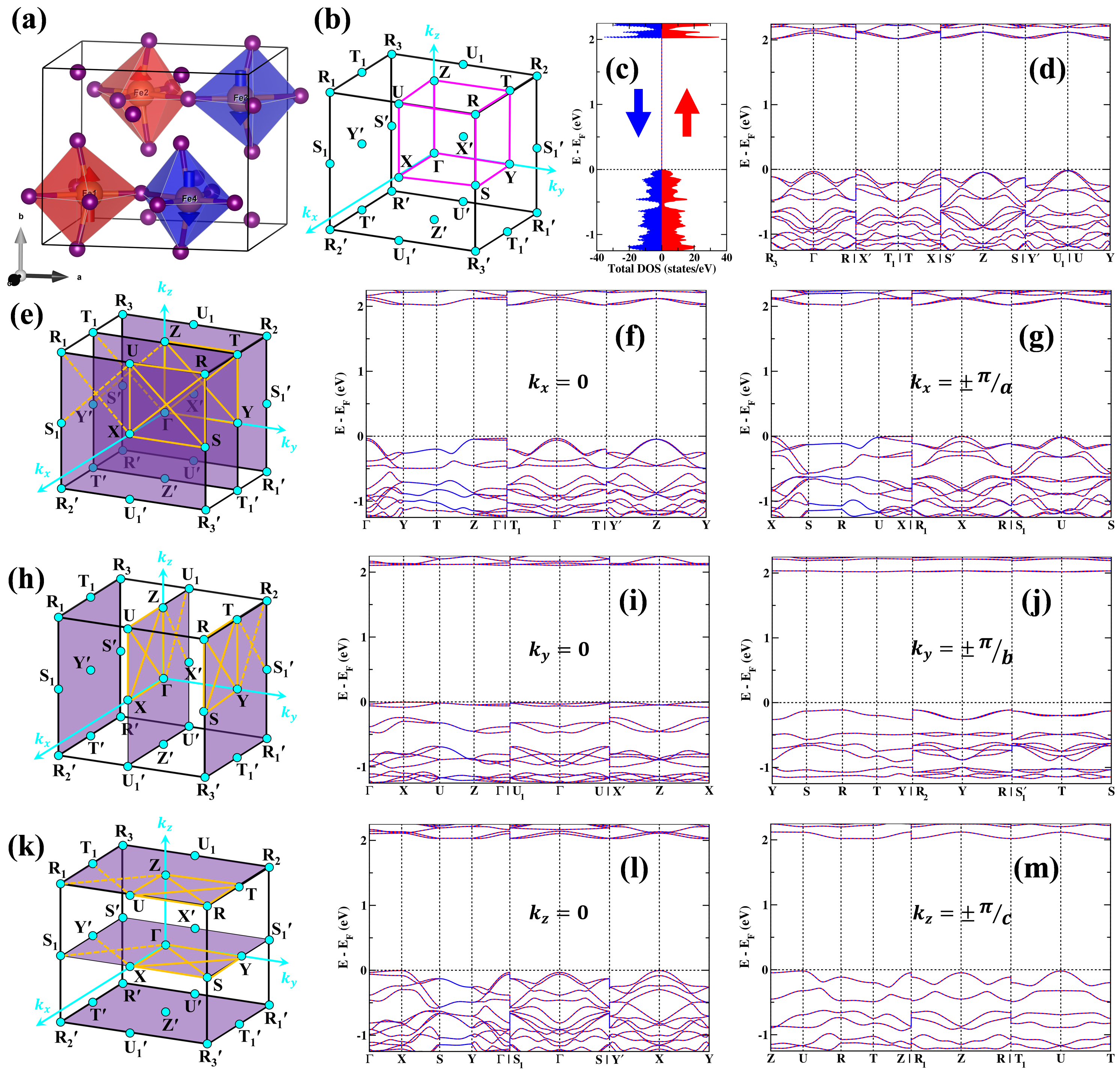}
    \caption{(a) The Fe sublattice with ($+$$+$$-$$-$)-type spin arrangement. The red octahedra denote the up-spin sublattice and the blue octahedra denote the down-spin sublattice. (b) \textit{Primitive} orthorhombic Brillouin zone (BZ) with the high-symmetry points shown as cyan dots and the irreducible Brillouin zone (IBZ) marked by magenta lines. (c) Spin-polarized electronic total density of states (DOS) of CaFePO$_5$. (d) Spin-polarized electronic band structure of CaFePO$_5$ along body diagonals of the IBZ showing spin-degeneracy, characteristic of the antiferromagnetic phase. (e) BZ with the $k_x=0$ and $k_x=\pm\frac{\pi}{a}$ planes highlighted in purple. The high-symmetry directions on the respective planes have been highlighted by orange lines and the spin-polarized electronic band structures of CaFePO$_5$ along these lines have been shown in (f) and (g). (h-j) The same for $k_y=0$ and $k_y=\pm\frac{\pi}{b}$ planes. (k-m) The same for $k_z=0$ and $k_z=\pm\frac{\pi}{c}$ planes. In (e), (h) and (k), the solid orange lines denote paths within the IBZ while the broken orange lines denote paths outside the IBZ.}
    \label{fig7}
\end{figure*}
The previous situation will become different if other collinear staggered spin arrangements of Fe atoms are taken into account. Let us consider the case when Fe1 \& Fe2 occupying $(\text{x},\sfrac{1}{4},\text{z})$ \& $(-\text{x}+\sfrac{1}{2},\sfrac{3}{4},\text{z}+\sfrac{1}{2})$ lattice points respectively become the up-spin atoms and Fe3 \& Fe4 occupying $(-\text{x},\sfrac{3}{4},-\text{z})$ \& $(\text{x}+\sfrac{1}{2},\sfrac{1}{4},-\text{z}+\sfrac{1}{2})$ lattice points respectively become the down-spin atoms as shown in \textcolor{black}{Fig.~\ref{fig7}(a)}. We denote this spin arrangement by ($+$$+$$-$$-$). The \textit{site-symmetry} group $\mathbf{W} \cong \bm{C_s}$ again maps every lattice point in 4c Wyckoff position to itself . But the space group operations exchanging the lattice points occupied by the up-spin atoms Fe1 \& Fe2 and the down-spin atoms Fe3 \& Fe4 respectively now differ:
\begin{eqnarray*}
    \text{Fe1: }(\text{x},\sfrac{1}{4},\text{z})\,&\overset{\tilde{C}_{2z}}{\underset{\tilde{m}_{yz}}\longleftrightarrow}&\,\text{Fe2: }(-\text{x}+\sfrac{1}{2},\sfrac{3}{4},\text{z}+\sfrac{1}{2}) \\
    \text{Fe3: }(-\text{x},\sfrac{3}{4},-\text{z})\,&\overset{\tilde{C}_{2z}}{\underset{\tilde{m}_{yz}}\longleftrightarrow}&\,\text{Fe4: }(\text{x}+\sfrac{1}{2},\sfrac{1}{4},-\text{z}+\sfrac{1}{2})
\end{eqnarray*}
The group of same-spin sublattice transformations that can be constructed using $\mathbf{W}$ and the space group operations $\tilde{C}_{2z}$ and $\tilde{m}_{yz}$ is $\mathbf{W}\,\cup\,\{\tilde{C}_{2z}\,,\,\tilde{m}_{yz}\} = \{\,\mathbb{I}\,,\,\tilde{m}_{zx}\,,\,\tilde{C}_{2z}\,,\,\tilde{m}_{yz}\,\} = \bm{\tilde{C}_{2v}^{z}} \cong \bm{C_{2v}^{7}}$ (also see Table~\ref{tab1}). Although this is indeed a \textit{halving} subgroup of $\bm{Pnma}$, but since $\inv\notin\bm{\tilde{C}_{2v}^{z}}$, the coset of $\bm{\tilde{C}_{2v}^{z}}$ in $\bm{Pnma}$ must contain $\inv$.
This means eq.~\eqref{eq1} becomes:
\begin{equation}
\text{\small{$\left[\mathfrak{C}_2\,||\,\mathcal{P}\right]\,\varepsilon_{j}(\sigma,\vec{k}) = \varepsilon_{j}(-\sigma,\!-\vec{k}) \!\Rightarrow\! \varepsilon_{j}(\sigma,\vec{k}) = \varepsilon_{j}(-\sigma,\!-\vec{k})$}}
    \label{eq11}
\end{equation}
Since for collinear order, there is always an inversion symmetry in the momentum space \cite{Smejkal2022a, Smejkal2022b}, the spin-polarized dispersion will be spin-degenerate throughout the first Brillouin zone and ($+$$+$$-$$-$)-type spin arrangement will yield AFM phase instead of AM phase. The SSG for ($+$$+$$-$$-$)-type spin arrangement at 4c Wyckoff position becomes $\bm{P\,^{1}n\,^{1}m\,^{\overline{1}}a}$ (S62.445) \cite{Litvin1977, Aroyo2006a, Chen2025}. \textcolor{black}{Figure~\ref{fig7}(b)} depicts the BZ along with the IBZ. First-principles calculations reveal a fully compensated Fe sublattice as evident from the spin-polarized total density of states (DOS) shown in \textcolor{black}{Fig.~\ref{fig7}(c)} with a spin magnetic moment of $\pm$4.14~$\mu_B$, slightly less than that of Fe$^{3+}$ indicating its covalent character. \textcolor{black}{Figure~\ref{fig7}(d)} shows spin-degeneracy in the electronic band structure of CaFePO$_5$ for ($+$$+$$-$$-$)-type spin arrangement along body diagonals of the IBZ, as expected of a conventional AFM. The spin-degeneracy along the other momentum directions as a consequence of the SSG symmetries, namely the $k_x=0,\pm\sfrac{\pi}{a}$, $k_y=0,\pm\sfrac{\pi}{b}$ and $k_z=0,\pm\sfrac{\pi}{c}$ planes, are shown in \textcolor{black}{Figs.~\ref{fig7}(e–m)}.

\subsubsection{(\texorpdfstring{$+$$-$$-$$+$}{+--+})-type spin arrangement} \label{Fe_pnnp}
\begin{figure*}
    \centering
    \includegraphics[width=\linewidth]{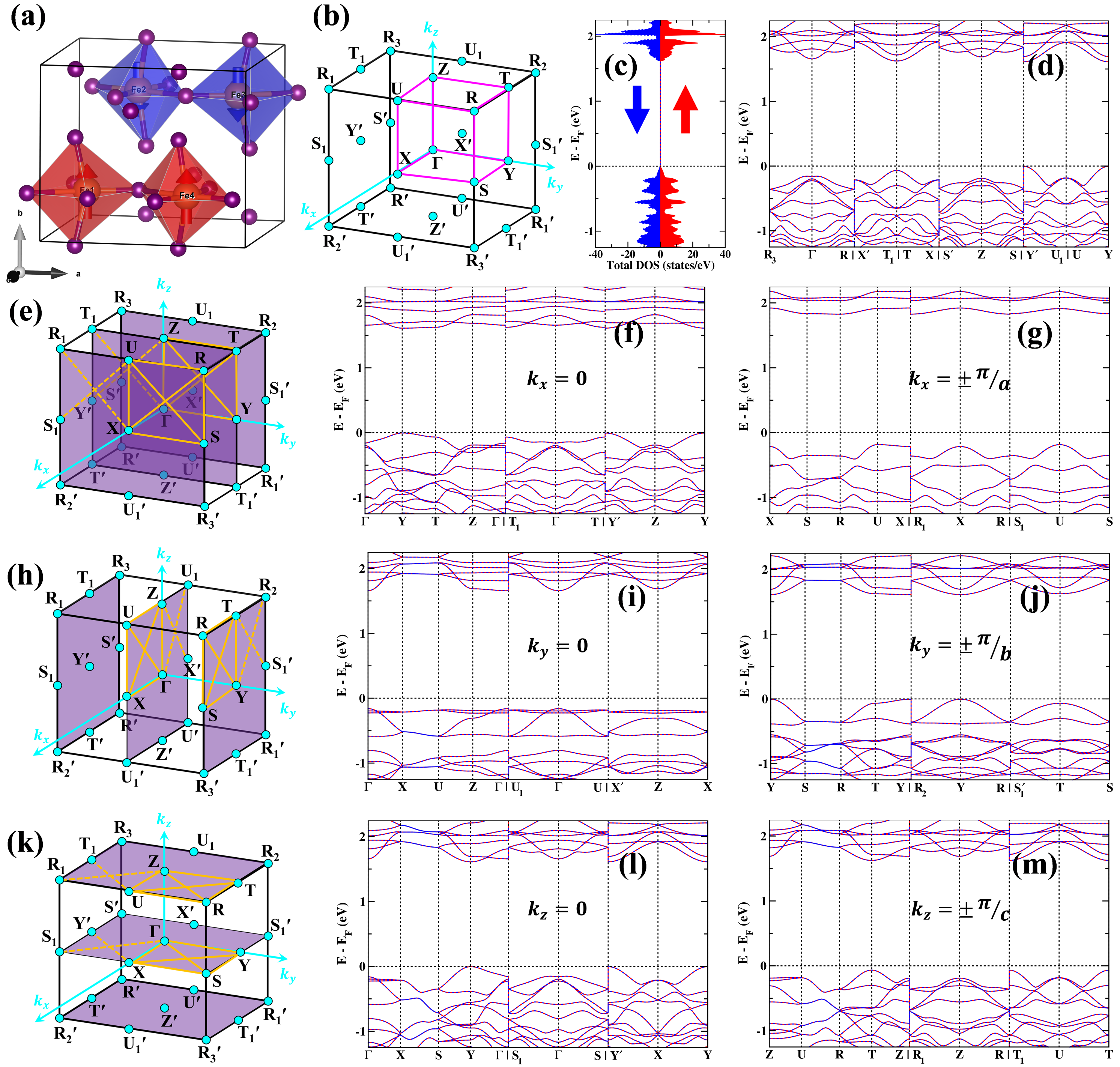}
    \caption{(a) The Fe sublattice with ($+$$-$$-$$+$)-type spin arrangement. The red octahedra denote the up-spin sublattice and the blue octahedra denote the down-spin sublattice. (b) \textit{Primitive} orthorhombic Brillouin zone (BZ) with the high-symmetry points shown as cyan dots and the irreducible Brillouin zone (IBZ) marked by magenta lines. (c) Spin-polarized electronic total density of states (DOS) of CaFePO$_5$. (d) Spin-polarized electronic band structure of CaFePO$_5$ along body diagonals of the IBZ showing spin-degeneracy, characteristic of the antiferromagnetic phase. (e) BZ with the $k_x=0$ and $k_x=\pm\frac{\pi}{a}$ planes highlighted in purple. The high-symmetry directions on the respective planes have been highlighted by orange lines and the spin-polarized electronic band structures of CaFePO$_5$ along these lines have been shown in (f) and (g). (h-j) The same for $k_y=0$ and $k_y=\pm\frac{\pi}{b}$ planes. (k-m) The same for $k_z=0$ and $k_z=\pm\frac{\pi}{c}$ planes. In (e), (h) and (k), the solid orange lines denote paths within the IBZ while the broken orange lines denote paths outside the IBZ.}
    \label{fig8}
\end{figure*}
Another case would be when Fe1 \& Fe4 occupying $(\text{x},\sfrac{1}{4},\text{z})$ \& $(\text{x}+\sfrac{1}{2},\sfrac{1}{4},-\text{z}+\sfrac{1}{2})$ lattice points respectively become the up-spin atoms and Fe2 \& Fe3 occupying $(-\text{x},\sfrac{3}{4},-\text{z})$ \& $(-\text{x}+\sfrac{1}{2},\sfrac{3}{4},\text{z}+\sfrac{1}{2})$ lattice points respectively become the down-spin atoms as shown in \textcolor{black}{Fig.~\ref{fig8}(a)}. Let us denote this spin arrangement by ($+$$-$$-$$+$). The \textit{site-symmetry} group $\mathbf{W} \cong \bm{C_s}$ again maps every lattice point in 4c Wyckoff position to itself . But the space group operations exchanging the lattice points occupied by the up-spin atoms Fe1 \& Fe4 and the down-spin atoms Fe2 \& Fe3 respectively differ again:
\begin{eqnarray*}
    \text{Fe1: }(\text{x},\sfrac{1}{4},\text{z})\,&\overset{\tilde{C}_{2x}}{\underset{\tilde{m}_{xy}}\longleftrightarrow}&\,\text{Fe4: }(\text{x}+\sfrac{1}{2},\sfrac{1}{4},-\text{z}+\sfrac{1}{2}) \\
    \text{Fe2: }(-\text{x},\sfrac{3}{4},-\text{z})\,&\overset{\tilde{C}_{2x}}{\underset{\tilde{m}_{xy}}\longleftrightarrow}&\,\text{Fe3: }(-\text{x}+\sfrac{1}{2},\sfrac{3}{4},\text{z}+\sfrac{1}{2})
\end{eqnarray*}
The group of same-spin sublattice transformations that can be constructed using $\mathbf{W}$ and the space group operations $\tilde{C}_{2x}$ and $\tilde{m}_{xy}$ is $\mathbf{W}\,\cup\,\{\tilde{C}_{2x}\,,\,\tilde{m}_{xy}\} = \{\,\mathbb{I}\,,\,\tilde{m}_{zx}\,,\,\tilde{C}_{2x}\,,\,\tilde{m}_{xy}\,\} = \bm{\tilde{C}_{2v}^{x}} \cong \bm{C_{2v}^{2}}$ (also see Table~\ref{tab1}) which is another \textit{halving} subgroup of $\bm{Pnma}$. This means eq.~\eqref{eq1} again becomes eq.~\eqref{eq11} and ($+$$-$$-$$+$)-type spin arrangement will yield AFM phase instead of AM phase. The SSG for ($+$$-$$-$$+$)-type spin arrangement at 4c Wyckoff position becomes $\bm{P\,^{\overline{1}}n\,^{1}m\,^{1}a}$ (S62.443) \cite{Litvin1977, Aroyo2006a, Chen2025}. \textcolor{black}{Figure~\ref{fig8}(b)} depicts the BZ along with the IBZ. First-principles calculations reveal a fully compensated Fe sublattice as evident from the spin-polarized total density of states (DOS) shown in \textcolor{black}{Fig.~\ref{fig8}(c)} with a spin magnetic moment of $\pm$4.18~$\mu_B$, slightly less than that of Fe$^{3+}$ indicating its covalent character. \textcolor{black}{Figure~\ref{fig8}(d)} shows spin-degeneracy in the electronic band structure of CaFePO$_5$ for ($+$$-$$-$$+$)-type spin arrangement along body diagonals of the IBZ, as expected of a conventional AFM. The spin-degeneracy along the other momentum directions as a consequence of the SSG symmetries, namely the $k_x=0,\pm\sfrac{\pi}{a}$, $k_y=0,\pm\sfrac{\pi}{b}$ and $k_z=0,\pm\sfrac{\pi}{c}$ planes, are shown in \textcolor{black}{Figs.~\ref{fig8}(e–m)}. \\

\noindent Recalling the discussions in Sec.~\ref{FLAM}, it should now be clear that 4c Wyckoff position is indeed \textit{partially compatible} because only for the specific ($+$$-$$+$$-$)-type spin arrangement, the AM phase is recovered. Any other fully compensated spin arrangement leads to the AFM phase.

\subsection{Magnetic ground state of \texorpdfstring{CuFePO$_5$}{CuFePO5}} \label{CuFePO5_neutron}

\begin{table*}[!htb]
    \begin{ruledtabular}
    \caption {Ground state energy of possible magnetic configurations of CuFePO$_5$ in the absence of SOC.}
    \label{tab2}
        \begin{tabular}{c c c c}
            \multirow{2}{*}{\textbf{\shortstack{Spin arrangement of \\ Cu sublattice}}} & \multirow{2}{*}{\textbf{\shortstack{Spin arrangement of \\ Fe sublattice}}} &  \multirow{2}{*}{\textbf{\shortstack{Relative ground state \\ energy per atom (meV)}}} & \multirow{2}{*}{\textbf{\shortstack{Maximum spin-splitting \\ (meV)}}} \\
            & & &\\
            \hline\hline
            A & \multirow{3}{*}{$+$$-$$+$$-$} & 8.37 & 153.10 \\
            C & & 0.00 & 173.00 \\
            G & & 8.39 & 144.10 \\
            \hline
            A & \multirow{3}{*}{$+$$+$$-$$-$} & 7.17 & 89.10 \\
            C & & 7.68 & 118.30 \\
            G & & 7.18 & 177.90 \\
            \hline
            A & \multirow{3}{*}{$+$$-$$-$$+$} & 14.11 & 121.20 \\
            C & & 14.57 & 74.70 \\
            G & & 14.13 & 146.70 \\
        \end{tabular}
    \end{ruledtabular}
\end{table*}

\begin{figure*}
    \centering
    \includegraphics[width=\linewidth]{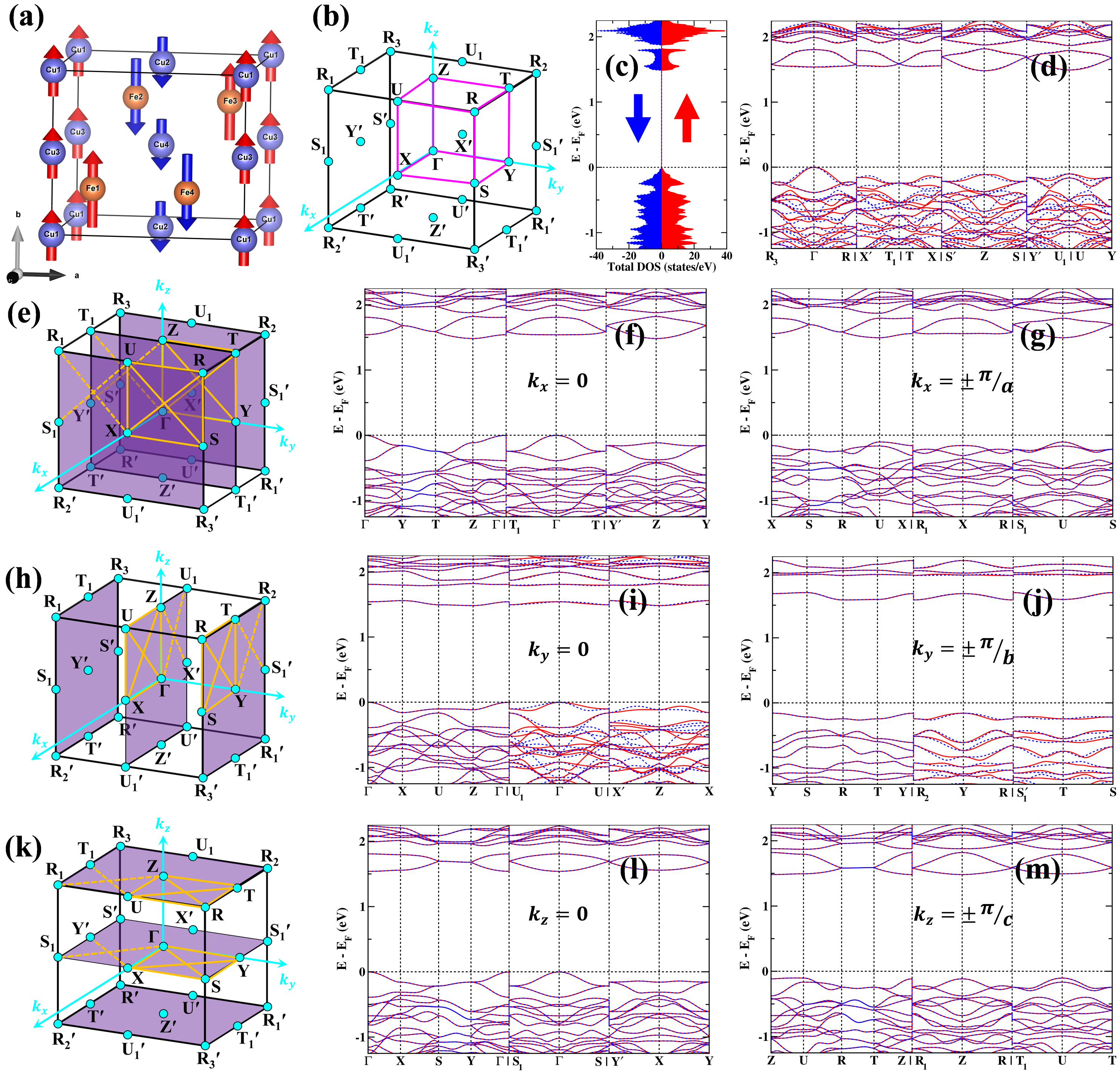}
    \caption{(a) The Cu and Fe sublattices with C-type and ($+$$-$$+$$-$)-type spin arrangement respectively. Only the magnetic atoms are shown for visual clarity. (b) \textit{Primitive} orthorhombic Brillouin zone (BZ) with the high-symmetry points shown as cyan dots and the irreducible Brillouin zone (IBZ) marked by magenta lines. (c) Spin-polarized electronic total density of states (DOS) of CuFePO$_5$. (d) Spin-polarized electronic band structure of CuFePO$_5$ along body diagonals of the IBZ showing spin-splitting, characteristic of the altermagnetic phase. (e) BZ with the $k_x=0$ and $k_x=\pm\frac{\pi}{a}$ planes highlighted in purple. The high-symmetry directions on the respective planes have been highlighted by orange lines and the spin-polarized electronic band structures of CuFePO$_5$ along these lines have been shown in (f) and (g). (h-j) The same for $k_y=0$ and $k_y=\pm\frac{\pi}{b}$ planes. (k-m) The same for $k_z=0$ and $k_z=\pm\frac{\pi}{c}$ planes. In (e), (h) and (k), the solid orange lines denote paths within the IBZ while the broken orange lines denote paths outside the IBZ.}
    \label{fig9}
\end{figure*}

In the previous subsections (Secs.~\ref{Cu_CuFePO5} and \ref{Fe_CuFePO5}), the fully compensated magnetic-sublattices of Cu and Fe were treated separately. In such settings, identifying the same-spin and opposite-spin sublattices posed no difficulty, because each magnetic-sublattice could be analyzed in isolation. The natural question now is: how should same-spin and opposite-spin sublattices be identified when both the magnetic species (Cu and Fe) are considered together ? 
Up till now, the literature on collinear magnetic phases formulated within the SSG framework has answered this question in the negative --- namely, that such a distinction of  same-spin and opposite-spin sublattices \textit{cannot} be defined in the FiM phase \cite{Smejkal2022a}. 
This is because the magnetic ion of one species can never be mapped onto a distinct magnetic species by any crystallographic space group operation (as discussed in Sec.~\ref{intro}). However, for the special class of AFiMs, we shall now provide an affirmative answer. In conventional fully compensated FiMs, the different magnetic species compensate each other meaning the moment of one magnetic species is nullified by the moment of another distinct magnetic species. Because AFiMs consist of fully compensated magnetic-sublattices for each individual magnetic species, the moment of one magnetic species will be nullified by the equal and opposite moment of the same magnetic species. This is true for every magnetic species present in an AFiM. Hence,
the SSG analysis for spin-sublattices associated with a distinct magnetic species remains well-defined. The same-spin and opposite-spin sublattice transformations act on each magnetic-sublattice \textit{independently}, without ever connecting different magnetic species and the full SSG symmetries of an AFiM are determined solely by those transformations that are \textit{common} to all magnetic-sublattices.

Within the $\bm{Pnma}$ space group, each of the 4a and 4c Wyckoff positions admits three distinct fully compensated collinear spin arrangements with magnetic propagation vector $\vec{\bm{q}}=0$, as discussed in Secs.~\ref{Cu_CuFePO5}~and~\ref{Fe_CuFePO5}. When both the Wyckoff positions are simultaneously occupied by magnetic species --- as in  CuFePO$_5$ --- these individual arrangements combine to yield nine possible configurations in total. For CuFePO$_5$, the relative ground state energy per atom and the corresponding maximum spin-splitting for all nine configurations are summarized in  Table~\ref{tab2}, with their electronic structures shown in \textcolor{black}{Figs.~\ref{fig9}–\ref{fig11}} and \textcolor{black}{supplementary figures \ref{fig1S}–\ref{fig6S}} in Supplementary Material (SM) \cite{supp}. 
The magnetic ground state of CuFePO$_5$ is in good agreement with earlier neutron diffraction studies \cite{Touaiher1994, Khayati2000, Khayati2001} where the Cu sublattice adopts a collinear C-type spin arrangement and the Fe sublattice exhibits a collinear ($+$$-$$+$$-$)-type spin arrangement as shown in \textcolor{black}{Fig.~\ref{fig9}(a)}. Our independent analysis of the Cu and Fe sublattices in the previous sections (see Secs.~\ref{Cu_C}~and~\ref{Fe_pnpn}) have identified the same \textit{halving} subgroup with the same orientation, namely $\hsgtwo = \bm{\tilde{C}_{2h}^{y}}$, for both magnetic-sublattices. This implies that all the \textit{real-space} operations in $\hsgtwo$ yield the same-spin sublattice transformations for both Cu and Fe sublattices, while the corresponding coset $\csg - \hsgtwo$ with $\csg=\bm{Pnma}$ generates the opposite-spin sublattice transformations. In other words, the \textit{real-space} operations in $\bm{\tilde{C}_{2h}^{y}}$ are \textit{common} to both Cu and Fe sublattices and the non-trivial spin group of CuFePO$_5$ can be constructed as follows:
\begin{equation}
    \ntsg = \ntsgspin \otimes \csg = [\,\mathfrak{I}\,||\,\bm{\tilde{C}_{2h}^{y}}\,] \cup [\,\mathfrak{C}_2\,||\,\csg-\bm{\tilde{C}_{2h}^{y}}\,]
    \label{eq12}
\end{equation}

Thus, C-type spin arrangement on the Cu sublattice and ($+$$-$$+$$-$)-type spin arrangement on the Fe sublattice in CuFePO$_5$ leads to SSG $\bm{P\,^{\overline{1}}n\,^{1}m\,^{\overline{1}}a}$ (S62.448) \cite{Litvin1977, Aroyo2006a, Chen2025}. From eqs.~\eqref{eq1}, \eqref{eq7} and \eqref{eq12}, all the resulting  spin-degenerate nodal lines and nodal planes across the BZ can be systematically mapped out using the \textit{isomorphic} spin point groups, as discussed in  Secs.~\ref{Cu_C} and \ref{Fe_pnpn}. \textcolor{black}{Figure~\ref{fig9}(b)} shows the BZ along with the IBZ. First-principles calculations reveal fully compensated nature of both Cu and Fe sublattices which is evident from the spin-polarized total DOS as shown in \textcolor{black}{Fig.~\ref{fig9}(c)}. The computed spin magnetic moment of Cu is $\pm$0.63~$\mu_B$ and that of Fe is $\pm$4.17~$\mu_B$ which is slightly less than that of Cu$^{2+}$ and Fe$^{3+}$ indicating their covalent nature. The resulting altermagnetic spin-splitting for these spin arrangements along the IBZ body diagonals is shown in \textcolor{black}{Fig.~\ref{fig9}(d)}. The spin-degeneracies and splittings on the high symmetry planes $k_x=0,\pm\sfrac{\pi}{a}$, $k_y=0,\pm\sfrac{\pi}{b}$ and $k_z=0,\pm\sfrac{\pi}{c}$, as enforced by the SSG symmetries, are displayed in \textcolor{black}{Figs.~\ref{fig9}(e–m)}.

There is, however an important subtlety. A \textit{real-space} operation in the parent crystallographic space group of an AFiM may map one same-spin site to another same-spin site within a magnetic-sublattice, while the same operation maps opposite-spin sites for another magnetic-sublattice. In such a case, the operation will not be a valid SSG symmetry of the AFiM, because it is not \textit{common} across all the different magnetic species. Consequently, the full SSG of the AFiM must exclude that \textit{real-space} operation, leading to a lowering of SSG symmetries relative to the crystallographic space group symmetries. We examine this scenario in detail in the next subsection. \\

\subsection{Alterferrimagnetic states of lower crystallographic symmetry} \label{CuFePO5_lowersym}

The neutron diffraction structure of CuFePO$_5$ is a relatively simple case of AFiM where both Cu and Fe sublattices have the same underlying SSG. As discussed in Sec.~\ref{CuFePO5_neutron} and listed in Table~\ref{tab2}, there are nine possible spin arrangements in CuFePO$_5$. Fixing ($+$$-$$+$$-$)-type spin arrangement at the 4c Wyckoff position, which yields AM phase within the Fe sublattice, there are three possibilities at the 4a Wyckoff position, out of which C-type spin arrangement within the Cu sublattice corresponds to the magnetic ground state of CuFePO$_5$ as explained in detail. The other two possibilities would be candidates for AFiMs with lower crystallographic symmetry than the parent crystallographic space group.

\begin{figure*}
    \centering
    \includegraphics[width=\linewidth]{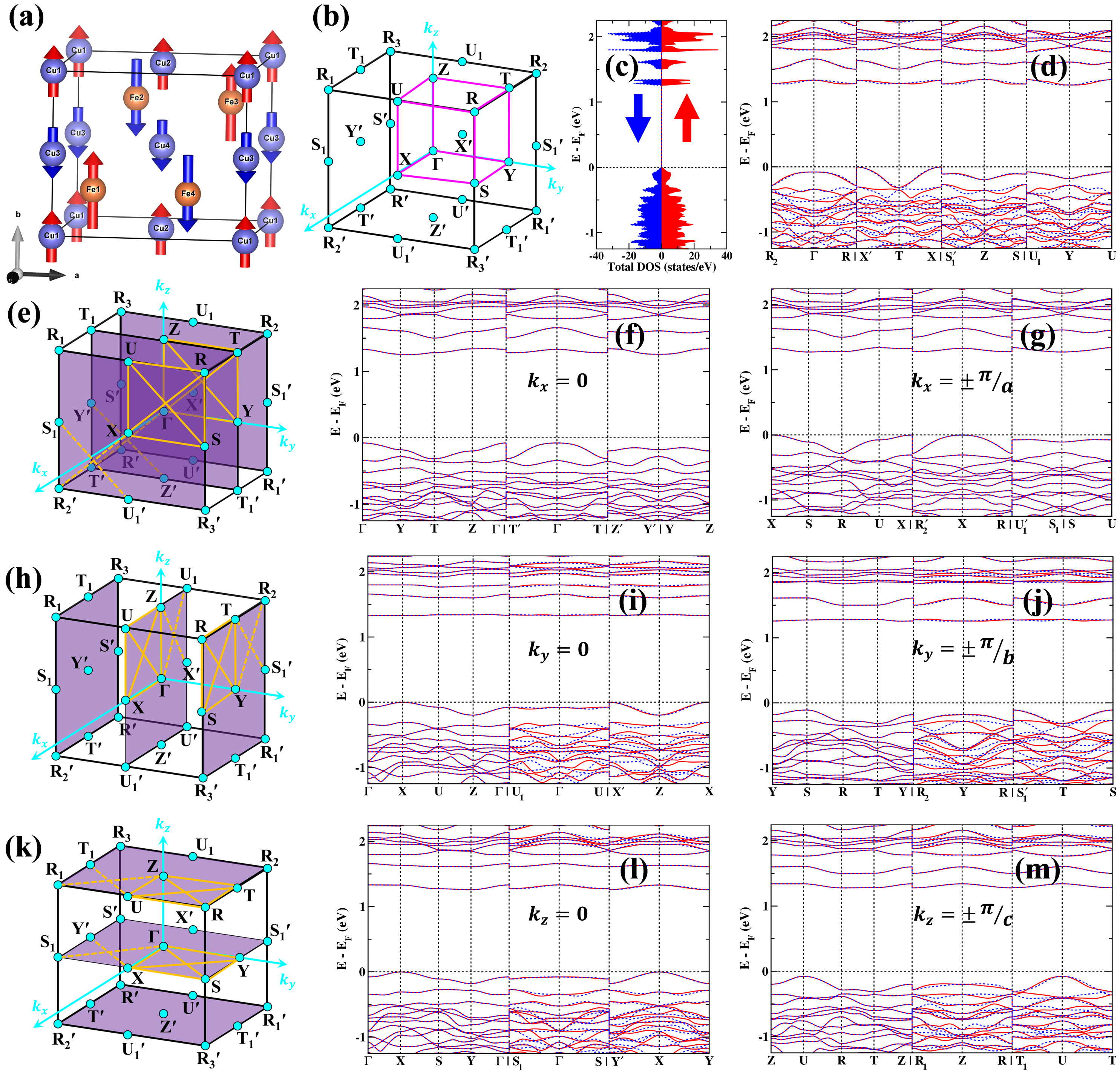}
    \caption{(a) The Cu and Fe sublattices with A-type and ($+$$-$$+$$-$)-type spin arrangement respectively. Only the magnetic atoms are shown for visual clarity. (b) \textit{Primitive} orthorhombic Brillouin zone (BZ) with the high-symmetry points shown as cyan dots and the irreducible Brillouin zone (IBZ) marked by magenta lines. (c) Spin-polarized electronic total density of states (DOS) of CuFePO$_5$. (d) Spin-polarized electronic band structure of CuFePO$_5$ along body diagonals of the IBZ showing spin-splitting, characteristic of the altermagnetic phase. (e) BZ with the $k_x=0$ and $k_x=\pm\frac{\pi}{a}$ planes highlighted in purple. The high-symmetry directions on the respective planes have been highlighted by orange lines and the spin-polarized electronic band structures of CuFePO$_5$ along these lines have been shown in (f) and (g). (h-j) The same for $k_y=0$ and $k_y=\pm\frac{\pi}{b}$ planes. (k-m) The same for $k_z=0$ and $k_z=\pm\frac{\pi}{c}$ planes. In (e), (h) and (k), the solid orange lines denote paths within the IBZ while the broken orange lines denote paths outside the IBZ.}
    \label{fig10}
\end{figure*}

\textcolor{black}{Figure~\ref{fig10}(a)} displays A-type spin arrangement of Cu sublattice and ($+$$-$$+$$-$)-type spin arrangement of Fe sublattice in CuFePO$_5$. From the parent crystallographic space group $\csg = \bm{Pnma}$, the same-spin sublattice transformations for A-type spin arrangement at the 4a Wyckoff position and ($+$$-$$+$$-$)-type spin arrangement at the 4c Wyckoff position form the \textit{halving} subgroups $\hsgone = \bm{\tilde{C}_{2h}^{z}}$ and $\hsgtwo = \bm{\tilde{C}_{2h}^{y}}$ respectively (derived and discussed in Secs.~\ref{Cu_A}, \ref{Cu_C} and \ref{Fe_pnpn}). The cosets $\csg-\hsgone$ and $\csg-\hsgtwo$ provide the respective opposite-spin sublattice transformations. Then, the \textit{common} same-spin and opposite-spin sublattice transformations are $\hsgone \cap \hsgtwo = \{\,\mathbb{I},\inv\,\}$ and  $(\csg-\hsgone) \cap (\csg-\hsgtwo) = \{\,\tilde{C}_{2x}, \tilde{m}_{yz}\,\}$ respectively. Then the reduced crystallographic space group is given by $\bm{\tilde{C}_{2h}^{x}} = \{\,\mathbb{I},\inv,\tilde{C}_{2x}, \tilde{m}_{yz}\,\}$, \textit{isomorphic} to $\bm{P2_1/n\,1\,1}$ or $\bm{P2_1/n}~(\bm{C_{2h}^{5}})$ \cite{Aroyo2006a, IUCr_volA2016}  
and a subgroup of $\csg$ (see Table~\ref{tab1}). The non-trivial spin group becomes:
\begin{equation}
    \ntsg = \ntsgspin \otimes \bm{\tilde{C}_{2h}^{x}} = [\,\mathfrak{I}\,||\,\{\,\mathbb{I},\inv\,\}\,] \cup [\,\mathfrak{C}_2\,||\,\{\,\tilde{C}_{2x}, \tilde{m}_{yz}\,\}\,]
    \label{eq13}
\end{equation}
Thus, A-type spin arrangement on the Cu sublattice and ($+$$-$$+$$-$)-type spin arrangement on the Fe sublattice in CuFePO$_5$ leads to SSG $\bm{P\,^{\overline{1}}2_1/^{\overline{1}}n}$ (S14.79) \cite{Litvin1977, Aroyo2006a, Chen2025}. Using Eqs.~\eqref{eq1}, \eqref{eq4}, \eqref{eq7} and \eqref{eq12}, all the resulting  spin-degenerate nodal lines and nodal planes across the BZ can be systematically mapped out based on the \textit{isomorphic} spin point groups, as discussed in  Secs.~\ref{Cu_A}, \ref{Cu_C} and \ref{Fe_pnpn}. \textcolor{black}{Figure~\ref{fig10}(b)} shows the BZ along with the IBZ. First-principles calculations reveal fully compensated nature of both Cu and Fe sublattices which is evident from the spin-polarized total DOS as shown in \textcolor{black}{Fig.~\ref{fig10}(c)}. 
The resulting altermagnetic spin-splitting for these spin arrangements along the IBZ body diagonals is shown in \textcolor{black}{Figure~\ref{fig10}(d)}. The spin-degeneracies and splittings on the high symmetry planes $k_x=0,\pm\sfrac{\pi}{a}$, $k_y=0,\pm\sfrac{\pi}{b}$ and $k_z=0,\pm\sfrac{\pi}{c}$, as enforced by the SSG symmetries, are displayed in \textcolor{black}{Figs.~\ref{fig10}(e–m)}.

\begin{figure*}
    \centering
    \includegraphics[width=\linewidth]{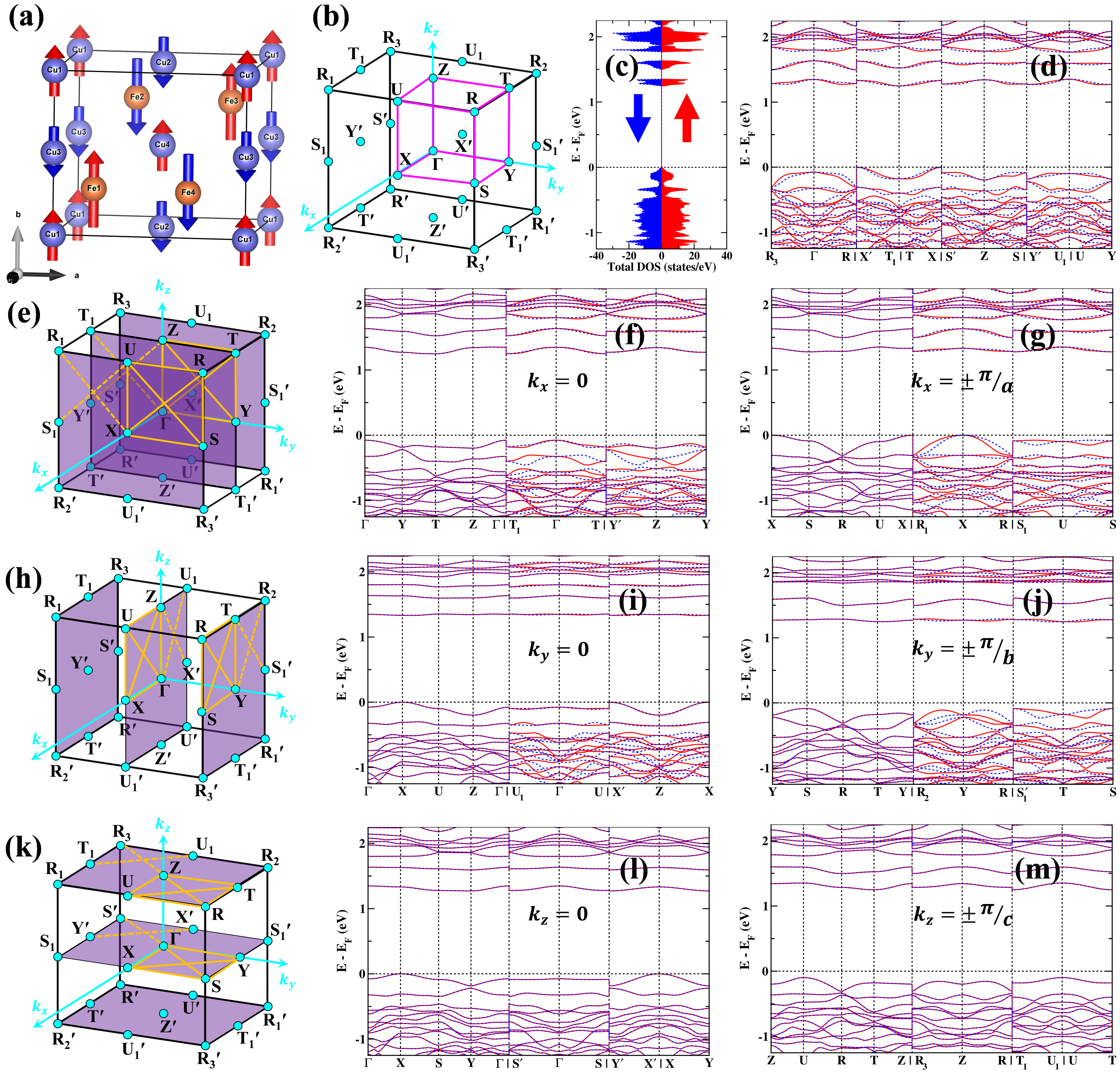}
    \caption{(a) The Cu and Fe sublattices with G-type and ($+$$-$$+$$-$)-type spin arrangement respectively. Only the magnetic atoms are shown for visual clarity. (b) \textit{Primitive} orthorhombic Brillouin zone (BZ) with the high-symmetry points shown as cyan dots and the irreducible Brillouin zone (IBZ) marked by magenta lines. (c) Spin-polarized electronic total density of states (DOS) of CuFePO$_5$. (d) Spin-polarized electronic band structure of CuFePO$_5$ along body diagonals of the IBZ showing spin-splitting, characteristic of the altermagnetic phase. (e) BZ with the $k_x=0$ and $k_x=\pm\frac{\pi}{a}$ planes highlighted in purple. The high-symmetry directions on the respective planes have been highlighted by orange lines and the spin-polarized electronic band structures of CuFePO$_5$ along these lines have been shown in (f) and (g). (h-j) The same for $k_y=0$ and $k_y=\pm\frac{\pi}{b}$ planes. (k-m) The same for $k_z=0$ and $k_z=\pm\frac{\pi}{c}$ planes. In (e), (h) and (k), the solid orange lines denote paths within the IBZ while the broken orange lines denote paths outside the IBZ.}
    \label{fig11}
\end{figure*}

\textcolor{black}{Figure~\ref{fig11}(a)} displays G-type spin arrangement of Cu sublattice and ($+$$-$$+$$-$)-type spin arrangement of Fe sublattice in CuFePO$_5$. The corresponding halving subgroups for the Cu sublattice at the 4a Wyckoff position and the Fe sublattice at the 4c Wyckoff position are $\hsgthree = \bm{\tilde{C}_{2h}^{x}}$ and $\hsgtwo = \bm{\tilde{C}_{2h}^{y}}$, respectively, as derived in Secs.~\ref{Cu_C}, \ref{Cu_G}, and \ref{Fe_pnpn}.
The cosets $\csg-\hsgthree$ and $\csg-\hsgtwo$ provide the respective opposite-spin sublattice transformations. Therefore, the \textit{common} same-spin and opposite-spin sublattice transformations are $\hsgthree \cap \hsgtwo = \{\,\mathbb{I},\inv\,\}$ and  $(\csg-\hsgthree) \cap (\csg-\hsgtwo) = \{\,\tilde{C}_{2z}, \tilde{m}_{xy}\,\}$ respectively. Together these operations define a reduced crystallographic space group $\bm{\tilde{C}_{2h}^{z}} = \{\,\mathbb{I},\inv,\tilde{C}_{2z}, \tilde{m}_{xy}\,\}$, which is \textit{isomorphic} to $\bm{P1\,1\,2_1/a}$ or $\bm{P2_1/a}~(\bm{C_{2h}^{5}})$ \cite{Aroyo2006a, IUCr_volA2016}  
and a subgroup of $\csg$ (see Table~\ref{tab1}). The corresponding non-trivial spin group becomes:
\begin{equation}
    \ntsg = \ntsgspin \otimes \bm{\tilde{C}_{2h}^{z}} = [\,\mathfrak{I}\,||\,\{\,\mathbb{I},\inv\,\}\,] \cup [\,\mathfrak{C}_2\,||\,\{\,\tilde{C}_{2z}, \tilde{m}_{xy}\,\}\,]
    \label{eq14}
\end{equation}
Thus, G-type spin arrangement on the Cu sublattice and ($+$$-$$+$$-$)-type spin arrangement on the Fe sublattice in CuFePO$_5$ leads to SSG $\bm{P\,^{\overline{1}}2_1/^{\overline{1}}a}$ (S14.79) \cite{Litvin1977, Aroyo2006a, Chen2025}. From eqs.~\eqref{eq1}, \eqref{eq7}, \eqref{eq10} and \eqref{eq14}, all the resulting  spin-degenerate nodal lines and nodal planes across the BZ can be systematically mapped out using the \textit{isomorphic} spin point groups, as discussed in  Secs.~\ref{Cu_C}, \ref{Cu_G} and \ref{Fe_pnpn}. \textcolor{black}{Figure~\ref{fig11}(b)} shows the BZ along with the IBZ. First-principles calculations reveal fully compensated nature of both Cu and Fe sublattices which is evident from the spin-polarized total DOS as shown in \textcolor{black}{Fig.~\ref{fig11}(c)}. The resulting altermagnetic spin-splitting for these spin arrangements along the IBZ body diagonals is shown in \textcolor{black}{Figure~\ref{fig11}(d)}. The spin-degeneracies and splittings on the high symmetry planes $k_x=0,\pm\sfrac{\pi}{a}$, $k_y=0,\pm\sfrac{\pi}{b}$ and $k_z=0,\pm\sfrac{\pi}{c}$, as enforced by the SSG symmetries, are displayed in \textcolor{black}{Figs.~\ref{fig11}(e–m)}.

It must be noted that the orientational degree of freedom of the symmetry operations in $\bm{Pnma}$ gives rise to an inherent ambiguity in denoting the crystallographic space group and spin space groups using conventional notations (also see Sec. Appendix~\ref{methods}). Nevertheless, it is this orientational degree of freedom which is responsible for tuning the symmetries and hence the electronic structure of CuFePO$_5$. The other six configurations of possible spin arrangements in CuFePO$_5$ have been discussed in SM \cite{supp}. Since in these configurations the spin arrangement of Fe sublattice yields the AFM phase, the spin arrangement of Cu sublattice shall be mostly responsible for the overall AFiM nature with extra spin-degeneracies arising from the spin only group $\sog$. The spin-splitting and degeneracies would be similar to what has been shown in Sec.~\ref{Cu_CuFePO5}. We note here that all the discussions on CuFePO$_5$ are equally valid for other members of the family of metal-oxide iron phosphates, namely NiFePO$_5$, CoFePO$_5$ and Fe$_2$PO$_5$ with similar spin-arrangements as CuFePO$_5$.

\section{Conclusion} \label{conc}

The fundamental lemma of altermagnetism provides a novel and general approach to discovering and categorizing altermagnets within the spin space group formalism. The central concept of \textit{halving} subgroups is now generalized to include the \textit{site-symmetry} groups associated with the Wyckoff position of a given crystallographic space group, giving rise to four distinct subcases depending on the underlying Bravais lattice. From the perspective of Material Sciences, this is of paramount importance because simply knowing the crystallographic space group and the Wyckoff positions of a synthesized material (information usually obtained during sample characterization) can give insights into the existence of altermagnetic phase, without looking up extensive tables of spin space groups or performing first-principles calculations to determine whether the material is altermagnetic or not. The FLAM as proposed by us has the constraint of being applicable to magnetic ordering in crystalline materials with propagation vector $\vec{\bm{q}} = 0$ such that the magnetic unit cell and the crystallographic unit cell coincide. This may be generalized to the case of the so-called supercell altermagnets \cite{Ubiergo2024} having a lower crystallographic symmetry. In fact for a material having some magnetic order with non-zero propagation vector, if one reduces the magnetic space group to the crystallographic space group by replacing the anti-unitary operations with the unitary ones, FLAM may still be able to correctly deduce the existence of AM phase within the spin space group formalism although verifying this requires more work.

An important consequence of FLAM is the emergence of a new class of ferrimagnets---termed \textit{Alterferrimagnets}---which can be viewed as a natural generalization of altermagnets. Whereas conventional FiMs lack the distinction of same-spin and opposite-spin sublattice transformations, AFiMs are formed from interpenetrating fully compensated magnetic-sublattices with well-defined same-spin and opposite-spin sublattice transformations similar to AMs. Generally in AMs, the parent crystallographic space group is \textit{isomorphic} to the non-trivial spin space group, i.e., the \textit{real-space} symmetries remain unchanged. But in AFiMs, the \textit{real-space} operations of the non-trivial spin space group form a proper or improper subgroup of the parent crystallographic space group. As a result, the spin space group landscape of AFiMs are richer than that of AMs. In this work, we have proposed MFePO$_5$ (M=Fe,Cu,Ni,Co) as a family of candidate AFiMs, but we anticipate that such phases are far more prevalent among fully compensated FiMs.
Future studies will focus on systematic searches for AFiMs and on exploring how their reduced SSG symmetries influence the electronic and transport properties. Owing to their inherent momentum-dependent spin-splitting, AFiMs are expected to exhibit direction-dependent anomalous transport responses. Moreover, the presence of multiple magnetic species within a single material provides an additional degree of tunability---absent in conventional AMs---which may be exploited for the design of novel spintronic devices.

\appendix
\setcounter{equation}{0}
\renewcommand{\thesubsection}{A\arabic{subsection}}
\renewcommand{\theequation}{\thesubsection.\arabic{equation}}

\section*{Appendix}
\subsection{Reviewing Spin Space Groups} \label{ssg}

Here we briefly review several key concepts of spin space groups, which have been used extensively in this work, and elucidate how altermagnetism has been defined in the literature on their basis \cite{Litvin1974, Litvin1977, Liu2022, Smejkal2022a, Smejkal2022b, Xiao2024, Chen2024, Jiang2024}. The crystalline Bravais lattice is described by \textit{polar} vectors residing in a three-dimensional (3D) Euclidean vector space, the position space. The reciprocal Bravais lattice is then described by \textit{polar} vectors residing in an \textit{isomorphic} 3D Euclidean vector space, the momentum space. The position and momentum spaces are collectively termed as the \textit{real-space}. In a crystalline magnetic material, the interacting magnetic moments of basis atoms, primarily arising due to spins of unpaired electrons, give rise to a magnetic order with a definite spin arrangement in the crystalline lattice. Treating the moments (spins) as \textit{axial} vectors in yet another 3D Euclidean vector space, termed the \textit{spin-space}, the spin arrangement dictates how spins in a magnetic material are spatially arranged, essentially providing a mapping between the \textit{spin-space} and the \textit{real-space}.  
Let us now define the SSGs in a mathematically rigorous manner. Formally, a spin space group $\ssg$ is obtained by taking the \textit{external} direct product of a group of transformations $\gspin$ residing in the \textit{spin-space} and another group $\gspace$ residing in the \textit{real-space}, i.e., $\ssg = \gspin \otimes \gspace$. Let us now define the subgroups $\gspinnorm \subseteq \gspin$ and $\gspacenorm \subseteq \gspace$ such that $\gspin$ and $\gspace$ can be respectively partitioned into disjoint sets called cosets. If $\gspinnorm$ and $\gspacenorm$ are further considered to be \textit{normal} in $\gspin$ and $\gspace$ (denoted by $\gspinnorm \triangleleft \gspin$ and $\gspacenorm \triangleleft \gspace$) respectively, then the set of corresponding disjoint cosets also form a group yielding the quotient groups $\gspin\!/\!\gspinnorm$ and $\gspace\!/\!\gspacenorm$. Choosing $\gspinnorm \triangleleft \gspin$ and $\gspacenorm \triangleleft \gspace$ such that the quotient groups are \textit{isomorphic} $(\gspin\!/\!\gspinnorm \cong \gspace\!/\!\gspacenorm)$, the \textit{isomorphic} coset decomposition of $\gspin$ and $\gspace$ is given by:
\begin{eqnarray*}
    \gspin =& \mathsf{s}_1\gspinnorm \cup \mathsf{s}_2\gspinnorm \cup ... \cup \mathsf{s}_n\gspinnorm &\, (\mathsf{s}_i\in\gspin \;\forall i = 1,...,n) \\
    \gspace =& \mathsf{x}_1\gspacenorm \cup \mathsf{x}_2\gspacenorm \cup ... \cup \mathsf{x}_n\gspacenorm &\, (\mathsf{x}_i\in\gspace \;\forall i = 1,...,n)
\end{eqnarray*}
Here, $n$ is the total number of distinct cosets corresponding to a subgroup and is referred to as the \textit{index} of the subgroup in the parent group. Pairing the elements of the cosets $\mathsf{s}_i\gspinnorm$ and $\mathsf{x}_i\gspacenorm$, the group elements of $\ssg$ are obtained using the \textit{isomorphism} theorem for groups \cite{Litvin1974} as an ordered pair $[\,\mathsf{s}_i\,||\,\mathsf{x}_i\,]$, where $\mathsf{s}_i \in \gspin$ and $\mathsf{x}_i \in \gspace$ ($\mathsf{s}_1$ and $\mathsf{x}_1$ are identities of $\gspin$ and $\gspace$ respectively). Then the full SSG can be written as:
\begin{gather}
    \text{\small{$\ssg = \gspin \!\otimes\! \gspace = [\gspinnorm||\gspacenorm] \cup [\mathsf{s}_2\gspinnorm||\mathsf{x}_2\gspacenorm] \cup\!...\!\cup [\mathsf{s}_n\gspinnorm||\mathsf{x}_n\gspacenorm]$}}
    \label{eqA1}
\end{gather}
Thus to construct a spin space group $\ssg$, it is imperative to find or choose subgroups $\gspinnorm$ and $\gspacenorm$ such that they are \textit{normal} in $\gspin$ and $\gspace$ respectively. It is only under this condition that the set of cosets will form quotient groups and the \textit{isomorphic} coset decomposition can be performed.

We now look at the structure of $\ssg$ in more detail \cite{Litvin1974, Litvin1977, Liu2022, Smejkal2022a, Smejkal2022b}.
In a SSG, there exists transformations of the \textit{spin-space} alone of the form $[\,\mathfrak{R}\,||\,\mathbb{I}\,]$. These transformations yield a \textit{normal} subgroup $\sog \triangleleft \ssg$ called the spin-only group. All other transformations which act on both \textit{real-space} and \textit{spin-space} form yet another \textit{normal} subgroup $\ntsg \triangleleft \ssg$ called the non-trivial spin group. They are of the general form $[\,\mathfrak{Q}\,||\,R\,]$ and do not contain any elements of the form $[\,\mathfrak{R}\,||\,\mathbb{I}\,]$ except the group identity $[\,\mathfrak{I}\,||\,\mathbb{I}\,]$. Since both $\sog$ and $\ntsg$ are \textit{normal} in $\ssg$, an alternate way of characterizing $\ssg$ is by the \textit{internal} direct product $\ssg = \sog \times \ntsg$ as described in Sec.~\ref{AM}. The spin-only group $\sog$ depends only on the type of spin arrangement in a magnetic crystal and is common to all SSGs since it is independent of any \textit{real-space} transformations. By contrast, the non-trivial spin group $\ntsg$ depends on the crystal itself via the crystallographic space group $\csg$ since $R \in \csg$ and distinguishes the different SSGs for a particular spin arrangement. Hence, finding the non-trivial spin groups is sufficient to enumerate and construct all possible SSGs. The primary definition of a SSG is that it must leave the spin arrangement in a crystal invariant and since both $\sog$ and $\ntsg$ are SSGs by themselves, this leads to the following interesting observations. The \textit{spin-space} transformations $\mathfrak{R}$ of $[\,\mathfrak{R}\,||\,\mathbb{I}\,] \in \sog$ can keep a spin arrangement invariant even without considering any \textit{real-space} transformations. These underlying \textit{spin-space} transformations form a \textit{normal} subgroup $\gspinnorm \triangleleft \gspin$, i.e., $\mathfrak{R} \in \gspinnorm$. By contrast, the \textit{spin-space} transformations $\mathfrak{Q}$ of $[\,\mathfrak{Q}\,||\,R\,] \in \ntsg$ are able to keep a spin arrangement invariant only in conjunction with the \textit{real-space} transformations $R$ since $[\,\mathfrak{R}\,||\,\mathbb{I}\,] \notin \ntsg \,\forall\, \mathfrak{R} \in \gspinnorm$, except $[\,\mathfrak{I}\,||\,\mathbb{I}\,]$. These underlying \textit{spin-space} transformations form yet another \textit{normal} subgroup $\ntsgspin \triangleleft \gspin$ such that $\gspinnorm \cap \ntsgspin = \{\mathfrak{I}\}$ and $\mathfrak{Q} \in \ntsgspin$. Thus, $\ntsg$ can be written in the form:
\begin{equation}
    \ntsg = \ntsgspin \otimes \csg
    \label{eqA2}
\end{equation}
The elements of $\ntsg$ are then obtained from an \textit{isomorphic} coset decomposition of the group of \textit{spin-space} transformations $\ntsgspin$ and the crystallographic space group $\csg$. The parent group of \textit{spin-space} transformations is then given by $\gspin = \gspinnorm \times \ntsgspin$ \cite{Litvin1974}.\\

\subsection{Methods} \label{methods}

We have primarily used the \textit{International Tables of Crystallography, Volume A} (ITA) convention \cite{Aroyo2006a, IUCr_volA2016} while denoting the crystallographic space groups, except in Table~\ref{tab1} where the standard convention \cite{IUCr_volA2016} has been used. The same convention has been extended to denote the spin space groups and thus may differ from the convention adopted in \cite{Litvin1977} or \cite{Chen2025}.

First-principles calculations within the framework of Kohn-Sham Density functional theory (KS-DFT) \cite{Hohenberg_Kohn1964, Kohn_Sham1965} were carried out using the Vienna Ab-initio Simulation Package (VASP) \cite{Kresse1993, Kresse1996a, Kresse1996b}, which is based on projector augmented wave (PAW) formalism \cite{Kresse1999, Blochl1994a}. The Perdew, Burke, and Ernzerhof (PBE) \cite{Perdew1996} exchange-correlation potential within the generalized gradient approximation (GGA) was employed.  Spin-polarized calculations have been performed without spin-orbit coupling (SOC). A Hubbard U of 3.0 eV was applied on $d$-orbitals of Cu and Fe using Dudarev's method \cite{Dudarev1998, Rohrbach2003} to capture the correlation effects. Due to the substantial energy difference of the magnetic configurations of CuFePO$_5$ relative to the ground state as listed in Table~\ref{tab2}, the use of Hubbard U is required to stabilize the other configurations, otherwise they converge only to the magnetic ground state Cu(C)–Fe($+$$-$$+$$-$). Brillouin zone (BZ) integration was performed on a 8$\times$10$\times$8 $\Gamma$-centered $k$-mesh using the Bl\"ochl-corrected linear tetrahedron method \cite{Blochl1994b} with the total energy convergence criteria set to $10^{-6}$ eV. A plane wave energy cutoff of 500 eV was used for all the calculations. For full magnetic compensation, the symmetric reduction of $k$-mesh within the BZ was switched off.

\begin{acknowledgments}
    CKB, AF and FB acknowledges support from the Italian Ministry of University and Research (MUR), financed by the European Union – Next Generation EU, through the PRIN 2022 project SUBLI ``Sustainable spin generators based on Van der Waals dichalcogenides,'' contract No. 2022M3WXE7 and from the PRIN 2022 project TOTEM ``Engineering topological quantum phases in hexagonal ternary compounds'' contract No. 2022HTPC2B. BD thanks IIT Bombay and Ministry of Education, Govt. of India, for financial support. 
\end{acknowledgments}
\unappendix
\bibliography{main}

\begin{thebibliography}{4}%
\makeatletter
\providecommand \@ifxundefined [1]{%
 \@ifx{#1\undefined}
}%
\providecommand \@ifnum [1]{%
 \ifnum #1\expandafter \@firstoftwo
 \else \expandafter \@secondoftwo
 \fi
}%
\providecommand \@ifx [1]{%
 \ifx #1\expandafter \@firstoftwo
 \else \expandafter \@secondoftwo
 \fi
}%
\providecommand \natexlab [1]{#1}%
\providecommand \enquote  [1]{``#1''}%
\providecommand \bibnamefont  [1]{#1}%
\providecommand \bibfnamefont [1]{#1}%
\providecommand \citenamefont [1]{#1}%
\providecommand \href@noop [0]{\@secondoftwo}%
\providecommand \href [0]{\begingroup \@sanitize@url \@href}%
\providecommand \@href[1]{\@@startlink{#1}\@@href}%
\providecommand \@@href[1]{\endgroup#1\@@endlink}%
\providecommand \@sanitize@url [0]{\catcode `\\12\catcode `\$12\catcode `\&12\catcode `\#12\catcode `\^12\catcode `\_12\catcode `\%12\relax}%
\providecommand \@@startlink[1]{}%
\providecommand \@@endlink[0]{}%
\providecommand \url  [0]{\begingroup\@sanitize@url \@url }%
\providecommand \@url [1]{\endgroup\@href {#1}{\urlprefix }}%
\providecommand \urlprefix  [0]{URL }%
\providecommand \Eprint [0]{\href }%
\providecommand \doibase [0]{https://doi.org/}%
\providecommand \selectlanguage [0]{\@gobble}%
\providecommand \bibinfo  [0]{\@secondoftwo}%
\providecommand \bibfield  [0]{\@secondoftwo}%
\providecommand \translation [1]{[#1]}%
\providecommand \BibitemOpen [0]{}%
\providecommand \bibitemStop [0]{}%
\providecommand \bibitemNoStop [0]{.\EOS\space}%
\providecommand \EOS [0]{\spacefactor3000\relax}%
\providecommand \BibitemShut  [1]{\csname bibitem#1\endcsname}%
\let\auto@bib@innerbib\@empty
\bibitem [{\citenamefont {Aroyo}\ \emph {et~al.}(2006)\citenamefont {Aroyo}, \citenamefont {Perez-Mato}, \citenamefont {Capillas}, \citenamefont {Kroumova}, \citenamefont {Ivantchev}, \citenamefont {Madariaga}, \citenamefont {Kirov},\ and\ \citenamefont {Wondratschek}}]{Aroyo2006a_sm}%
  \BibitemOpen
  \bibfield  {author} {\bibinfo {author} {\bibfnamefont {M.~I.}\ \bibnamefont {Aroyo}}, \bibinfo {author} {\bibfnamefont {J.~M.}\ \bibnamefont {Perez-Mato}}, \bibinfo {author} {\bibfnamefont {C.}~\bibnamefont {Capillas}}, \bibinfo {author} {\bibfnamefont {E.}~\bibnamefont {Kroumova}}, \bibinfo {author} {\bibfnamefont {S.}~\bibnamefont {Ivantchev}}, \bibinfo {author} {\bibfnamefont {G.}~\bibnamefont {Madariaga}}, \bibinfo {author} {\bibfnamefont {A.}~\bibnamefont {Kirov}},\ and\ \bibinfo {author} {\bibfnamefont {H.}~\bibnamefont {Wondratschek}},\ }\href {https://doi.org/10.1524/zkri.2006.221.1.15} {\bibfield  {journal} {\bibinfo  {journal} {Z. Kristallogr. Cryst. Mater.}\ }\textbf {\bibinfo {volume} {221}},\ \bibinfo {pages} {15} (\bibinfo {year} {2006})}\BibitemShut {NoStop}%
\bibitem [{\citenamefont {Arnold}\ \emph {et~al.}(2016)\citenamefont {Arnold}, \citenamefont {Aroyo}, \citenamefont {Bertaut}, \citenamefont {Burzlaff}, \citenamefont {Chapuis}, \citenamefont {Fischer}, \citenamefont {Flack}, \citenamefont {Glazer}, \citenamefont {Grimmer}, \citenamefont {Gruber}, \citenamefont {Hahn}, \citenamefont {Klapper}, \citenamefont {Koch}, \citenamefont {Konstantinov}, \citenamefont {Kopsk{\`y}}, \citenamefont {Litvin}, \citenamefont {Looijenga-Vos}, \citenamefont {M{\"u}ller}, \citenamefont {Momma}, \citenamefont {Shmueli}, \citenamefont {Souvignier}, \citenamefont {Spence}, \citenamefont {de~Wolff}, , \citenamefont {Wondratschek},\ and\ \citenamefont {Zimmermann}}]{IUCr_volA2016_sm}%
  \BibitemOpen
  \bibfield  {author} {\bibinfo {author} {\bibfnamefont {H.}~\bibnamefont {Arnold}}, \bibinfo {author} {\bibfnamefont {M.~I.}\ \bibnamefont {Aroyo}}, \bibinfo {author} {\bibfnamefont {E.~F.}\ \bibnamefont {Bertaut}}, \bibinfo {author} {\bibfnamefont {H.}~\bibnamefont {Burzlaff}}, \bibinfo {author} {\bibfnamefont {G.}~\bibnamefont {Chapuis}}, \bibinfo {author} {\bibfnamefont {W.}~\bibnamefont {Fischer}}, \bibinfo {author} {\bibfnamefont {H.~D.}\ \bibnamefont {Flack}}, \bibinfo {author} {\bibfnamefont {A.~M.}\ \bibnamefont {Glazer}}, \bibinfo {author} {\bibfnamefont {H.}~\bibnamefont {Grimmer}}, \bibinfo {author} {\bibfnamefont {B.}~\bibnamefont {Gruber}}, \bibinfo {author} {\bibfnamefont {T.}~\bibnamefont {Hahn}}, \bibinfo {author} {\bibfnamefont {H.}~\bibnamefont {Klapper}}, \bibinfo {author} {\bibfnamefont {E.}~\bibnamefont {Koch}}, \bibinfo {author} {\bibfnamefont {P.}~\bibnamefont {Konstantinov}}, \bibinfo {author} {\bibfnamefont {V.}~\bibnamefont {Kopsk{\`y}}}, \bibinfo {author} {\bibfnamefont {D.~B.}\
  \bibnamefont {Litvin}}, \bibinfo {author} {\bibfnamefont {A.}~\bibnamefont {Looijenga-Vos}}, \bibinfo {author} {\bibfnamefont {U.}~\bibnamefont {M{\"u}ller}}, \bibinfo {author} {\bibfnamefont {K.}~\bibnamefont {Momma}}, \bibinfo {author} {\bibfnamefont {U.}~\bibnamefont {Shmueli}}, \bibinfo {author} {\bibfnamefont {B.}~\bibnamefont {Souvignier}}, \bibinfo {author} {\bibfnamefont {J.~C.~H.}\ \bibnamefont {Spence}}, \bibinfo {author} {\bibfnamefont {P.~M.}\ \bibnamefont {de~Wolff}}, , \bibinfo {author} {\bibfnamefont {H.}~\bibnamefont {Wondratschek}},\ and\ \bibinfo {author} {\bibfnamefont {H.}~\bibnamefont {Zimmermann}},\ }\href {https://doi.org/10.1107/97809553602060000114} {\emph {\bibinfo {title} {{International tables for crystallography volume A: Space-group symmetry}}}},\ \bibinfo {edition} {2nd}\ ed.,\ edited by\ \bibinfo {editor} {\bibfnamefont {M.~I.}\ \bibnamefont {Aroyo}}\ (\bibinfo  {publisher} {Wiley Online Library},\ \bibinfo {year} {2016})\BibitemShut {NoStop}%
\bibitem [{\citenamefont {Litvin}(1977)}]{Litvin1977_sm}%
  \BibitemOpen
  \bibfield  {author} {\bibinfo {author} {\bibfnamefont {D.~B.}\ \bibnamefont {Litvin}},\ }\href {https://doi.org/10.1107/S0567739477000709} {\bibfield  {journal} {\bibinfo  {journal} {Acta Cryst. A}\ }\textbf {\bibinfo {volume} {33}},\ \bibinfo {pages} {279} (\bibinfo {year} {1977})}\BibitemShut {NoStop}%
\bibitem [{\citenamefont {Chen}\ \emph {et~al.}(2025)\citenamefont {Chen}, \citenamefont {Liu}, \citenamefont {Liu}, \citenamefont {Yu}, \citenamefont {Ren}, \citenamefont {Li}, \citenamefont {Zhang},\ and\ \citenamefont {Liu}}]{Chen2025_sm}%
  \BibitemOpen
  \bibfield  {author} {\bibinfo {author} {\bibfnamefont {X.}~\bibnamefont {Chen}}, \bibinfo {author} {\bibfnamefont {Y.}~\bibnamefont {Liu}}, \bibinfo {author} {\bibfnamefont {P.}~\bibnamefont {Liu}}, \bibinfo {author} {\bibfnamefont {Y.}~\bibnamefont {Yu}}, \bibinfo {author} {\bibfnamefont {J.}~\bibnamefont {Ren}}, \bibinfo {author} {\bibfnamefont {J.}~\bibnamefont {Li}}, \bibinfo {author} {\bibfnamefont {A.}~\bibnamefont {Zhang}},\ and\ \bibinfo {author} {\bibfnamefont {Q.}~\bibnamefont {Liu}},\ }\href {https://doi.org/10.1038/s41586-025-08715-7} {\bibfield  {journal} {\bibinfo  {journal} {Nature}\ }\textbf {\bibinfo {volume} {640}},\ \bibinfo {pages} {349} (\bibinfo {year} {2025})}\BibitemShut {NoStop}%
\end{thebibliography}%
\clearpage


\resumetoc

\title{Supplementary Material for `The Fundamental Lemma of Altermagnetism: Emergence of Alterferrimagnetism'}

\maketitle
\onecolumngrid
\tableofcontents
\def\thefootnote{$\bigstar$}\footnotetext{These authors contributed equally to this work.}
\def\thefootnote{\arabic{footnote}}
\clearpage

\setcounter{section}{0}
\setcounter{subsection}{0}
\setcounter{equation}{0}
\setcounter{figure}{0}
\setcounter{table}{0}
\makeatletter
\renewcommand{\bibnumfmt}[1]{[S#1]}
\renewcommand{\citenumfont}[1]{S#1}
\renewcommand{\thesection}{S\Roman{section}}
\renewcommand{\thesubsection}{\thesection\Alph{subsection}}
\renewcommand{\thefigure}{S\arabic{figure}}
\renewcommand{\theequation}{S.\arabic{equation}}
\renewcommand{\thetable}{S\arabic{table}}

\begin{bibunit}[apsrev4-2]

    \noindent Here, we present further information on the other possible spin arrangements in CuFePO$_5$. We follow the \textit{International Tables of Crystallography, Volume A} (ITA) settings \cite{Aroyo2006a_sm, IUCr_volA2016_sm} while denoting the crystallographic space groups. The same convention is extended to denote the spin space groups to avoid possible ambiguities.

    \section{Alterferrimagnets with symmetries of Altermagnets}
    
    In the main text, it has been mentioned that there are nine possible magnetic configurations corresponding to different spin arrangements in CuFePO$_5$ at the 4a and 4c Wyckoff positions (see Table~\ref{tab1} of main manuscript), and all three configurations corresponding to the ($+$$-$$+$$-$)-type spin arrangement at the 4c Wyckoff position have been discussed in detail (see Sec.~\ref{AFiM} of main manuscript). Now, the remaining six configurations shall be discussed here. Since the Wyckoff postion 4c is \textit{partially compatible} with the AM phase, these remaining configurations also give rise to alterferrimagnetic states with lower crystallographic symmetry in the non-trivial spin group $\ntsg$, as discussed in Sec.~\ref{CuFePO5_lowersym} of main manuscript, mostly determined by the spin arrangement at the \textit{fully compatible} Wyckoff position 4a. Interestingly, the alterferrimagnetic states discussed here show spin-splitting similar to that of the pure altermagnetic states discussed in Sec.~\ref{Cu_CuFePO5} of main manuscript. This is a consequence of the symmetries of the spin only group $\sog$ which gives rise to additional spin-degeneracies when the full SSG $\ssg$ of these alterferrimagnetic states is taken into account.
    
    \subsection{(\texorpdfstring{$+$$+$$-$$-$}{++--})-type spin arrangement at 4c Wyckoff position}
    
    The ($+$$+$$-$$-$)-type spin arrangement at the 4c Wyckoff position yields the AFM phase and hence the Fe sublattice in CuFePO$_5$ is expected to show spin-degeneracy throughout the Brillouin zone (also see Sec.~\ref{Fe_ppnn} of main manuscript). Then the three possible spin arrangements of the Cu sublattice at the 4a Wyckoff position gives rise to SSG symmetries similar to those discussed in Sec.~\ref{Cu_CuFePO5} of main manuscript.
    
    \subsubsection{A-type spin arrangement at 4a Wyckoff position}
    \begin{figure*}[p]
        \centering
        \includegraphics[width=\linewidth]{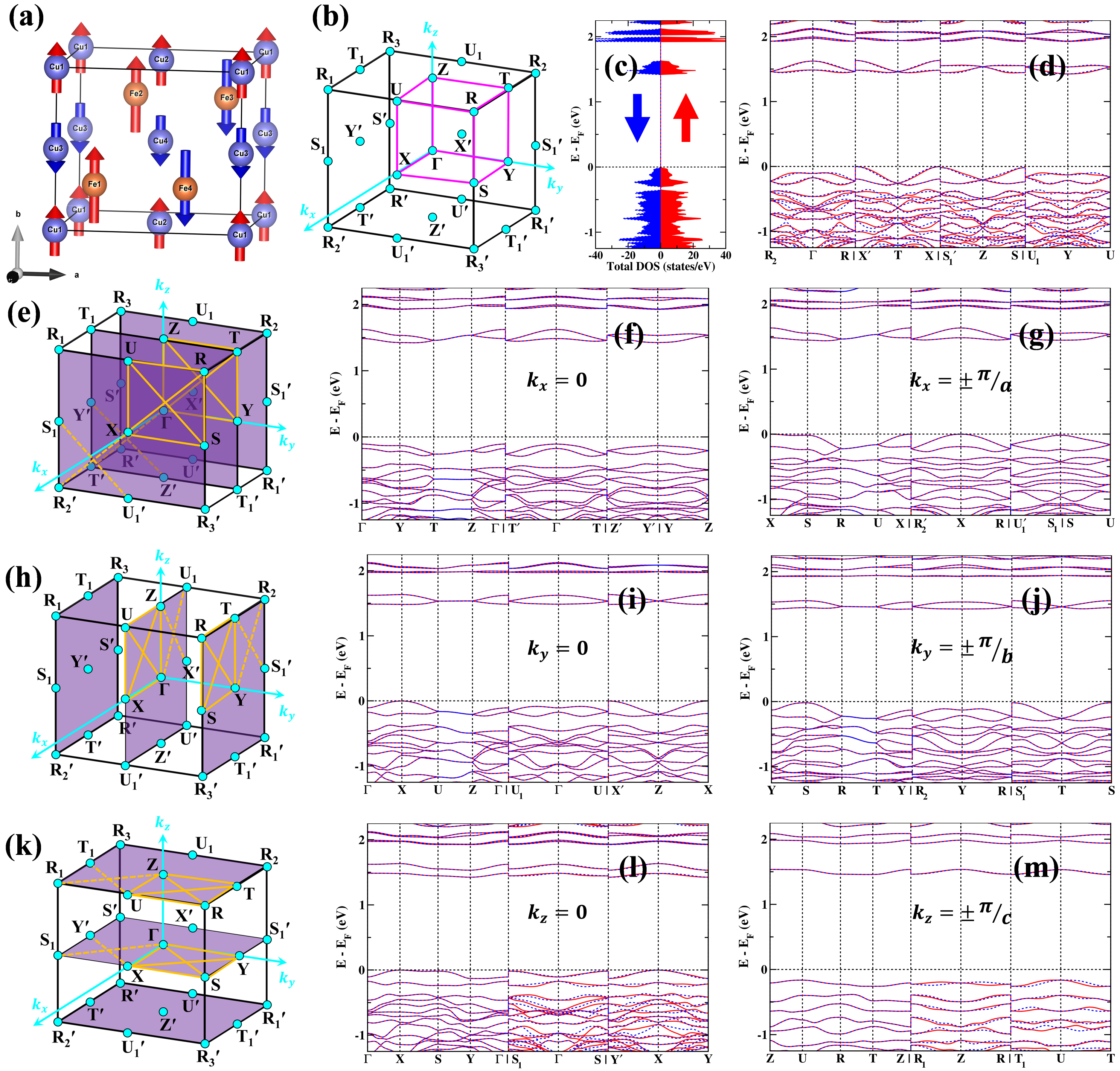}
        \caption{(a) The Cu and Fe sublattices with A-type and Fe($+$$+$$-$$-$)-type spin arrangement respectively. Only the magnetic atoms are shown for visual clarity. (b) \textit{Primitive} orthorhombic Brillouin zone (BZ) with the high-symmetry points shown as cyan dots and the irreducible Brillouin zone (IBZ) marked by magenta lines. (c) Spin-polarized electronic total density of states (TDOS) of CuFePO$_5$. (d) Spin-polarized electronic band structure of CuFePO$_5$ along body diagonals of the IBZ showing spin-splitting, characteristic of the altermagnetic phase. (e) BZ with the $k_x=0$ and $k_x=\pm\frac{\pi}{a}$ planes highlighted in purple. The high-symmetry directions on the respective planes have been highlighted by orange lines and the spin-polarized electronic band structures of CuFePO$_5$ along these lines have been shown in (f) and (g). (h-j) The same for $k_y=0$ and $k_y=\pm\frac{\pi}{b}$ planes. (k-m) The same for $k_z=0$ and $k_z=\pm\frac{\pi}{c}$ planes. In (e), (h) and (k), the solid orange lines denote paths within the IBZ while the broken orange lines denote paths outside the IBZ.}
        \label{fig1S}
    \end{figure*}
    \textcolor{black}{Figure~\ref{fig1S}(a)} displays A-type spin arrangement of Cu sublattice and ($+$$+$$-$$-$)-type spin arrangement of Fe sublattice in CuFePO$_5$. From the parent crystallographic space group $\csg = \bm{Pnma}$, the same-spin sublattice transformations for A-type spin arrangement at the 4a Wyckoff position and ($+$$+$$-$$-$)-type spin arrangement at the 4c Wyckoff position form the \textit{halving} subgroups $\hsgone = \bm{\tilde{C}_{2h}^{z}}$ and $\hsgfour = \bm{\tilde{C}_{2v}^{z}}$ respectively (derived and discussed in Secs.~\ref{Cu_A} and \ref{Fe_ppnn} of main manuscript). The cosets $\csg-\hsgone$ and $\csg-\hsgfour$ provide the respective opposite-spin sublattice transformations. Then, the \textit{common} same-spin and opposite-spin sublattice transformations are $\hsgone \cap \hsgfour = \{\,\mathbb{I}, \tilde{C}_{2z}\,\}$ and  $(\csg-\hsgone) \cap (\csg-\hsgfour) = \{\,\tilde{C}_{2y}, \tilde{C}_{2x}\,\}$ respectively. Then the reduced crystallographic space group is given by $\bm{\tilde{D}_{2}} = \{\,\mathbb{I}, \tilde{C}_{2z}, \tilde{C}_{2y}, \tilde{C}_{2x}\,\}$, \textit{isomorphic} to $\bm{P2_{1}2_{1}2_{1}}~(\bm{D_{2}^{4}})$ \cite{Aroyo2006a_sm, IUCr_volA2016_sm} and a subgroup of $\csg$. The non-trivial spin group becomes:
    \begin{equation}
        \ntsg = \ntsgspin \otimes \bm{\tilde{D}_{2}} = [\,\mathfrak{I}\,||\,\{\,\mathbb{I}, \tilde{C}_{2z}\,\}\,] \cup [\,\mathfrak{C}_2\,||\,\{\,\tilde{C}_{2y}, \tilde{C}_{2x}\,\}\,]
        \label{eq1S}
    \end{equation}
    Thus, A-type spin arrangement on the Cu sublattice and ($+$$+$$-$$-$)-type spin arrangement on the Fe sublattice in CuFePO$_5$ leads to SSG $\bm{P\,^{\overline{1}}2_{1}\,^{\overline{1}}2_{1}\,^{1}2_{1}}$ (19.27) \cite{Litvin1977_sm, Chen2025_sm}. From Eqs.~\eqref{eq1}, \eqref{eq4} and \eqref{eq11} of main manuscript and eq.~\eqref{eq1S}, all the resulting  spin-degenerate nodal lines and nodal planes across the BZ can be systematically mapped out using the \textit{isomorphic} spin point groups, as discussed in  Sec.~\ref{AFiM} of main manuscript. \textcolor{black}{Figure~\ref{fig1S}(b)} shows the BZ along with the IBZ. First-principles calculations reveal fully compensated nature of both Cu and Fe sublattices which is evident from the spin-polarized total DOS as shown in \textcolor{black}{Fig.~\ref{fig1S}(c)}. 
    The resulting altermagnetic spin-splitting for these spin arrangements along the IBZ body diagonals is shown in \textcolor{black}{Figure~\ref{fig1S}(d)}. The spin-degeneracies and splittings on the high symmetry planes $k_x=0,\pm\sfrac{\pi}{a}$, $k_y=0,\pm\sfrac{\pi}{b}$ and $k_z=0,\pm\sfrac{\pi}{c}$, as enforced by the SSG symmetries, are displayed in \textcolor{black}{Figs.~\ref{fig1S}(e–m)}.
    
    \subsubsection{C-type spin arrangement at 4a Wyckoff position}
    \begin{figure*}[p]
        \centering
        \includegraphics[width=\linewidth]{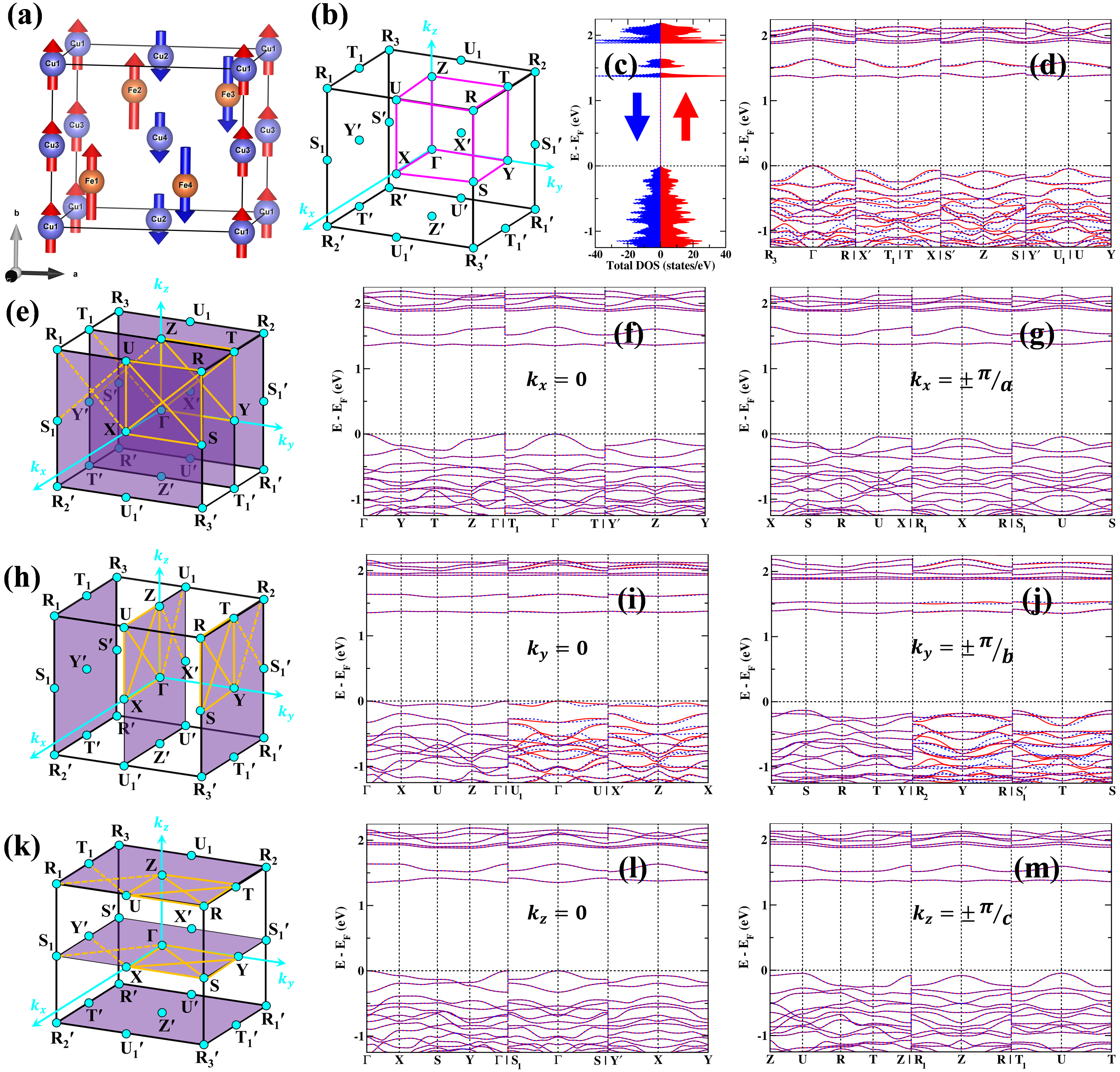}
        \caption{(a) The Cu and Fe sublattices with C-type and Fe($+$$+$$-$$-$)-type spin arrangement respectively. Only the magnetic atoms are shown for visual clarity. (b) \textit{Primitive} orthorhombic Brillouin zone (BZ) with the high-symmetry points shown as cyan dots and the irreducible Brillouin zone (IBZ) marked by magenta lines. (c) Spin-polarized electronic total density of states (TDOS) of CuFePO$_5$. (d) Spin-polarized electronic band structure of CuFePO$_5$ along body diagonals of the IBZ showing spin-splitting, characteristic of the altermagnetic phase. (e) BZ with the $k_x=0$ and $k_x=\pm\frac{\pi}{a}$ planes highlighted in purple. The high-symmetry directions on the respective planes have been highlighted by orange lines and the spin-polarized electronic band structures of CuFePO$_5$ along these lines have been shown in (f) and (g). (h-j) The same for $k_y=0$ and $k_y=\pm\frac{\pi}{b}$ planes. (k-m) The same for $k_z=0$ and $k_z=\pm\frac{\pi}{c}$ planes. In (e), (h) and (k), the solid orange lines denote paths within the IBZ while the broken orange lines denote paths outside the IBZ.}
        \label{fig2S}
    \end{figure*}
    \textcolor{black}{Figure~\ref{fig2S}(a)} displays C-type spin arrangement of Cu sublattice and ($+$$+$$-$$-$)-type spin arrangement of Fe sublattice in CuFePO$_5$. From the parent crystallographic space group $\csg = \bm{Pnma}$, the same-spin sublattice transformations for A-type spin arrangement at the 4a Wyckoff position and ($+$$+$$-$$-$)-type spin arrangement at the 4c Wyckoff position form the \textit{halving} subgroups $\hsgtwo = \bm{\tilde{C}_{2h}^{y}}$ and $\hsgfour = \bm{\tilde{C}_{2v}^{z}}$ respectively (derived and discussed in Secs.~\ref{Cu_C} and \ref{Fe_ppnn} of main manuscript). The cosets $\csg-\hsgtwo$ and $\csg-\hsgfour$ provide the respective opposite-spin sublattice transformations. Then, the \textit{common} same-spin and opposite-spin sublattice transformations are $\hsgtwo \cap \hsgfour = \{\,\mathbb{I}, \tilde{m}_{zx}\,\}$ and  $(\csg-\hsgtwo) \cap (\csg-\hsgfour) = \{\,\tilde{m}_{xy}, \tilde{C}_{2x}\,\}$ respectively. Then the reduced crystallographic space group is given by $\bm{\tilde{C}_{2v}^{x}} = \{\,\mathbb{I}, \tilde{m}_{zx}, \tilde{m}_{xy}, \tilde{C}_{2x}\,\}$, \textit{isomorphic} to $\bm{P2_{1}ma}~(\bm{C_{2v}^{2}})$ (also see Table~\ref{tab1} of main manuscript) \cite{Aroyo2006a_sm, IUCr_volA2016_sm} and a subgroup of $\csg$. The non-trivial spin group becomes:
    \begin{equation}
        \ntsg = \ntsgspin \otimes \bm{\tilde{C}_{2v}^{x}} = [\,\mathfrak{I}\,||\,\{\,\mathbb{I}, \tilde{m}_{zx}\,\}\,] \cup [\,\mathfrak{C}_2\,||\,\{\,\tilde{m}_{xy}, \tilde{C}_{2x}\,\}\,]
        \label{eq2S}
    \end{equation}
    Thus, C-type spin arrangement on the Cu sublattice and ($+$$+$$-$$-$)-type spin arrangement on the Fe sublattice in CuFePO$_5$ leads to SSG $\bm{P\,^{\overline{1}}2_{1}\,^{1}m\,^{\overline{1}}a}$ (26.69) \cite{Litvin1977_sm, Chen2025_sm}. From Eqs.~\eqref{eq1}, \eqref{eq7} and \eqref{eq11} of main manuscript and eq.~\eqref{eq2S}, all the resulting  spin-degenerate nodal lines and nodal planes across the BZ can be systematically mapped out using the \textit{isomorphic} spin point groups, as discussed in  Sec.~\ref{AFiM} of main manuscript. \textcolor{black}{Figure~\ref{fig2S}(b)} shows the BZ along with the IBZ. First-principles calculations reveal fully compensated nature of both Cu and Fe sublattices which is evident from the spin-polarized total DOS as shown in \textcolor{black}{Fig.~\ref{fig2S}(c)}. 
    The resulting altermagnetic spin-splitting for these spin arrangements along the IBZ body diagonals is shown in \textcolor{black}{Figure~\ref{fig2S}(d)}. The spin-degeneracies and splittings on the high symmetry planes $k_x=0,\pm\sfrac{\pi}{a}$, $k_y=0,\pm\sfrac{\pi}{b}$ and $k_z=0,\pm\sfrac{\pi}{c}$, as enforced by the SSG symmetries, are displayed in \textcolor{black}{Figs.~\ref{fig2S}(e–m)}.
    
    \subsubsection{G-type spin arrangement at 4a Wyckoff position}
    \begin{figure*}[p]
        \centering
        \includegraphics[width=\linewidth]{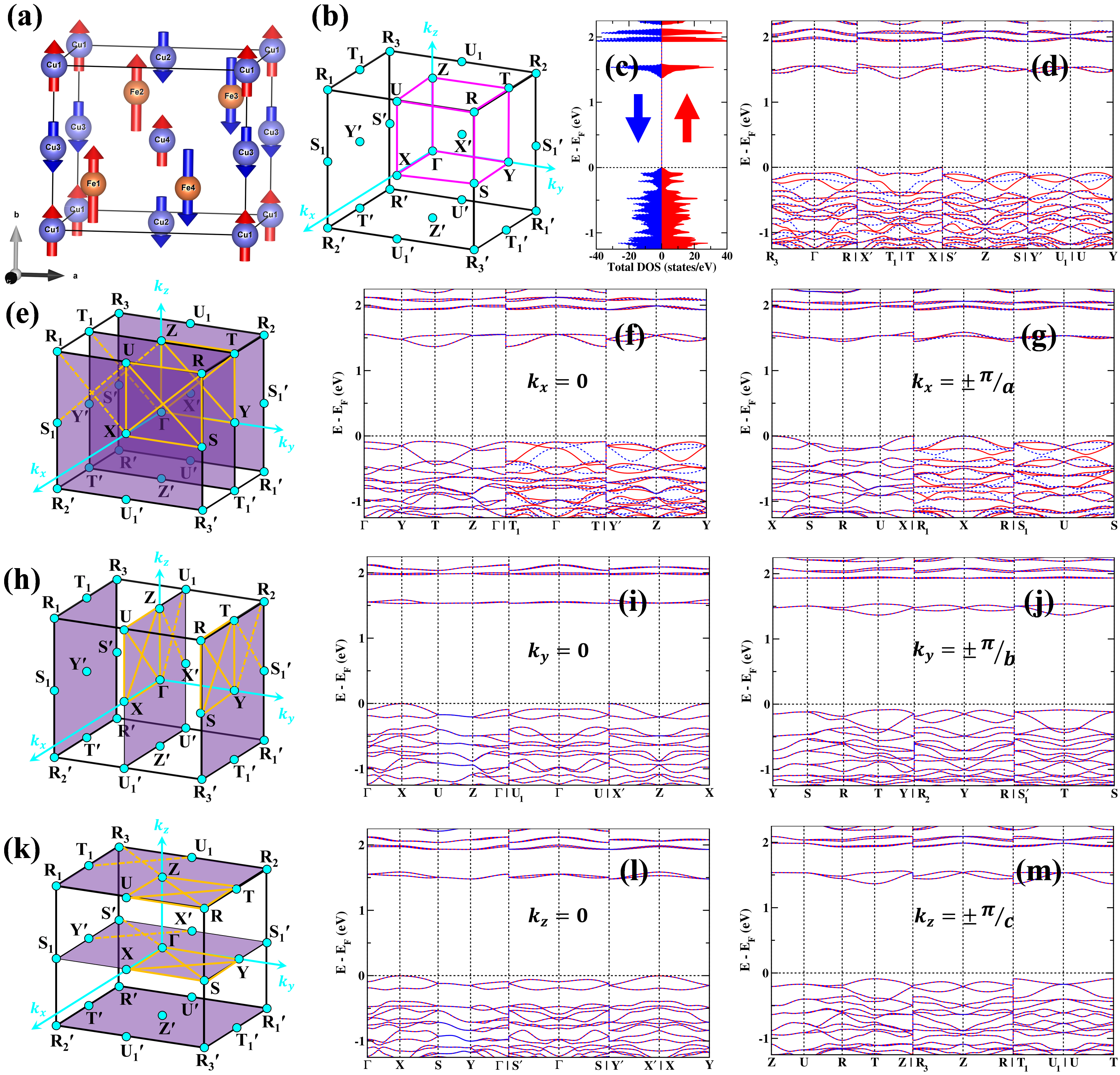}
        \caption{(a) The Cu and Fe sublattices with G-type and Fe($+$$+$$-$$-$)-type spin arrangement respectively. Only the magnetic atoms are shown for visual clarity. (b) \textit{Primitive} orthorhombic Brillouin zone (BZ) with the high-symmetry points shown as cyan dots and the irreducible Brillouin zone (IBZ) marked by magenta lines. (c) Spin-polarized electronic total density of states (TDOS) of CuFePO$_5$. (d) Spin-polarized electronic band structure of CuFePO$_5$ along body diagonals of the IBZ showing spin-splitting, characteristic of the altermagnetic phase. (e) BZ with the $k_x=0$ and $k_x=\pm\frac{\pi}{a}$ planes highlighted in purple. The high-symmetry directions on the respective planes have been highlighted by orange lines and the spin-polarized electronic band structures of CuFePO$_5$ along these lines have been shown in (f) and (g). (h-j) The same for $k_y=0$ and $k_y=\pm\frac{\pi}{b}$ planes. (k-m) The same for $k_z=0$ and $k_z=\pm\frac{\pi}{c}$ planes. In (e), (h) and (k), the solid orange lines denote paths within the IBZ while the broken orange lines denote paths outside the IBZ.}
        \label{fig3S}
    \end{figure*}
    \textcolor{black}{Figure~\ref{fig3S}(a)} displays G-type spin arrangement of Cu sublattice and ($+$$+$$-$$-$)-type spin arrangement of Fe sublattice in CuFePO$_5$. From the parent crystallographic space group $\csg = \bm{Pnma}$, the same-spin sublattice transformations for A-type spin arrangement at the 4a Wyckoff position and ($+$$+$$-$$-$)-type spin arrangement at the 4c Wyckoff position form the \textit{halving} subgroups $\hsgthree = \bm{\tilde{C}_{2h}^{x}}$ and $\hsgfour = \bm{\tilde{C}_{2v}^{z}}$ respectively (derived and discussed in Secs.~\ref{Cu_G} and \ref{Fe_ppnn} of main manuscript). The cosets $\csg-\hsgthree$ and $\csg-\hsgfour$ provide the respective opposite-spin sublattice transformations. Then, the \textit{common} same-spin and opposite-spin sublattice transformations are $\hsgthree \cap \hsgfour = \{\,\mathbb{I}, \tilde{m}_{yz}\,\}$ and  $(\csg-\hsgthree) \cap (\csg-\hsgfour) = \{\,\tilde{m}_{xy}, \tilde{C}_{2y}\,\}$ respectively. Then the reduced crystallographic space group is given by $\bm{\tilde{C}_{2v}^{y}} = \{\,\mathbb{I}, \tilde{m}_{yz}, \tilde{m}_{xy}, \tilde{C}_{2y}\,\}$, \textit{isomorphic} to $\bm{Pn2_{1}a}~(\bm{C_{2v}^{9}})$ (also see Table~\ref{tab1} of main manuscript) \cite{Aroyo2006a_sm, IUCr_volA2016_sm} and a subgroup of $\csg$. The non-trivial spin group becomes:
    \begin{equation}
        \ntsg = \ntsgspin \otimes \bm{\tilde{C}_{2v}^{y}} = [\,\mathfrak{I}\,||\,\{\,\mathbb{I}, \tilde{m}_{yz}\,\}\,] \cup [\,\mathfrak{C}_2\,||\,\{\,\tilde{m}_{xy}, \tilde{C}_{2y}\,\}\,]
        \label{eq3S}
    \end{equation}
    Thus, G-type spin arrangement on the Cu sublattice and ($+$$+$$-$$-$)-type spin arrangement on the Fe sublattice in CuFePO$_5$ leads to SSG $\bm{P\,^{1}n\,^{\overline{1}}2_{1}\,^{\overline{1}}a}$ (33.147) \cite{Litvin1977_sm, Chen2025_sm}. From Eqs.~\eqref{eq1}, \eqref{eq10} and \eqref{eq11} of main manuscript and eq.~\eqref{eq3S}, all the resulting  spin-degenerate nodal lines and nodal planes across the BZ can be systematically mapped out using the \textit{isomorphic} spin point groups, as discussed in  Sec.~\ref{AFiM} of main manuscript. \textcolor{black}{Figure~\ref{fig3S}(b)} shows the BZ along with the IBZ. First-principles calculations reveal fully compensated nature of both Cu and Fe sublattices which is evident from the spin-polarized total DOS as shown in \textcolor{black}{Fig.~\ref{fig3S}(c)}. 
    The resulting altermagnetic spin-splitting for these spin arrangements along the IBZ body diagonals is shown in \textcolor{black}{Figure~\ref{fig3S}(d)}. The spin-degeneracies and splittings on the high symmetry planes $k_x=0,\pm\sfrac{\pi}{a}$, $k_y=0,\pm\sfrac{\pi}{b}$ and $k_z=0,\pm\sfrac{\pi}{c}$, as enforced by the SSG symmetries, are displayed in \textcolor{black}{Figs.~\ref{fig3S}(e–m)}.

    \subsection{(\texorpdfstring{$+$$-$$-$$+$}{+--+})-type spin arrangement at 4c Wyckoff position}
    
    The Fe($+$$-$$-$$+$)-type spin arrangement at the 4c Wyckoff position yields the AFM phase and hence the Fe sublattice in CuFePO$_5$ is again expected to show spin-degeneracy throughout the Brillouin zone (also see Sec.~\ref{Fe_pnnp} of main manuscript). Then the three possible spin arrangements of the Cu sublattice at the 4a Wyckoff position gives rise to SSG symmetries similar to those discussed in Sec.~\ref{Cu_CuFePO5} of main manuscript.
    
    \subsubsection{A-type spin arrangement at 4a Wyckoff position}
    \begin{figure*}[p]
        \centering
        \includegraphics[width=\linewidth]{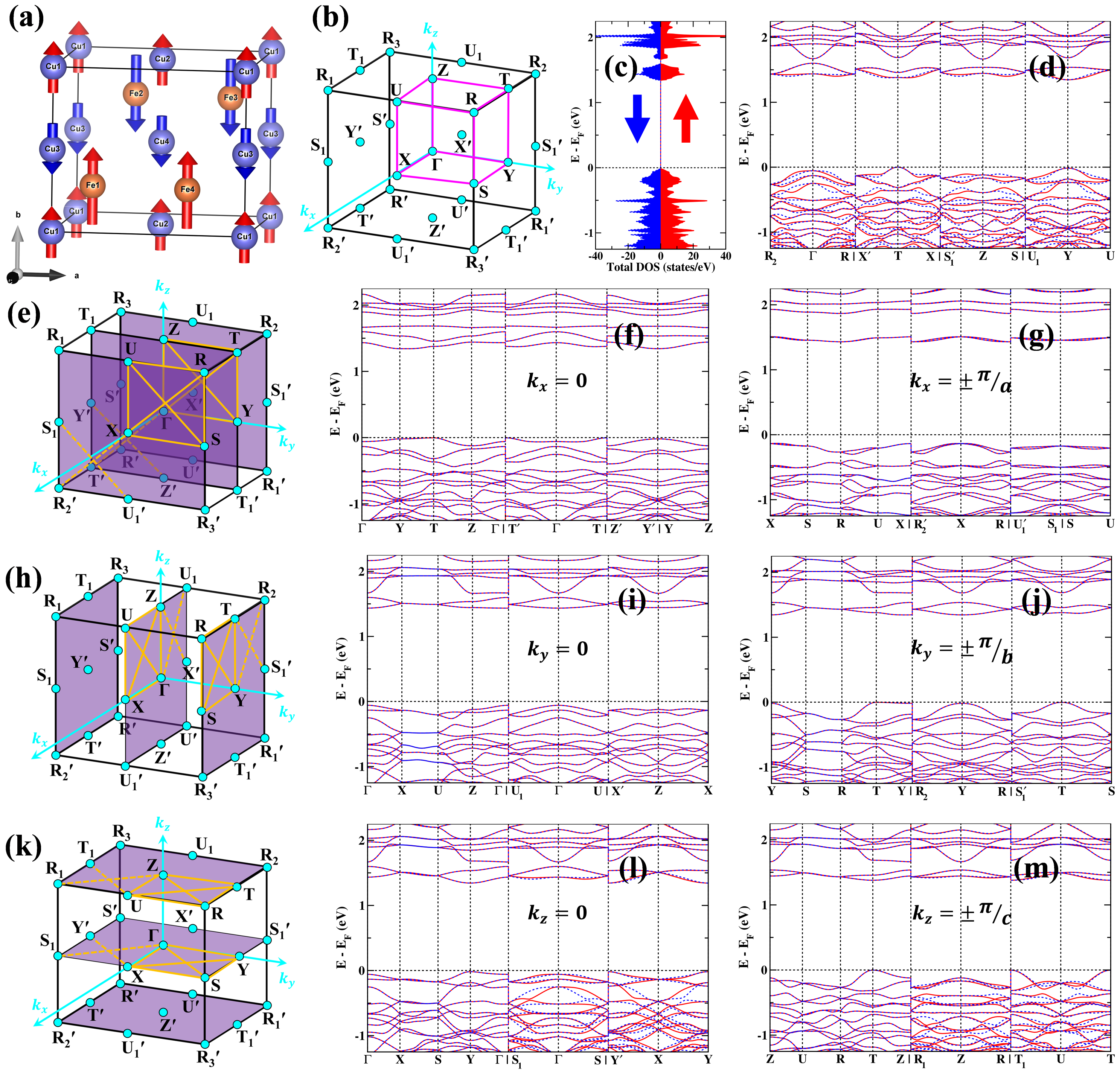}
        \caption{(a) The Cu and Fe sublattices with A-type and Fe($+$$-$$-$$+$)-type spin arrangement respectively. Only the magnetic atoms are shown for visual clarity. (b) \textit{Primitive} orthorhombic Brillouin zone (BZ) with the high-symmetry points shown as cyan dots and the irreducible Brillouin zone (IBZ) marked by magenta lines. (c) Spin-polarized electronic total density of states (TDOS) of CuFePO$_5$. (d) Spin-polarized electronic band structure of CuFePO$_5$ along body diagonals of the IBZ showing spin-splitting, characteristic of the altermagnetic phase. (e) BZ with the $k_x=0$ and $k_x=\pm\frac{\pi}{a}$ planes highlighted in purple. The high-symmetry directions on the respective planes have been highlighted by orange lines and the spin-polarized electronic band structures of CuFePO$_5$ along these lines have been shown in (f) and (g). (h-j) The same for $k_y=0$ and $k_y=\pm\frac{\pi}{b}$ planes. (k-m) The same for $k_z=0$ and $k_z=\pm\frac{\pi}{c}$ planes. In (e), (h) and (k), the solid orange lines denote paths within the IBZ while the broken orange lines denote paths outside the IBZ.}
        \label{fig4S}
    \end{figure*}
    \textcolor{black}{Figure~\ref{fig4S}(a)} displays A-type spin arrangement of Cu sublattice and ($+$$-$$-$$+$)-type spin arrangement of Fe sublattice in CuFePO$_5$. From the parent crystallographic space group $\csg = \bm{Pnma}$, the same-spin sublattice transformations for A-type spin arrangement at the 4a Wyckoff position and ($+$$-$$-$$+$)-type spin arrangement at the 4c Wyckoff position form the \textit{halving} subgroups $\hsgone = \bm{\tilde{C}_{2h}^{z}}$ and $\hsgfive = \bm{\tilde{C}_{2v}^{x}}$ respectively (derived and discussed in Secs.~\ref{Cu_A} and \ref{Fe_pnnp} of main manuscript). The cosets $\csg-\hsgone$ and $\csg-\hsgfive$ provide the respective opposite-spin sublattice transformations. Then, the \textit{common} same-spin and opposite-spin sublattice transformations are $\hsgone \cap \hsgfive = \{\,\mathbb{I}, \tilde{m}_{xy}\,\}$ and  $(\csg-\hsgone) \cap (\csg-\hsgfive) = \{\,\tilde{m}_{yz}, \tilde{C}_{2y}\,\}$ respectively. Then the reduced crystallographic space group is given by $\bm{\tilde{C}_{2v}^{y}} = \{\,\mathbb{I}, \tilde{m}_{xy}, \tilde{m}_{yz}, \tilde{C}_{2y}\,\}$, \textit{isomorphic} to $\bm{Pn2_{1}a}~(\bm{C_{2v}^{9}})$ (also see Table~\ref{tab1} of main manuscript) \cite{Aroyo2006a_sm, IUCr_volA2016_sm} and a subgroup of $\csg$. The non-trivial spin group becomes:
    \begin{equation}
        \ntsg = \ntsgspin \otimes \bm{\tilde{C}_{2v}^{y}} = [\,\mathfrak{I}\,||\,\{\,\mathbb{I}, \tilde{m}_{xy}\,\}\,] \cup [\,\mathfrak{C}_2\,||\,\{\,\tilde{m}_{yz}, \tilde{C}_{2y}\,\}\,]
        \label{eq4S}
    \end{equation}
    Thus, A-type spin arrangement on the Cu sublattice and ($+$$-$$-$$+$)-type spin arrangement on the Fe sublattice in CuFePO$_5$ leads to SSG $\bm{P\,^{\overline{1}}n\,^{\overline{1}}2_{1}\,^{1}a}$ (33.146) \cite{Litvin1977_sm, Chen2025_sm}. From Eqs.~\eqref{eq1}, \eqref{eq4} and \eqref{eq11} of main manuscript and eq.~\eqref{eq4S}, all the resulting  spin-degenerate nodal lines and nodal planes across the BZ can be systematically mapped out using the \textit{isomorphic} spin point groups, as discussed in  Sec.~\ref{AFiM} of main manuscript. \textcolor{black}{Figure~\ref{fig4S}(b)} shows the BZ along with the IBZ. First-principles calculations reveal fully compensated nature of both Cu and Fe sublattices which is evident from the spin-polarized total DOS as shown in \textcolor{black}{Fig.~\ref{fig4S}(c)}. 
    The resulting altermagnetic spin-splitting for these spin arrangements along the IBZ body diagonals is shown in \textcolor{black}{Figure~\ref{fig4S}(d)}. The spin-degeneracies and splittings on the high symmetry planes $k_x=0,\pm\sfrac{\pi}{a}$, $k_y=0,\pm\sfrac{\pi}{b}$ and $k_z=0,\pm\sfrac{\pi}{c}$, as enforced by the SSG symmetries, are displayed in \textcolor{black}{Figs.~\ref{fig4S}(e–m)}.
    
    \subsubsection{C-type spin arrangement at 4a Wyckoff position}
    \begin{figure*}[p]
        \centering
        \includegraphics[width=\linewidth]{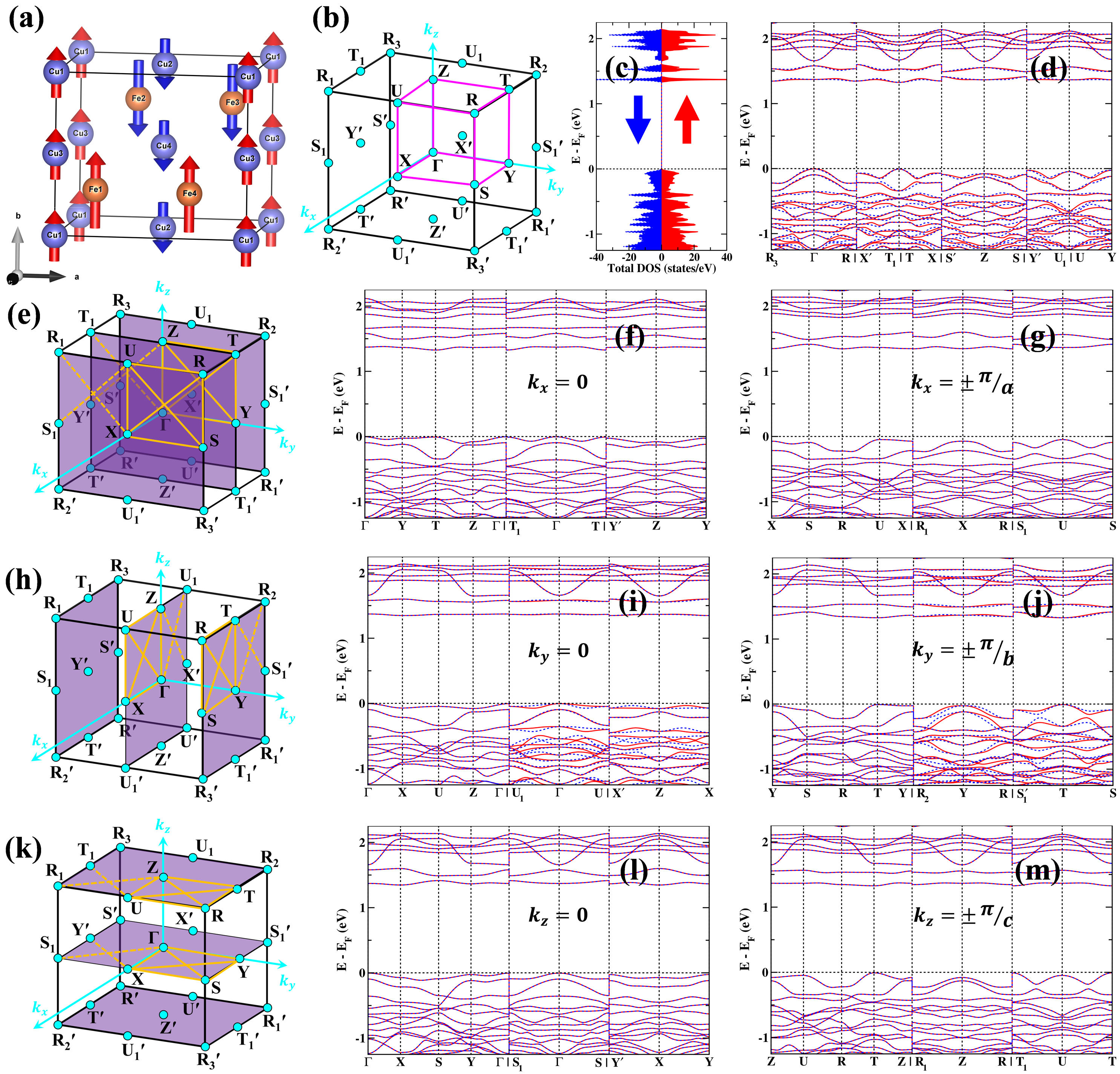}
        \caption{(a) The Cu and Fe sublattices with C-type and Fe($+$$-$$-$$+$)-type spin arrangement respectively. Only the magnetic atoms are shown for visual clarity. (b) \textit{Primitive} orthorhombic Brillouin zone (BZ) with the high-symmetry points shown as cyan dots and the irreducible Brillouin zone (IBZ) marked by magenta lines. (c) Spin-polarized electronic total density of states (TDOS) of CuFePO$_5$. (d) Spin-polarized electronic band structure of CuFePO$_5$ along body diagonals of the IBZ showing spin-splitting, characteristic of the altermagnetic phase. (e) BZ with the $k_x=0$ and $k_x=\pm\frac{\pi}{a}$ planes highlighted in purple. The high-symmetry directions on the respective planes have been highlighted by orange lines and the spin-polarized electronic band structures of CuFePO$_5$ along these lines have been shown in (f) and (g). (h-j) The same for $k_y=0$ and $k_y=\pm\frac{\pi}{b}$ planes. (k-m) The same for $k_z=0$ and $k_z=\pm\frac{\pi}{c}$ planes. In (e), (h) and (k), the solid orange lines denote paths within the IBZ while the broken orange lines denote paths outside the IBZ.}
        \label{fig5S}
    \end{figure*}
    \textcolor{black}{Figure~\ref{fig5S}(a)} displays C-type spin arrangement of Cu sublattice and ($+$$-$$-$$+$)-type spin arrangement of Fe sublattice in CuFePO$_5$. From the parent crystallographic space group $\csg = \bm{Pnma}$, the same-spin sublattice transformations for A-type spin arrangement at the 4a Wyckoff position and ($+$$-$$-$$+$)-type spin arrangement at the 4c Wyckoff position form the \textit{halving} subgroups $\hsgtwo = \bm{\tilde{C}_{2h}^{y}}$ and $\hsgfive = \bm{\tilde{C}_{2v}^{x}}$ respectively (derived and discussed in Secs.~\ref{Cu_C} and \ref{Fe_pnnp} of main manuscript). The cosets $\csg-\hsgtwo$ and $\csg-\hsgfive$ provide the respective opposite-spin sublattice transformations. Then, the \textit{common} same-spin and opposite-spin sublattice transformations are $\hsgtwo \cap \hsgfive = \{\,\mathbb{I}, \tilde{m}_{zx}\,\}$ and  $(\csg-\hsgtwo) \cap (\csg-\hsgfive) = \{\,\tilde{m}_{yz}, \tilde{C}_{2z}\,\}$ respectively. Then the reduced crystallographic space group is given by $\bm{\tilde{C}_{2v}^{z}} = \{\,\mathbb{I}, \tilde{m}_{zx}, \tilde{m}_{yz}, \tilde{C}_{2z}\,\}$, \textit{isomorphic} to $\bm{Pnm2_{1}}~(\bm{C_{2v}^{7}})$ (also see Table~\ref{tab1} of main manuscript) \cite{Aroyo2006a_sm, IUCr_volA2016_sm} and a subgroup of $\csg$. The non-trivial spin group becomes:
    \begin{equation}
        \ntsg = \ntsgspin \otimes \bm{\tilde{C}_{2v}^{z}} = [\,\mathfrak{I}\,||\,\{\,\mathbb{I}, \tilde{m}_{zx}\,\}\,] \cup [\,\mathfrak{C}_2\,||\,\{\,\tilde{m}_{yz}, \tilde{C}_{2z}\,\}\,]
        \label{eq5S}
    \end{equation}
    Thus, C-type spin arrangement on the Cu sublattice and ($+$$-$$-$$+$)-type spin arrangement on the Fe sublattice in CuFePO$_5$ leads to SSG $\bm{P\,^{\overline{1}}n\,^{1}m\,^{\overline{1}}2_{1}}$ (31.126) \cite{Litvin1977_sm, Chen2025_sm}. From Eqs.~\eqref{eq1}, \eqref{eq7} and \eqref{eq11} of main manuscript and eq.~\eqref{eq5S}, all the resulting  spin-degenerate nodal lines and nodal planes across the BZ can be systematically mapped out using the \textit{isomorphic} spin point groups, as discussed in  Sec.~\ref{AFiM} of main manuscript. \textcolor{black}{Figure~\ref{fig5S}(b)} shows the BZ along with the IBZ. First-principles calculations reveal fully compensated nature of both Cu and Fe sublattices which is evident from the spin-polarized total DOS as shown in \textcolor{black}{Fig.~\ref{fig5S}(c)}. 
    The resulting altermagnetic spin-splitting for these spin arrangements along the IBZ body diagonals is shown in \textcolor{black}{Figure~\ref{fig5S}(d)}. The spin-degeneracies and splittings on the high symmetry planes $k_x=0,\pm\sfrac{\pi}{a}$, $k_y=0,\pm\sfrac{\pi}{b}$ and $k_z=0,\pm\sfrac{\pi}{c}$, as enforced by the SSG symmetries, are displayed in \textcolor{black}{Figs.~\ref{fig5S}(e–m)}.
    
    \subsubsection{G-type spin arrangement at 4a Wyckoff position}
    \begin{figure*}[p]
        \centering
        \includegraphics[width=\linewidth]{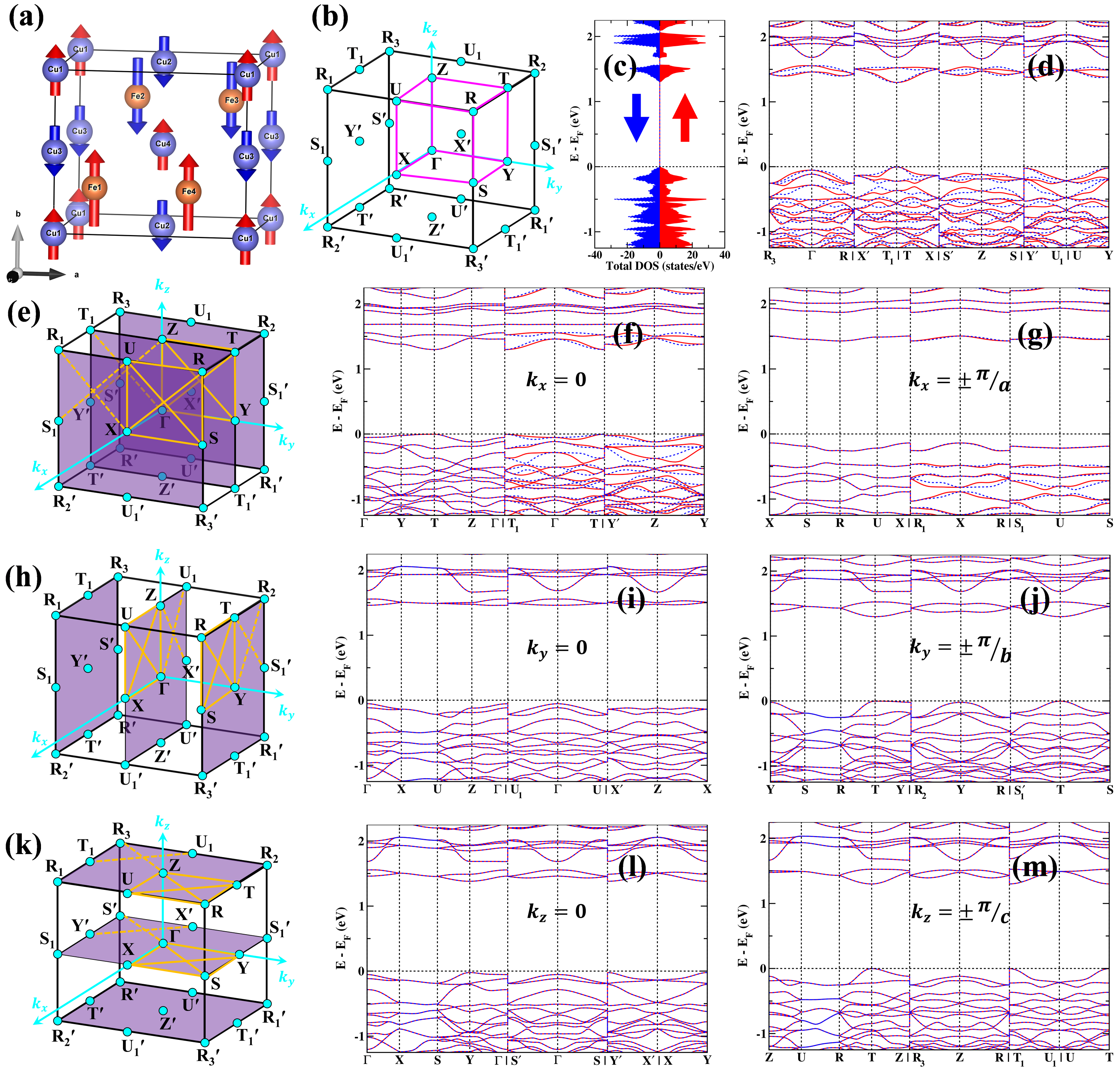}
        \caption{(a) The Cu and Fe sublattices with G-type and Fe($+$$-$$-$$+$)-type spin arrangement respectively. Only the magnetic atoms are shown for visual clarity. (b) \textit{Primitive} orthorhombic Brillouin zone (BZ) with the high-symmetry points shown as cyan dots and the irreducible Brillouin zone (IBZ) marked by magenta lines. (c) Spin-polarized electronic total density of states (TDOS) of CuFePO$_5$. (d) Spin-polarized electronic band structure of CuFePO$_5$ along body diagonals of the IBZ showing spin-splitting, characteristic of the altermagnetic phase. (e) BZ with the $k_x=0$ and $k_x=\pm\frac{\pi}{a}$ planes highlighted in purple. The high-symmetry directions on the respective planes have been highlighted by orange lines and the spin-polarized electronic band structures of CuFePO$_5$ along these lines have been shown in (f) and (g). (h-j) The same for $k_y=0$ and $k_y=\pm\frac{\pi}{b}$ planes. (k-m) The same for $k_z=0$ and $k_z=\pm\frac{\pi}{c}$ planes. In (e), (h) and (k), the solid orange lines denote paths within the IBZ while the broken orange lines denote paths outside the IBZ.}
        \label{fig6S}
    \end{figure*}
    \textcolor{black}{Figure~\ref{fig6S}(a)} displays G-type spin arrangement of Cu sublattice and ($+$$-$$-$$+$)-type spin arrangement of Fe sublattice in CuFePO$_5$. From the parent crystallographic space group $\csg = \bm{Pnma}$, the same-spin sublattice transformations for A-type spin arrangement at the 4a Wyckoff position and ($+$$-$$-$$+$)-type spin arrangement at the 4c Wyckoff position form the \textit{halving} subgroups $\hsgthree = \bm{\tilde{C}_{2h}^{x}}$ and $\hsgfive = \bm{\tilde{C}_{2v}^{x}}$ respectively (derived and discussed in Secs.~\ref{Cu_G} and \ref{Fe_pnnp} of main manuscript). The cosets $\csg-\hsgthree$ and $\csg-\hsgfive$ provide the respective opposite-spin sublattice transformations. Then, the \textit{common} same-spin and opposite-spin sublattice transformations are $\hsgthree \cap \hsgfive = \{\,\mathbb{I}, \tilde{C}_{2x}\,\}$ and  $(\csg-\hsgthree) \cap (\csg-\hsgfive) = \{\,\tilde{C}_{2y}, \tilde{C}_{2z}\,\}$ respectively. Then the reduced crystallographic space group is given by $\bm{\tilde{D}_{2}} = \{\,\mathbb{I}, \tilde{C}_{2x}, \tilde{C}_{2y}, \tilde{C}_{2z}\,\}$, \textit{isomorphic} to $\bm{P2_{1}2_{1}2_{1}}~(\bm{D_{2}^{4}})$ (also see Table~\ref{tab1} of main manuscript) \cite{Aroyo2006a_sm, IUCr_volA2016_sm} and a subgroup of $\csg$. The non-trivial spin group becomes:
    \begin{equation}
        \ntsg = \ntsgspin \otimes \bm{\tilde{D}_{2}} = [\,\mathfrak{I}\,||\,\{\,\mathbb{I}, \tilde{C}_{2x}\,\}\,] \cup [\,\mathfrak{C}_2\,||\,\{\,\tilde{C}_{2y}, \tilde{C}_{2z}\,\}\,]
        \label{eq6S}
    \end{equation}
    Thus, G-type spin arrangement on the Cu sublattice and ($+$$-$$-$$+$)-type spin arrangement on the Fe sublattice in CuFePO$_5$ leads to SSG $\bm{P\,^{1}2_{1}\,^{\overline{1}}2_{1}\,^{\overline{1}}2_{1}}$ (19.27) \cite{Litvin1977_sm, Chen2025_sm}. From Eqs.~\eqref{eq1}, \eqref{eq10} and \eqref{eq11} of main manuscript and eq.~\eqref{eq6S}, all the resulting  spin-degenerate nodal lines and nodal planes across the BZ can be systematically mapped out using the \textit{isomorphic} spin point groups, as discussed in  Sec.~\ref{AFiM} of main manuscript. \textcolor{black}{Figure~\ref{fig6S}(b)} shows the BZ along with the IBZ. First-principles calculations reveal fully compensated nature of both Cu and Fe sublattices which is evident from the spin-polarized total DOS as shown in \textcolor{black}{Fig.~\ref{fig6S}(c)}.  
    The resulting altermagnetic spin-splitting for these spin arrangements along the IBZ body diagonals is shown in \textcolor{black}{Figure~\ref{fig6S}(d)}. The spin-degeneracies and splittings on the high symmetry planes $k_x=0,\pm\sfrac{\pi}{a}$, $k_y=0,\pm\sfrac{\pi}{b}$ and $k_z=0,\pm\sfrac{\pi}{c}$, as enforced by the SSG symmetries, are displayed in \textcolor{black}{Figs.~\ref{fig6S}(e–m)}.

    \putbib[sm]
    
\end{bibunit}

\end{document}